\documentclass{JHEP}

\usepackage{latexsym}
\usepackage{graphics}

\def\comment #1{}
\def\cf {{\it cf. }}

\def\refer #1{{(\ref{#1})}}
\def\fullref #1{\ref{#1} (p.\pageref{#1})}

\def\bra #1{\left\langle{#1}\right|}
\def\ket #1{\left|{#1}\right\rangle}
\def\bracket #1#2{\left\langle{#1}|{#2}\right\rangle}

\def\hermitianMetricOperator {{\hat \eta}}

\def\of #1{\!\left({#1}\right)}

\def\N {\mathrm{I\!N}}
\def\R {\mathrm{I\!R}}

\def\grad {\nabla}
\def\gradOp {\hat\grad}

\def\set #1{\left\lbrace{#1}\right\rbrace}

\def\brackets #1{\left[{#1}\right]}
\def\braces #1{\left\lbrace{#1}\right\rbrace}

\def\order #1{\mathcal{O}\of{#1}}

\def\commutator #1#2{\brackets{{#1},{#2}}}
\def\antiCommutator #1#2{\braces{{#1},{#2}}}

\def\adjoint #1{{{{#1}^{\dag}}}}

\def\defas {:=}
\def\shallbe {\stackrel{!}{=}}
\def\equalby #1{\stackrel{\refer{#1}}{=}}

\def\onbComp {\mathrm{e}}
\def\onbAnnihilator {\mbox{$\hat \onbComp$}}
\def\onbCreator {\mbox{$\adjoint{\onbAnnihilator}$}}
\def\coordAnnihilator {\mbox{$\hat \mathrm{c}$}}
\def\coordCreator {\adjoint{\coordAnnihilator}}

\def\numberOperator {{\hat N}}

\def\manifold {\mathcal{M}}

\def\extd {\mathbf{{d}}}

\def\coextd {\adjoint{\extd}}

\def\clifford {{\mbox{$\hat \gamma$}}}

\def\Dirac {\mathbf{D}}

\def\eigenspace #1#2 {\mathrm{eig}\of{#1,#2}}

\def\fatomega {\mbox{\boldmath$\omega$}}

\def\fatDelta {\mbox{\boldmath$\Delta$}}

\newlength{\skiplength}
\def\skiph #1{\settowidth{\skiplength}{#1}\hspace{\skiplength}}

\def\inner {\!\cdot\!}

\def\coxeter {g^{\tt v}}

\title{Covariant Hamiltonian evolution in supersymmetric quantum systems}
\author{Urs Schreiber \\ Universit{\"a}t Duisburg-Essen \\ Essen, 45117, Germany\\
   E-mail: \email{Urs.Schreiber@uni-essen.de}}
\abstract{
We develop a general formalism for covariant Hamiltonian
evolution of supersymmetric (field) theories
by making use of the fact that these
can be represented on the
exterior bundle over their bosonic configuration space 
as generalized Dirac-K{\"ah}ler
systems of the form $(\mathbf d \pm \mathbf d^\dag)\ket{\psi} = 0$. 
By using suitable deformations of the supersymmetry generators
we find covariant Hamiltonians for target spaces with general
gravitational and Kalb-Ramond field backgrounds and discuss their
perturbation theory. 

As an example, these results are applied to the study of 
curvature corrections of superstring spectra for ${\rm AdS}_3\times {\rm S}^3$ 
close to its pp-wave limit.
}

\begin{document}

\setcounter{equation}{0}

\tableofcontents

\newpage
\section{Introduction}

Supersymmetric quantum theory and differential
(possibly non-commutative) geometry are in some sense two aspects of the same thing,
as has been
emphasized long ago
in articles such as \cite{Witten:1982} and 
\cite{FroehlichGrandjeanRecknagel:1996,FroehlichGrandjeanRecknagel:1997}.
 
This correspondence becomes however manifest only in the 
\emph{Schr{\"o}dinger representation} of quantum theory, 
 where states are expressed as functionals over
configuration space and operators act by (functional) multiplication and 
(functional) differentiation.\footnote{
  For a list of references on field theory in Schr{\"o}dinger representation see 
  for instance
  \cite{CorichiCortezQuevedo:2002}.
} For field theoretic applications
this representation is usually considered more awkward than the common Fock space representation,
however it certainly has also advantages, at least on the conceptual level.
In particular
Hamiltonians (and Hamiltonian \emph{constraints} for that matter) can be
identified with (generalized) \emph{Laplace operators} 
on configuration space, which makes manifest the connection between
quantum (field) theory and the geometry of configuration space. 
Supersymmetric quantum (field) theory furthermore
provides the corresponding \emph{Dirac operators}. 
In \cite{FroehlichGrandjeanRecknagel:1996,FroehlichGrandjeanRecknagel:1997}
it was stressed that the quantum theory provides thus nothing else but
\emph{spectral data} of configuration space.

Some illustrative examples of this correspondence have
been made explicit in \cite{Schreiber:2001}. The quantum theory of a
scalar field plus superpartner, as well as 
$N=1$, $d=4$ supergravity were represented on their respective configuration
space by generalized Dirac-K{\"a}hler equations. Schematically these look like
\begin{eqnarray}
  \left(\extd \pm \coextd\right)\ket{\psi} &=& 0
  \,.
\end{eqnarray}
Here $\extd$ denotes the (generalized) exterior derivative
on configuration space and $\coextd$ is its adjoint with respect to the
Hodge inner product on differential forms over configuration space.
The state $\ket{\psi}$ is a section of this exterior bundle, which
itself must be regarded as the superspace over the original bosonic configuration
space. 

But also the superconformal constraints of Type II strings are of this 
Dirac-K{\"a}hler form,
if $\extd$ is taken to be a generalized and deformed exterior derivative
over loop space, the configuration space of the string. For superstrings
in gravitational and Kalb-Ramond backgrounds this was perhaps first noticed in
\cite{Chamseddine:1997}. That the same holds true for all massless
NS backgrounds and that the deformation can always be written as 
a similarity transformation $\extd \to e^{-W}\extd e^{W}$ 
is shown in \cite{Schreiber:2004}. 
There are indications that this
statement indeed generalizes to all kinds of background fields
\cite{Giannakis:2002}.

The motivation for studying the Dirac-K{\"a}hler
representation of supersymmetric systems, and of superstrings
in particular, comes from its emphasis of the role of generalized
Dirac operators in these theories. The program of Connes' Noncommutative
Geometry \cite{Connes:1994} in the sense of a theory of spectral
geometry
shows that a great deal of information is encoded and can
be extracted from generalized Dirac operators. The relation between
supersymmetric physics and Noncommutative Geometry has been
particularly emphasized in 
\cite{FroehlichGrandjeanRecknagel:1996,FroehlichGrandjeanRecknagel:1997}
and the identification of the superstring's worldsheet supercharge with
a Dirac operator in a spectral triple has been used in 
\cite{Lizzi:1999,LizziSzabo:1998,FroehlichGawedzki:1993} 
to study stringy symmetries and dualities. However, with the advent of 
the M=Matrix proposal \cite{BanksFischlerShenkerSusskind:1996}
the emphasis on Noncommutative Geometry in the string theory literature
has shifted from its \emph{spectral} aspect, and hence the role of the
Dirac operator, towards its \emph{noncommutative} aspect and the role
of the algebra (for instance \cite{ConnesDouglasSchwarz:1998}).

One purpose of this paper and its companion \cite{Schreiber:2004} is
to demonstrate that an emphasis on the Dirac-Ramond operator of 
string theory, generalized to arbitrary backgrounds, 
allows to address current questions in string theory in an interesting
and worthwhile way. In particular in this paper we shall try to
address the general issue of computation of superstring spectra
in nontrivial and not exactly solvable backgrounds from
a perspective that puts the worldsheet supercharge and its role
as a Dirac operator on loop space in the center of attention. 
We describe a covariant formalism for the construction of superstring
Hamiltonians associated with arbitrary timelike Killing vectors of
target space. These Hamiltonians are to be regarded as NSR-string analogues
to the lightcone Hamiltonian of the Green-Schwarz string.

The latter has recently been used in the study of curvature
corrections to string spectra on ${\rm AdS}$ backgrounds. Motivated by
the insight \cite{BerensteinMaldacenaNastase:2002} that the string spectrum
on the pp-wave limit of such backgrounds corresponds directly to 
certain states in the dual field theory, as predicted by the 
AdS/CFT correspondence, several attempts are being made to extend this
result to higher order corrections by perturbatively computing the
spectrum on the string theory side 
\cite{ParnachevSahakyan:2002,ParnachevSahakyan:2002b,CallanLeeMcLoughlinSchwarzSwansonWu:2003}.

In all these approaches lightcone gauge needs to be fixed. As always, this
choice of gauge brings with it some simplifications but also a couple of
technical subtleties \cite{CallanLeeMcLoughlinSchwarzSwansonWu:2003}. In general
it is restricted to backgrounds that possess a lightlike Killing vector. 
In general, the quantum theory of the Green-Schwarz string is poorly understood.

The purpose of the following discussion is to analyze the question whether 
it is possible to construct appropriate Hamiltonians and their
perturbation theory not in the context of the Green-Schwarz string but 
in that of the
covariantly quantized NSR superstring with manifest worldsheet supersymmetry.
The first part of the paper, \S\fullref{section: covariant parameter evolution},
approaches this question in a general way by developing a covariant
Hamiltonian perturbation theory for any supersymmetric systems which are governed
by constraints of the Dirac-K{\"a}hler type. In the second part
\S\fullref{curvature corrections to superstring spectra} this general
theory is applied to the superstring by making use of results given
in \cite{Schreiber:2004}. As a first example application we test our
perturbation theory in a well-understood context, namely the pp-wave
limit of ${\rm AdS}_3 \times {\rm S}^3$, which can be regarded as a toy
example for the more interesting ${\rm AdS}_5 \times {\rm S}^5$
case \cite{ParnachevSahakyan:2002}.

The key ideas of the perturbation theory developed here are the following:

Given a (pseudo-)Riemannian manifold $(\manifold,g)$ the inhomogeneous differential forms
$\ket{\phi}$ over it form an inner product space ${\cal H}$ with respect to the
Hodge inner product $\bracket{\cdot}{\cdot}$. 
We consider physical systems modeled on such a space of states and governed by
constraints of the form
\begin{eqnarray}
  \label{summary: constraints}
  \Dirac^{(A)}_\pm \ket{\phi} &=& 0
  \,,
\end{eqnarray}
where $\Dirac_\pm^{(A)}$ are Dirac operators on ${\cal H}$, which are obtained from the
ordinary Dirac(-K\"ahler) operators $\Dirac_\pm = \extd \pm \coextd$ by
a deformation 
\begin{eqnarray}
  \extd^{(A)} &\defas& A^{-1}\extd\, A\,, \hspace{0.4cm} \coextd^{(A)}\defas (\extd^{(A)})^\dag
  \nonumber\\
  \Dirac_\pm^{(A)} &\defas& \extd^{(A)}  \pm \coextd^{(A)}
  \,,
\end{eqnarray}
where $A$ is any invertible operator on ${\cal H}$.
The relevance of these assumptions for string theory
lies in the fact that the RNS superstring 
in various backgrounds can be 
rewritten this way when $\manifold$ is taken to be loop space over spacetime.

After introducing analogous deformations of the form creation operators
$\coordCreator^\mu \defas dx^\mu \wedge$ and the associated Clifford
generators $\clifford_\pm \defas \coordCreator \pm \coordAnnihilator$
by setting
\begin{eqnarray}
  \coordCreator^{(A)} &\defas& A^{\dag -1}\coordCreator A^\dag
  \nonumber\\
  \clifford_\pm^{(A)} &\defas& \coordCreator^{(A)} \pm \coordAnnihilator^{(A)}
\end{eqnarray}
it is easy to see that the Lie derivative operator ${\cal L}_{v_0}$ along a
timelike Killing vector $v_0$ of $(\manifold,g)$ can be expressed as
\begin{eqnarray}
  {\cal L}_{v_0}
  &=&
  \frac{1}{4}
  \left(
    \antiCommutator{\clifford_+^{(A)}}{\Dirac^{(A)}_-}
    -
    \antiCommutator{\clifford_-^{(A)}}{\Dirac^{(A)}_+}
  \right)
  \,.
\end{eqnarray}
It follows that the constraints \refer{summary: constraints} imply 
a Schr\"odinger equation
\begin{eqnarray}
  i{\cal L}_{v_0}\ket{\phi}
  &=&
  {\bf H}^{(A)}\ket{\phi}
\end{eqnarray}
of evolution along the parameter $t_{v_0}$,  where the Hamiltonian is given by\footnote{
In the context of classical electromagnetism, to which the present formalism also applies
(\cf appendix \fullref{p-form electromagnetism}), this
operator is sometimes known as the Maxwell operator generating time evolution of the
electromagnetic field.
}
\begin{eqnarray}
  \label{summary: Hamiltonian}
  {\bf H}^{(A)}
  &\defas&
  \frac{i}{4}
  \left(
    \commutator{\clifford_-^{(A)}}{\Dirac^{(A)}_+}
    -
    \commutator{\clifford_+^{(A)}}{\Dirac^{(A)}_-}
  \right)
  \nonumber\\
  &=&
  \frac{i}{2}
  \left(
    \clifford_-^{(A)}\Dirac^{(A)}_+
    -
    \clifford_+^{(A)}\Dirac^{(A)}_-
  \right)
  +
  i {\cal L}_0
  \,.
\end{eqnarray}
For the deformations considered here there is a Krein space operator
$\hermitianMetricOperator$ 
(with $\hermitianMetricOperator^\dag = \hermitianMetricOperator$
and $\hermitianMetricOperator^2 = 1$) which
serves to define a positive definite scalar product\footnote{
The term $\delta\of{t_{v_0}}$ restricts integration to a hypersurfaces orthogonal
to the flow lines of the Killing vector $v_0$.}
$\bracket{\cdot}{\cdot}_\hermitianMetricOperator \defas 
\bra{\cdot}  \delta\of{t_{v_0}} \hermitianMetricOperator \ket{\cdot}$
on physical states 
with respect to which ${\bf H}^{(A)}$ is hermitian:\footnote{
The Hamiltonian ${\bf H}^{(A)}$ needs furthermore to commute with $t_{v_0}$.
For certain deformations this requires slight modifactions of the general argument
summarized here.
}
\begin{eqnarray}
  ({\bf H}^{(A)})^{\dag_\hermitianMetricOperator}
  \;\defas\;
  \hermitianMetricOperator ({\bf H}^{(A)})^\dag \hermitianMetricOperator
  \;=\;
  {\bf H}^{(A)}
  \,.
\end{eqnarray}

This allows to perform quantum mechanical perturbation theory
in a fully covariant framework.
Even though no light cone gauge is required 
a particularly simple
formula for the first order energy shift of a given state under a given perturbation of
the constraints \refer{summary: constraints} is obtained when $v_0$ is of the form
$
v_0 = e^\gamma p + e^{-\gamma} k 
$
for $p$ and $k$ two lightlike Killing vectors: In the limit $\gamma \gg 1$
we find for the first order shift of the light cone energy associated with $p$
the expression
\begin{eqnarray}
  e^{-\gamma}
  \bra{\phi^{(0)}}
  \hermitianMetricOperator
  \,
  ({\bf H}^{(A)})^{(1)}
  \ket{\phi^{(0)}}
  &\to&
  \bra{\phi^{(0)}}
  \hermitianMetricOperator
  \,
  i({\cal L}_{p})^{(1)}
  \ket{\phi^{(0)}}
\end{eqnarray} 
(where the $n$-th order perturbation of an object $O$ is written as $O^{(n)}$).

$\,$\\

The structure of this paper is as follows:

The central results concerning covariant Hamiltonian evolution in
supersymmetric systems are developed in \S\fullref{section: covariant parameter evolution}. 
First some
existing material on supersymmetric quantum theory in differential geometric
formulation is recalled in a coherent fashion in  
\S\fullref{on supersymmetric quantum theory in geometrical formulation}, 
where we
also elaborate on the general structure of deformations of the
supersymmetry algebra and on the differential geometric meaning of the
operators appearing in quantum SWZW models. 

These facts are then used in \S\fullref{subsection: covariant parameter evolution} 
to construct the general formalism
for covariant Hamiltonian evolution in backgrounds with arbitrary metric and 
NS 2-form fields. Finally the associated perturbation theory is developed.

In \S\fullref{curvature corrections to superstring spectra} 
these techniques are applied to the perturbative calculation of
superstring spectra. 
First \S\fullref{loop space and B-field background} 
reviews the loop space formulation
which puts the NSR string in the required Dirac-K{\"a}hler form. In
\S\fullref{review of AdS3} 
basic data of the ${\rm AdS}_3 \times {S}^3$ background
as well as its pp-wave Penrose limit are listed, which are then inserted
in \S\fullref{covariant calculation of AdS spectrum} 
into the perturbation formalism developed before. The
obtained perturbative spectrum of strings in this scenario is then
compared in \S\fullref{exact calculation of the spectrum} to the exact result 
and the implications of the
calculation are discussed in \S\fullref{discussion of perturbative result}.

Further details are given in the appendices:

Appendix \fullref{differential geometry} collects various objects that play 
a role in the formulation
of differential geometry in terms of operators on the exterior bundle,
which is the technical basis for the formulation of supersymmetric quantum
theory used here.
Most of this material is elementary and mainly meant to set up
notation and concepts, but also some not so widely known facts 
are derived and emphasized,
which are crucial for the developments in 
\S\fullref{section: covariant parameter evolution}. 

In appendix \fullref{proofs} proofs are given which are omitted 
for the sake of brevity in the main text.

Appendix \fullref{p-form electromagnetism} 
illustrates our formalism in terms of a well-known example to which 
it happens to apply, too, namely that of
classical electromagnetism in differential form language.

Finally appendix \fullref{Lie groups and algebras} 
lists some standard facts about Lie algebras that
are needed for the discussion of SWZW models in 
\S\fullref{WZW models}.

\newpage
\section{Covariant parameter evolution for supersymmetric quantum systems}
\label{section: covariant parameter evolution}

\subsection{On supersymmetric quantum theory in geometrical formulation}
\label{on supersymmetric quantum theory in geometrical formulation}

\subsubsection{Introduction}
\label{supersymmetric systems}
Let the (pseudo-)Riemannian manifold $(\manifold,g)$ be the configuration space of some
physical system. A supersymmetric extension of this system has as
configuration space the superspace $\cal S\manifold$ over $\manifold$, which can be
identified with 
$\Omega^1\of{\manifold}$, the 1-form bundle over $\manifold$. An
arbitrary quantum state of the supersymmetric system is therefore
a superfunction on $\Omega^1\of{\manifold}$, which is an inhomogeneous
differential form over $\manifold$, i.e.
an element of $\Gamma\of{\Omega\of{\manifold}}$, the space of sections of
the total form bundle $\Omega\of{\manifold} = \oplus_{p=0}^D \Omega^p\of{\manifold}$.

Before proceeding we briefly list some of the notation which will be used
frequently in the following. The details are given in 
\S\fullref{differential geometry}:

\begin{eqnarray}
  \begin{array}{rclll}
    \Gamma\of{\Omega\of{\manifold}} & & & \mbox{the space of sections of the exterior bundle}
     \\
    \bracket{\cdot}{\cdot} & & & 
       \mbox{the Hodge inner product on $\Gamma\of{\Omega\of{\manifold}}$}
       & \refer{Hodge inner product} \\
    \coordCreator^\mu &=& dx^\mu \wedge & 
     \mbox{operator of exterior multiplication by $dx^\mu$}
     & \refer{operator of exterior multiplication}\\
    \coordAnnihilator^\mu &=& dx^\mu \rightharpoonup & 
       \mbox{operator of interior multiplication}
      & \refer{ext multiplication CAR}\\
    \clifford_\pm^\mu &=& \coordCreator^\mu \pm \coordAnnihilator^\mu &
       \mbox{the associated Clifford algebra generators} &
      \refer{definition of the Clifford generators} \\
    \gradOp_\mu &=& \partial^c_\mu - \Gamma_\mu{}^\alpha{}_\beta 
      \coordCreator^\beta\coordAnnihilator_\alpha
     & \mbox{covariant derivative on $\Gamma\of{\Omega\of{\manifold}}$} &
     \refer{definition of action of covGradOp}
       \refer{definition partial derivative operators} \\
    \extd &=& \coordCreator^\mu\gradOp_\mu & 
        \mbox{the exterior derivative on $\Gamma\of{\Omega\of{\manifold}}$} 
     & \refer{definition of the exterior derivative}
     \\
    \coextd &=& - \coordAnnihilator^\mu \gradOp_\mu & 
       \mbox{its adjoint with respect to $\bracket{\cdot}{\cdot}$}
    & \refer{definition exterior coderivative in intro}
    \\
    \Dirac_\pm &=& \extd \pm \coextd & \mbox{the associated Dirac operators}
    & \refer{definition Dirac operator}\\
   {\cal L}_v &=& \antiCommutator{\extd}{v^\mu \coordAnnihilator_\mu}
   & \mbox{Lie derivative along $v$} & \refer{definition Lie derivative operator}\\
   \hermitianMetricOperator &&& \mbox{a Krein space operator}
     & \refer{naive and simple hermitian metric operator}
       \refer{general hermitian metric operator} \\
   \bracket{\cdot}{\cdot}_\hermitianMetricOperator &=& 
     \bra{\cdot}\hermitianMetricOperator \ket{\cdot} &
    \mbox{a scalar product on $\Gamma\of{\Omega\of{\manifold}}$}
     & \refer{modified inner product}
  \end{array}
  \nonumber
\end{eqnarray}

The systems of interest here will have semi-Riemannian configuration space
metric $g$ and be governed by sets of equations that are generalizations of
\begin{eqnarray}
  \label{constraints of susy system}
  \Dirac_+ \omega \;=\; 0 \;=\;  \Dirac_- \omega\,,\hspace{1cm}\omega\in
    \Gamma\of{\Omega\of{\manifold}}
\end{eqnarray}
(\cf \refer{definition Dirac operator}).
In the case that 
$\fatDelta = \Dirac_+^2$ (\cf \refer{definition Laplace-Beltrami operator}) is taken as a generator of ``time''-translations
in the system's parameter space the relations
\begin{eqnarray}
  \antiCommutator{\Dirac_+}{\Dirac_+} &=& 2\fatDelta
\end{eqnarray}
may be regarded as the $(N=1)$-supersymmetry algebra in 1 dimension. 
Similarly
\begin{eqnarray}
  \antiCommutator{\Dirac^i}{\Dirac^j} &=& 2\delta^{ij}\fatDelta
  \,,
\end{eqnarray}
where $i,j\in\set{1,2}$ and $\Dirac^1 \defas \Dirac_+$, $\Dirac^2\defas i\Dirac_-$,
is the 1-dimensional supersymmetry algebra with $N=2$, which is of course equivalent to
\begin{eqnarray}
  \label{d extd algebra once again}
  \antiCommutator{\extd}{\extd} &=& 0
  \nonumber \\
  \antiCommutator{\coextd}{\coextd} &=& 0
  \nonumber\\
  \antiCommutator{\extd}{\coextd} &=& \fatDelta
  \,.
\end{eqnarray}
This algebra gives us 1+0 dimensional supersymmetric field theory, i.e.
supersymmetric quantum mechanics.

Is there a deformation of $\extd$, $\coextd$ that turns this algebra into the
2-dimensional $(N=1)$-algebra?
Recall the central observation from the last part of \cite{Witten:1982}:

When choosing, as usual, the unitary representation 
of the Clifford algebra $\mathrm{Cl}\of{1,1}$
given by
\begin{eqnarray}
  \gamma^0_{AB} &=& 
  \left[
     \begin{array}{rr}
       0 & 1\\
       -1 & 0
     \end{array}
  \right]
  \nonumber\\
  \gamma^1_{AB} &=& 
  \left[
     \begin{array}{rr}
       0 & 1\\
       1 & 0
     \end{array}
  \right]
  \,,
\end{eqnarray}
and supercharges $Q$ that are real,
\begin{eqnarray}
  \label{d = 2 N = 1 susy algbra}
  \bar Q &=& Q^\mathrm{T} \gamma^0
  \,,
\end{eqnarray}
then the ``$QQ=P$'' bracket looks like
\begin{eqnarray}
  \antiCommutator{Q_A}{Q_B} 
  &=&
  -
  (\gamma^\mu\gamma^0)_{AB} P_\mu
  \nonumber\\
  &=&
  \left[
    \begin{array}{cc}
      P_0 + P_1 &      0 \\
             0      &  P_0 - P_1
    \end{array}
  \right]_{AB}
  \,.
\end{eqnarray}
In terms of the linear combinations
\begin{eqnarray}
  d_k &\defas& \frac{1}{\sqrt{2}}\left(Q_1 - i Q_2\right)
  \nonumber\\
  d^\ast_k &\defas& \frac{1}{\sqrt{2}}\left(Q_1 + i Q_2\right)
\end{eqnarray}
this is equivalent to
\begin{eqnarray}
  \antiCommutator{d_k}{d_k} &=& P_1
  \nonumber\\
  \antiCommutator{d^\ast_k}{d^\ast_k} &=& P_1
  \nonumber\\
  \antiCommutator{d_k}{d^\ast_k} &=& P_0
  \,.
\end{eqnarray}
This is almost of the form \refer{d extd algebra once again}, except that the 
$\antiCommutator{\extd}{\extd}$ and $\antiCommutator{\coextd}{\coextd}$ brackets
pick up a non-zero value equal to the generator $P_1$ of spatial translations. One
way to realize this deformation is the following:

\paragraph{Deformations by Killing vectors.}

In the presence of a Killing vector $k = k^\mu\partial_\mu$, one can
consider a deformation $\extd_k$ of the exterior derivative defined by
\begin{eqnarray}
  \label{k-deformed exterior derivatives}
  \extd_k 
   &\defas&
  \extd + i\coordAnnihilator_\mu k^\mu
  \,.
\end{eqnarray}
The adjoint operator is then
\begin{eqnarray}
  \label{k-deformed coexterior derivatives}
  \coextd_k
  &\defas&
  \coextd - i\coordCreator_\mu k^\mu
  \,.
\end{eqnarray}
By the definition of the Lie-derivative \refer{definition Lie derivative operator} one finds
\begin{eqnarray}
  \label{extdk squares to a Lies}
  \extd^2 &=& i{\cal L}_k
  \,,
\end{eqnarray}
and, since $k$ is Killing, by \refer{adjoint definition of Lie derivative} also
\begin{eqnarray}
  \label{coextdk squares to a Lies}  
  \coextd^2 &=& i{\cal L}_k
  \,.
\end{eqnarray}

Defining
\begin{eqnarray}
  \label{k deformed Dirac operators}
  \Dirac_{k,\pm} &=& \extd_k \pm \coextd_k
  \nonumber\\
  &=&
  \gamma_{\mp}^\mu \left(\gradOp_\mu \mp i k_\mu\right)
\end{eqnarray}
one has, with $A,B\in \set{+,-}$ and $s_\pm \defas \pm 1$,
\begin{eqnarray}
  \antiCommutator
    {\Dirac_{k,A}}
    {\Dirac_{k,B}}
  &=&
  2\delta_{AB}\left(s_A \fatDelta_k + i{\cal L}_k\right)
  \,,
\end{eqnarray}
where the deformed Laplace-Beltrami operator is
\begin{eqnarray}
  \label{Killing deformed Laplace Beltrami}
  \fatDelta_k
  &\defas&
  \antiCommutator{\extd_k}{\coextd_k}
  \nonumber\\
  &=&
  \fatDelta
  +
  k^2
  +
  i
  \left(
    \antiCommutator{\coextd}{\coordAnnihilator_\mu k^\mu}
    -
    \antiCommutator{\extd}{\coordCreator_\mu k^\mu}
  \right)
  \nonumber\\
  &=&
  \fatDelta
  +
  k^2
  -
  i
  (\partial_{[\mu} k_{\nu]})
  \left(
    \coordCreator^\mu\coordCreator^\nu
    +
    \coordAnnihilator^\mu\coordAnnihilator^\nu
  \right)
  \,.
\end{eqnarray}

Note that the deformed exterior differential operators still satisfy the
duality relation \refer{duality relation d extd}:
\begin{eqnarray}
  \coextd_k &=& -\bar\star\, \extd_k \,\bar\star
  \,.
\end{eqnarray}

This gives us the algeba of $d=2$, $N=1$ supersymmetry, which is necessary
to describe the manifestly worldsheet supersymmetric string. It is now
of interest how the generators of this algebra may be deformed in order
to incorporate the effect of various background fields, without 
affecting the structure of the algebra itself.

\subsubsection{Deformations of the supersymmetry generators.}
\label{deformations of the supersymmetry generators}

The construction of supersymmetric quantum theories usually
involves choosing a bosonic Lagrangian,	
replacing its fields with appropriate superfields, and integrating out the 
Grassmannian variables to obtain the supersymmetric Lagrangian of the
component fields, which may finally be quantized.
An alternative way
to obtain new supersymmetric quantum theories,
which shall be studied here,  is to pick a given one (for instance a simple, free theory)
and then deform its symmetry generators (for instance so as to introduce
interaction and potentials) in a way that
preserves the supersymmetry algebra. When working in the Schr\"odinger representation 
this may
radically reduce the computational effort and increase transparency, 
as will be demonstrated here.

Furthermore, more importantly for the purposes of 
\S\fullref{subsection: covariant parameter evolution}, 
such deformations
of the supersymmetry generators allow to deform other operators
analogously such that results derived in the undeformed case can be rather
straightforwardly adapted to the deformed case. This will be essential for the
construction of the covariant Hamiltonian for $b$-field backgrounds in
\S\fullref{Parameter evolution in the presence of a B-field}.

Below it is shown that this strategy involves a generalization of
the deformations already considered in the first part of \cite{Witten:1982}.
Applications of this method to the study of actual physical systems have been
rare, one example being \cite{BeneGraham:1994} \cite{Schreiber:2001} (and references given there), where
the method is applied to the study of supersymmetric quantum cosmology.
In \cite{Schreiber:2004} and \S\fullref{loop space and B-field background} it is shown 
that it is also applied with some profit to the fundamental string.

$\,$\\

We start by discussing deformations of the 1-dimensional supersymmetry algebra:

\paragraph{The case $D=1$, $N=2$.}

Recall from \refer{d extd algebra once again} that the $D=1,N=2$ supersymmetry algebra may be
represented by operators $\extd$ and $\coextd$ 
(usually, but not necessarily, similar or equal to  the exterior derivative and co-derivative),
which satisfy
\begin{eqnarray}
  \label{extd coext algebra still another time}
  \antiCommutator{\extd}{\extd} &=& 0
  \nonumber\\
  \antiCommutator{\coextd}{\coextd} &=& 0
  \nonumber\\
  \antiCommutator{\extd}{\coextd} &=& \fatDelta
  \,,
\end{eqnarray}
as well as
\begin{eqnarray}
  \label{adjoint relation between extd and coextd}
  (\extd)^\dag &=& \coextd
\end{eqnarray}
and therefore
\begin{eqnarray}
  \fatDelta^\dag &=& \fatDelta
  \,.
\end{eqnarray}

Given any such an algebra, we are now looking for a 1-parameter family of algebra homomorphisms
$h_\epsilon$, $\epsilon\in \R$, which are continuously connected to the identity
(i.e. $h_0$ is the identity operation) and which map these operators to 
\begin{eqnarray}
  \extd^\epsilon &\defas& h_\epsilon\of{\extd}
  \nonumber\\
  \coextd^\epsilon &\defas& h_\epsilon\of{\coextd}
  \nonumber\\
  \fatDelta^\epsilon &\defas& h_\epsilon\of{\fatDelta}
  \,,
\end{eqnarray}
in a way that preserves the relations \refer{extd coext algebra still another time} 
and \refer{adjoint relation between extd and coextd}.
It is very easy to see which kinds of $h_\epsilon$ are possible:

By assumption of continuity we have
\begin{eqnarray}
  \extd^\epsilon &=& \extd + \epsilon {\bf X} + \order{\epsilon^2}
  \,,
\end{eqnarray}
where {\bf X} is some operator to be determined. The algebra requires that
\begin{eqnarray}
  0 &=& (\extd^\epsilon)^2
  \nonumber\\
  &=&
  \epsilon\antiCommutator{\extd}{{\bf X}} + \order{\epsilon^2}
  \,,
\end{eqnarray}
and therefore that $\extd$ anticommutes with its first order deformation:
\begin{eqnarray}
  \label{extd anticommutes with its first order deformation}
  \antiCommutator{\extd}{{\bf X}} &=& 0
  \,.
\end{eqnarray}
Since $\extd$ is nilpotent ${\bf X}$ is locally ``exact''
\begin{eqnarray}
  \label{first order deformation of extd chosen extd-exact}
  {\bf X} &=& \commutator{\extd}{{\bf W}}
  \,,
\end{eqnarray}
where ${\bf W}$ is any even graded operator. Assuming that ${\bf X}$ is of this form we have
\begin{eqnarray}
  \frac{d}{d\epsilon}\extd^\epsilon
  &=&
  \commutator{\extd}{{\bf W}}
  \nonumber\\
  \Rightarrow
  \skiph{$\frac{d}{d\epsilon}\extd^\epsilon$}
  \extd^\epsilon &=&
  \exp\of{-\epsilon {\bf W}}\,\extd\,\exp\of{\epsilon {\bf W}}
  \,.
\end{eqnarray}
We call 
\begin{eqnarray}
  \label{the deformation operator}
  {\mathbf A} &\defas& \exp\of{\mathbf W}
\end{eqnarray}
the deformation operator.
The other deformed operators follow from this by
\begin{eqnarray}
  \coextd^\epsilon &\defas& (\extd^\epsilon)^\dag
  \nonumber\\
  &=&
  \exp\of{\epsilon {\bf W}^\dag}\,\coextd\,\exp\of{-\epsilon {\bf W}^\dag}
  \nonumber\\
  \fatDelta^\epsilon &\defas&
  \antiCommutator{\extd^\epsilon}{\coextd^\epsilon}
  \,.
\end{eqnarray}
Note that if ${\bf W}$ is antihermitian the deformation is a pure gauge
transformation.

\paragraph{Examples.}

\begin{enumerate}
\item
One example is the famous special case where ${\bf W} = W$ is the
operator of multiplication by the real function $W$, which has been used 
in \cite{Witten:1982} to study Morse theory. If $W$ were taken to be
purley imaginary, then ${\bf W}$ would be anti-hermitian and hence
correspond to a pure phase shift symmetry that could be gauged away.

\item
Consider the operators
\begin{eqnarray}
  \extd^0 &=& \coordCreator^a \partial_a
  \nonumber\\
  \coextd^0 &=& -\coordAnnihilator^a \partial_a 
\end{eqnarray}
on flat space $(\manifold,\eta)$ (where $\eta$ is the flat metric). 
Now pick a non-trivial metric $g$ 
which in the $\partial_a$-basis satisfies ${\rm det}\of{g} = 1$ and 
pick (locally) an associated vielbein $e^\mu{}_a$. Now there is an invertible
linear operator $\mathbf A$ defined by
\begin{eqnarray}
  {\mathbf A} \coordCreator^{a_1}\cdots \coordCreator^{a_p}\ket{1}
  &\defas&
  e^{\mu=a_1}{}_{b_1}\coordCreator^{b_1}\cdots 
  e^{\mu=a_p}{}_{b_p}
  \coordCreator^{b_p} \ket{1}
  \,.
\end{eqnarray}
In fact, when we regard $e^\mu{}_a$ as a matrix $e$ and let $\ln e$ be the
logarithm of that matrix, then $\mathbf A$ can be written in the form
\refer{the deformation operator} as
\begin{eqnarray}
  {\mathbf A} &=& \exp\of{\coordCreator\inner (\ln e)^{\rm T}\!\inner \coordAnnihilator}
  \,.
\end{eqnarray}
A little reflection shows that
\begin{eqnarray}
  \extd &=& {\mathbf A} \extd^0 {\mathbf A}^{-1}
\end{eqnarray}
is the operator representation of the exterior derivative 
on $(\manifold,g)$ and hence 
\begin{eqnarray}
  \coextd &=&  {\mathbf A}^{-1} \coextd^0  {\mathbf A}^\dag
\end{eqnarray}
is its adjoint.
(Of course $\extd$ as an abstract operator is independent of the
metric on $\manifold$, but its representation in terms of operators
$\coordCreator^a$ and $\partial_a$ is not. Compare \S\fullref{differential operators}.)

This way we can understand the metric field on the manifold as inducing a
deformation of the supersymmetry generators of flat space. This will be seen
to be a general phenomenon. In \S\fullref{example background b field} it is shown how 
similarly a Kalb-Ramond field background is represented by an exponential deformation
operator.

\item
  The above restriction to ${\rm det}\of{g} = 1$ ensures that the 
inner product $\bracket{\cdot}{\cdot}$ and hence the adjoint operation
${(\cdot)}^\dag$ itself receives no deformation (this follows from
equations 
\refer{Hodge inner product}
and
\refer{adjoint relation of partial}
that are given in the appendix).
Alternatively one can allow a conformal factor $e^{2\phi}$ but still keep the
undeformed Hodge inner product. This describes a dilaton background:
\begin{eqnarray}
  {\bf A} &=& \exp\of{\phi \coordCreator^\mu \coordAnnihilator_\mu}
  \;=\;\exp\of{\phi \numberOperator}
  \,.
\end{eqnarray}
(If instead the inner product is accordingly modified this $\bf A$
induces an ordinary conformal transformation. This is discussed in 
\S\fullref{conformal (Weyl) transformations})

\end{enumerate}

The example of central importance for the following is the case
${\bf A} = \exp\of{\frac{1}{2}b_{\mu\nu}\coordCreator^\mu\coordCreator^\mu}$,
which induces a Kalb-Ramond field background. This is discussed in more detail
for the $D=2$ supersymmetry algebra in \S\fullref{example background b field}.

Together with the metric and dilaton backgrounds discussed above this shows
that all the massless NS-NS backgrounds of the superstring find their natural
realization in terms of deformations of the supersymmetry generators.

$\,$\\

The results for the 1-dimensional supersymmetry algebra straightforwardly
carry over to two dimensions:

\paragraph{The case $D=2$, $N=1$.}
Now the supersymmetry algebra looks like (\cf \refer{extdk squares to a Lies},
\refer{coextdk squares to a Lies}, \refer{Killing deformed Laplace Beltrami})
\begin{eqnarray}
  \label{N=1 D=2 susy algebra}
  \extd_k^2 &=& i{\cal L}_k
  \nonumber\\
  \coextd_k^2 &=& i{\cal L}_k
  \nonumber\\
  \antiCommutator{\extd_k}{\coextd_k} &=& {\fatDelta}_k
  \,.
\end{eqnarray}
We shall restrict attention to homomorphimsm $h_\epsilon$ that leave
the element ${\cal L}_k$ invariant\footnote{
When applied to the string, ${\cal L}_k$ will be the generator of reparameterizations
along the string. Because the string must be reparameterization invariant in
any background this generator must be preserved by the deformation.
}:
\begin{eqnarray}
  h_\epsilon\of{{\cal L}_k} &\shallbe& {\cal L}_k\,,\hspace{1cm}\forall\,\epsilon
  \,.
\end{eqnarray}
The analysis then closely parallels that of the $(D=1,N=2)$-case:

Setting again
\begin{eqnarray}
  \extd_k^\epsilon &=& \extd_k + \epsilon{\bf X} + \order{\epsilon^2}
\end{eqnarray}
one obtains from
\begin{eqnarray}
  i{\cal L}_k &\shallbe& (\extd^\epsilon_k)^2
  \nonumber\\
  &=&
  i{\cal L}_k + \epsilon\antiCommutator{\extd_k}{{\bf X}} + \order{\epsilon^2}
\end{eqnarray}
the already familiar condition
\begin{eqnarray}
  \antiCommutator{\extd_k}{{\bf X}} &=& 0
  \,.
\end{eqnarray}
But now $\extd_k$ is nilpotent only modulo ${\cal L}_k$, so this is solved in
analogy with \refer{first order deformation of extd chosen extd-exact} by
setting
\begin{eqnarray}
  {\bf X} &=& \commutator{\extd_k}{{\bf W}}
\end{eqnarray}
subject to the condition that
\begin{eqnarray}
  \commutator{{\cal L}_k}{{\bf W}} &=& 0
  \,.
\end{eqnarray}
As before, the family of homomorphisms is therefore given by
\begin{eqnarray}
  \label{algebra homomorphism for d=2 N= 1}
  \extd_k^\epsilon &\defas&
  \exp\of{-\epsilon{\bf W}}\extd_k\exp\of{\epsilon{\bf W}}
  \,,\hspace{1.5cm}\commutator{{\bf W}}{{\cal L}_k} = 0
  \nonumber\\
  \coextd_k^\epsilon &\defas&
  \exp\of{\epsilon{\bf W}^\dag}\coextd_k\exp\of{-\epsilon{\bf W}^\dag}  
  \nonumber\\
  {\cal L}_k^\epsilon &\defas& {\cal L}_k
  \nonumber\\
  \fatDelta_k^\epsilon &\defas&
  \antiCommutator{\extd_k^\epsilon}{\coextd_k^\epsilon}
  \,.
\end{eqnarray}
Note that 
\begin{eqnarray}
  \label{deformed extdk as deformed extd plus deformed annihilator}
  \extd_k^\epsilon &=& \extd^\epsilon + h_\epsilon\of{ik_\mu \coordAnnihilator^\mu}
  \nonumber\\
  \coextd_k^\epsilon &=& \coextd^\epsilon - h_\epsilon\of{ik_\mu \coordCreator^\mu}
  \,.
\end{eqnarray}

Such deformations of the $D=2$ $N=1$ supersymmetry algebra will play an
important role in the following constructions. We will show that choosing
$\mathbf W$ to be a 2-form (a ``$b$-field'') gives the supersymmetry constraints associated
with a Kalb-Ramond field background. The fact that these relatively
complicated constraints can be obtained from algebraically simple deformations 
of the form \refer{algebra homomorphism for d=2 N= 1} will make it possible
to systematically generalize results pertaining to vanishing 2-form backgrounds
to non-vanishing 2-form backgrounds. In particular this will allow us
to adapt the construction of the Hamiltonian generator for pure metric
backgrounds derived in \S\fullref{Target space Killing evolution} 
to that for $g$- and $b$-field 
backgrounds in 
\S\fullref{Parameter evolution in the presence of a B-field}. 
This task would have been rather unfeasible
in terms of the complicated expanded form of the supersymmetry
generators (see below).

\subsubsection{Deformation by background $B$-field.}
\label{example background b field}

In this section it is shown how a Kalb-Ramond background gives rise to a deformation 
as discussed above.

Consider the case where on $\manifold$ there is, in addition
to the metric $g$
(admitting the Killing vector $k$) an antisymmetric 2-form field
\begin{eqnarray}
  b &\defas& \frac{1}{2}b_{\mu\nu}dx^\mu\wedge dx^\nu
\end{eqnarray}
with field strength
\begin{eqnarray}
  h_{\mu\nu\rho} &\defas& (\extd b)_{\mu\nu\rho}\;=\; 3\partial_{[\mu}b_{\nu\rho]}
  \,.
\end{eqnarray}

In order to couple this background field to our system \refer{N=1 D=2 susy algebra}, 
it is natural to set in \refer{algebra homomorphism for d=2 N= 1}
\begin{eqnarray}
  \label{definition of b-deformation exponent}
  {\bf W}^{(b)} &\defas& \frac{1}{2}b_{\mu\nu}\coordCreator^\mu\coordCreator^\nu
  \,.
\end{eqnarray}
For this choice one finds
\begin{eqnarray}
  \label{susy generators for B field background}
  \extd_k^{\epsilon=1}\;\defas\; \extd_k^{(b)}
  &=&
  \extd_k
  +
  \frac{1}{6}\coordCreator^\mu\coordCreator^\nu\coordCreator^\rho h_{\mu\nu\rho}
  +
  i k^\mu b_{\mu\nu}\coordCreator^\nu
  \nonumber\\
  &=&
  \coordCreator^\mu \gradOp_\mu
  +
  \frac{1}{6}\coordCreator^\mu\coordCreator^\nu\coordCreator^\rho h_{\mu\nu\rho}
  +
  ik^\mu\left(
    g_{\mu\nu}\coordAnnihilator^\nu + b_{\mu\nu} \coordCreator^\nu
  \right)
  \nonumber\\
  \coextd_k^{\epsilon=1}\;\defas\; \coextd_k^{(b)}
  &=&
  \coextd_k
  -
  \frac{1}{6}\coordAnnihilator^\mu\coordAnnihilator^\nu\coordAnnihilator^\rho h_{\mu\nu\rho}
  -
  i k^\mu b_{\mu\nu}\coordAnnihilator^\nu
  \nonumber\\
  &=&
  -\coordAnnihilator^\mu \gradOp_\mu
  - \frac{1}{6}\coordAnnihilator^\mu\coordAnnihilator^\nu\coordAnnihilator^\rho h_{\mu\nu\rho}
  -ik^\mu
  \left(
    g_{\mu\nu}\coordCreator^\nu + b_{\mu\nu} \coordAnnihilator^\nu    
  \right)
  \,.
\end{eqnarray}
The first part of these expressions, the one coming from the deformation of
the exterior derivative itself,
was already considered in \cite{FroehlichGrandjeanRecknagel:1997}
(p. 25) as an example for a supersymmetric quantum theory involving torsion.
We here note that the $b$-field deformation of the full $D=2$ supersymmetry algebra
\refer{k-deformed exterior derivatives} and \refer{k-deformed coexterior derivatives}
in addition gives the terms proportional to $k^\mu$ on the right of
\refer{susy generators for B field background}. It turns out that these
are precisely the terms needed to identify the generators in
\refer{susy generators for B field background} with the supersymmetry generators of
the $D=2$, $N=1$ nonlinear supersymmetric sigma model 
which describes superstring propagation in the respective
$b$-field background, \cf \S\fullref{loop space and B-field background}.  
We thus have found an algebraic way to
derive the constraints of the NSR superstring in gravitational
and Kalb-Ramond backgrounds. Knowledge of the deformation operator
$\exp\of{\frac{1}{2}b_{\mu\nu}\coordCreator^\mu \coordCreator^\nu}$
allows us to algebraically relate these constraints to the
ordinary $\extd_k$ and $\coextd_k$ operators. 

In order to further analyze the result \refer{susy generators for B field background}
note that the corresponding Dirac operators are
\begin{eqnarray}
  \label{b-deformed Dirac operators}
  \Dirac_{k\mp}^{(b)}
  &=&
  \clifford_\pm^\mu
  \left(
  \gradOp^{(b)}_\mu
  -
  i(b_{\mu\nu} \mp g_{\mu\nu})k^\nu
  \right)
  -
  \frac{1}{12}
  h_{abc}
  \clifford_\pm^a\clifford_\pm^b\clifford_\pm^c
  \,.
\end{eqnarray}
Here we have identified
$\gradOp^{(b)}_\mu$ 
as a deformation of the covariant derivative operator
\begin{eqnarray}
  \label{b-deformed covariant derivative operator}
  \gradOp_\mu^{(b)}
  &\defas&
  \partial_\mu
  +
  \frac{1}{4}\omega^+_{\mu ab}
  \clifford^{a+}\clifford^{b+}
  -
  \frac{1}{4}\omega^-_{\mu ab}
  \clifford^{a-}\clifford^{b-}
  \nonumber\\
  &=&
  \gradOp_\mu
  +
  \frac{1}{4}
  h_{\mu ab}
  \left(
    \onbCreator^a \onbCreator^b + \onbAnnihilator^a \onbAnnihilator^b
  \right)
  \,,
\end{eqnarray}
which involves  connections with torsion $\pm \frac{1}{2}h$
\begin{eqnarray}
  \omega^\pm_{abc}
  &\defas&
  \omega_{abc}
  \pm
  \frac{1}{2}
  h_{abc}
\end{eqnarray}
(\cf \S\fullref{Dirac, Laplace-Beltrami, and spinors} and \S\fullref{Torsion}),
and
which acts on the Clifford algebras $\clifford^{\pm}$ as the covariant
derivative associated with the connections with torsion $\omega^\pm$, respectively:
\begin{eqnarray}
  \label{action of b-deformed covariant derivative on pm cliffords}
  \commutator{\nabla^{(b)}_\mu}
 {v_a \clifford^{a\pm}}
  &=&
  \left(
    \nabla^\pm_\mu v_a
  \right)
  \clifford^{a\pm}
  \,.
\end{eqnarray}

Its commutators give the torsion deformed curvature operator
(\cf \refer{definition Riemann operator})
\begin{eqnarray}
  \label{torsion deformed curvature operator}
  {\bf R}^{(h)}_{\mu\nu}
  &\defas&
  \commutator{\gradOp_\mu^{(b)}}{\gradOp_\nu^{(b)}}
  \nonumber\\
  &=&
  {\bf R}_{\mu\nu}
  + 
  \frac{1}{8}
  (\nabla_{[\mu}h_{\nu] ab} )
  \left(
    \onbCreator^a \onbCreator^b + \onbAnnihilator^a \onbAnnihilator^b
  \right)
  + 
  \frac{1}{4}
  h_{\mu ac}h_{\nu}{}^c{}_b \,\onbCreator^a\onbAnnihilator^b
  \,. 
\end{eqnarray}
This expression vanishes iff $\frac{1}{2}h$ is the parallelizing torsion
(see \refer{Riemann tensor with torsion} in appendix \fullref{Torsion} )
\emph{and} 
$\nabla_{[\mu}h_{\nu] ab} = 0$. 
This is of course true if $h_a{}^b{}_c$ are the structure constants
of a group manifold, which is an interesting special case to which
we now turn:

\subsubsection{SWZW models}
\label{WZW models}

For $\manifold$ a Lie group manifold and $h = db$ twice
its parallelizing torsion
the above construction reduces to that of super Wess-Zumino-Novikov-Witten models.
These are of course well known, but because
SWZW models will play an important role as exactly solvable backgrounds
from which our perturbation theory may proceed and since
we will need the special
representation \refer{total WZW currents are deformed Lie derivatives}, 
to be derived below,
of the currents in terms of Lie derivatives
on spinors, this section spells out some aspects of SWZW models
in terms of the formalism used here.\footnote{
A standard text on the ordinary $2D$ WZW model is \cite{DiFrancescoMathieuSenechal:1997}.
The original supersymmetric extension of the WZW model was given in
\cite{VecchiaKnizhnikPetersenRossi:1985}.
WZW models with extended supersymmetry are discussed in 
\cite{SpindelSevrinTroostVanProeyen:1988} and
\cite{SpindelSevrinTroostVanProeyen:1988b}. We mostly follow the treatment 
in \cite{FroehlichGawedzki:1993}. }

Another purpose of this section is to put the general construction of
Hamiltonian generators in \S\fullref{Target space Killing evolution} 
into perspective: 
As discussed below (see 
\refer{definition of bsosnic current as superpartner of fermion in general formalism} and 
\refer{total WZW currents are deformed Lie derivatives}) 
the anticommutator of the Dirac operators \refer{b-deformed Dirac operators} with
the Clifford generators associated with the invariant vielbein field $e_a$ of the
group manifold gives the ``total current'' operators, which are,
however, essentially (up to a spurious term proportional to $k^\mu$)
Lie derivative operators along $e_a$. Accordingly the respective
\emph{commutator} gives the associated ``Hamiltonian'' (by the
general scheme that will be discussed in 
\S\fullref{Target space Killing evolution}). 
Except for the spurious term this is
hence already almost what we are looking for. The constructions in
\S\fullref{Parameter evolution in the presence of a B-field} 
may therefore also be regarded as a generalization of the concept of
``currents'' on group manifolds to more general backgrounds.

$\,$\\

Equation \refer{action of b-deformed covariant derivative on pm cliffords} 
shows that a special case of high symmetry is one where
there is a $b$-field with field strength $h$ and a metric $g$ such that
two vielbein fields $e^\pm_a$ exist, which are parallel with respect to the
connections with torsion:
\begin{eqnarray}
  \nabla^\pm_\mu
  e^\pm_a
  &=&
  0
  \,.
\end{eqnarray}
According to a general fact about Lie groups
(see \refer{parallelity of the invariant vielbeins} in appendix
\fullref{Lie groups and algebras}), 
this is true when $g$ is the Killing metric on a group manifold and
$h_{abc}$ is proportional to the structure constants of that group:
\begin{eqnarray}
  \commutator{\nabla_\mu^{(b)}}{\left(e_a^{\pm}\right)^\mu \clifford_{\mu\pm}}
  &=&
  0
  \,.
\end{eqnarray}  
Note that, by 
\refer{Lie connection and torsion},
there exists a 3-form $h_{\mu\nu\lambda}$ such that
\begin{eqnarray}
  \label{vanishing of the connections with torsion in WZW models}
  \omega^\pm[e^\pm]_{a b c} &=& 0
  \,.
\end{eqnarray}
Inserting this into 
\refer{b-deformed covariant derivative operator} 
gives the $b$-deformed covariant derivative
operator
\begin{eqnarray}
   \label{b-deformed cov der operator in invariant bases}
  e^\sigma_a{}^\mu\gradOp^{(b)}_{\mu}
   &=&
  e^{\sigma}_a{}^\mu
  \left(
    \partial^\sigma_{\mu}
  -\sigma
  \frac{1}{4}
  \omega^{-\sigma}[e^\sigma]_{\mu b c}
  \clifford_{-\sigma}^{b}
  \clifford_{-\sigma}^{c}
  \right)
  \,,
\end{eqnarray}
where $\sigma = +1$ or $\sigma = -1$
and $\partial^\sigma_{\mu}$ is the partial derivative operator that
commutes with $e^\sigma_a{}^\mu \clifford_\mu$ (but, in general, 
not with $e^{-\sigma}_a{}^\mu\clifford_\mu$),
\cf \refer{proprties of partial op}.)

This expression makes it manifest that this covariant derivative
commutes with all the Clifford generators associated with 
the vielbein fields $e^\pm$:
\begin{eqnarray}
  \commutator
    {\gradOp^{(b)}_{\mu}}
    {e_a^\sigma{}^\mu\clifford_\sigma^{\mu}}
  &=&
  0\,.
\end{eqnarray}
Because of the relation
\begin{eqnarray}
  \commutator
    { e^\sigma_a{}^\mu\gradOp^{(b)}_{\mu} }
    {e^\sigma_b{}^\nu\gradOp^{(b)}_{\nu} }
   &=&
   f_{a}{}^{c}{}_{b} e^\sigma_c{}^\mu\gradOp^{(b)}_{\mu} 
\end{eqnarray}
\begin{eqnarray}
  \commutator
    { e^\sigma_a{}^\mu\gradOp^{(b)}_{\mu} }
    { e^{-\sigma}_b{}^\nu\gradOp^{(b)}_{\nu} }
   &=&
   \commutator{e^\sigma_a}{e^{-\sigma}_b}
   \nonumber\\
   &=&
   0
\end{eqnarray}
it now follows that the $\pm$-components of the model completely decouple:
First of all we have
\begin{eqnarray}
  \antiCommutator{\Dirac_{k,\sigma}^{(b)}}{
    e^\sigma{}^a{}_\mu\clifford_\sigma^{\mu}}
  &=&
  0
  \nonumber\\
  \commutator{\Dirac_{k,\sigma}^{(b)}}{e^\sigma_a{}^\mu \gradOp_{\mu}^{(b)}}
  &=&
  0
  \,.  
\end{eqnarray}
The remaining non-vanishing anticommutator defines the
``total currents''
which are hence the superpartners of the fermions $\clifford_\sigma^{a}$:
\begin{eqnarray}
  \label{definition of bsosnic current as superpartner of fermion in general formalism}
  J^{\sigma}_{a}
  &\defas&
  \antiCommutator
  {\Dirac_{k,{-\sigma}}^{(b)}}
  {  \frac{\sigma}{2}
    e_a^\sigma{}^\mu \clifford_{\mu,\sigma}}
  \nonumber\\
  &=&
  J^{\rm bos \sigma}_{a}
  +
  J^{\rm fer \sigma}_{a}
  \,,
\end{eqnarray}
where the bosonic currents
$J^{\rm bos}$ and the fermionic currents $J^{\rm fer}$ are defined by
\begin{eqnarray}
  \label{first general definition of the currents}
  J^{\rm bos \sigma}_{a}
  &\defas&
  e^\sigma_a{}^\mu
  \left(
    \gradOp^{(b)}_{\mu}
    -
    i \left(b_{\mu \nu} -\sigma g_{\mu \nu}\right)k^{\nu}
  \right)
  \nonumber\\
  J^{\rm fer \sigma}_{a}
  &\defas&
  -
  e^\sigma_a{}^\mu
  \frac{1}{4}h_{\mu b c}\clifford_\sigma^{b}\clifford_\sigma^{c}
  \nonumber\\
  &=&
  \sigma
  e^\sigma_a{}^\mu
  \frac{1}{2}\omega[e^\sigma]_{\mu b c}\clifford_\sigma^{b}\clifford_\sigma^{c}
  \,.
\end{eqnarray}

 Using  \refer{b-deformed cov der operator in invariant bases} 
and \refer{first general definition of the currents} 
one finds that the total current is, up to the $k$-dependent term,
the Lie derivative operator (\cf \refer{Lie deriv operator in ONB form}) 
along $e^\sigma_a$:
\begin{eqnarray}
  \label{total WZW currents are deformed Lie derivatives}
  J^{\sigma}_{a}
  &=&
  \underbrace{
  e^\sigma_a{}^\mu
  \left(
  \partial^\sigma_{\mu}
  +
  \frac{1}{2}\omega[e^\sigma]_{\mu b c}
  \left(
    \clifford_+^{b}\clifford_+^{c}
    -
    \clifford_-^{b}\clifford_-^{c}
  \right)
  \right)
  }_{
    = {\cal L}_{e^\sigma_a}
  }
  -
  i
  e^\sigma_a{}^\mu 
  \left(b_{\mu \nu} \mp g_{\mu \nu}\right)k^{\nu}
  \,.
\end{eqnarray}
In particular, since all the $e^\sigma_a$ are Killing vectors, this Lie derivative
operator splits into two Lie derivative operators on the $\pm$-spinors,
as shown in \refer{Killing spinor Lie derivative} of 
\S\fullref{Dirac, Laplace-Beltrami, and spinors}.

This is what should be compared with the general expression 
\refer{b-deformation of Killing vector by clifford anticommutators}
of the Lie derivative in terms of anticommutators of fermions with the supercharges
that is derived below in \S\fullref{Parameter evolution in the presence of a B-field}.
In fact, in the case where the spurious term vanishes
\begin{eqnarray}
  e^\sigma_a{}^\mu 
  \left(b_{\mu \nu} \mp g_{\mu \nu}\right) = 0  
\end{eqnarray}
for some index ${}_a$, the covariant Hamiltonian constructed there is
precisely the Hamiltonian associated with the time parameter flowing along
$e^\sigma_a$. 

We have emphasized the applicability of the present formalism to the
superstring. But it should be stressed that it is in fact more general.
Indeed we have not even specified the precise form of the Killing
vector $k$, yet. For any such $k$ we get from
\refer{commutator of biliniear fermionic currents on goups manifolds}
for the fermionic currents the 
commutation relations
\begin{eqnarray}
  \label{algebra of fermionic currents}
  \commutator{J^{\rm fer \sigma}_{a}}{J^{\rm fer\sigma}_{b}}
  &=&
  f_{a}{}^{c}{}_{b}
  J^{\rm fer \sigma}_{c}
  \nonumber\\
  \commutator{J^{\rm fer\pm}_{a^\pm}}{J^{\rm fer\pm}_{b^\mp}}
  &=&
  0
  \,,
\end{eqnarray}
i.e. a representation of the Lie algebra.
Furthermore we generally get for the $b$-deformed Laplace-Beltrami operator
on SWZW backgrounds the simple expression
\begin{eqnarray}
  \left(\Dirac^{(b)}_{-\sigma}\right)^2
  &=&
  \sigma
  \left(
  g^{a b}
  \gradOp^{(b)}_{a}
  \gradOp^{(b)}_{b}
  -
  \frac{1}{12}\coxeter d
  \right)
 \,,
\end{eqnarray}
which is, up to a sign and a scalar shift, the quadratic Casimir of
the group.
This operator manifestly commutes with the fermions
\begin{eqnarray}
  \commutator{\left(\Dirac^{(b)}\right)^2}{e^\sigma_{a\mu}\clifford_\sigma^{\mu}}
  &=& 
  0
\end{eqnarray}
from which it follows that the total currents commute with $\Dirac_\pm^{(b)}$:
\begin{eqnarray}
  \commutator{\Dirac^{(b)}_{-\sigma}}{J^\sigma_{a}}
  &=&
  0
  \,.
\end{eqnarray}

However the bosonic analog of \refer{algebra of fermionic currents}, namely
\begin{eqnarray}
  \label{current algebra of the formal bosonic currents}
  \commutator
   {J^{\rm bos \sigma}_{a}}
   {J^{\rm bos \sigma}_{b}}
   &=&
   f_{a}{}^{c}{}_{b}
   J^{\rm bos \sigma}_{c}
   \nonumber\\
  \commutator
   {J^{\rm bos \sigma}_{a^{\sigma}}}
   {J^{\rm bos \,-\sigma}_{b^{-\sigma}}}
   &=&
   0
\end{eqnarray}
holds only for special $k$ (in particular for the trivial case $k=0$). 
In \S\fullref{loop space and B-field background}
it is discussed
how the $k$-vector field generating reparameterizations on loop space 
has this property and hence how the familiar $D=2$ SWZW model is
reobtained from the present approach.

SWZW backgrounds of course play an important role in practice because due to their
high symmetry they allow exact solutions. They are therefore natural starting points
for any perturbation theory in the background fields. Thus the typical perturbative
calculation along the lines of \S\fullref{Perturbation of background fields} 
and \S\fullref{qm perturbation theory} below
will start with an exactly known spectrum on an SWZW background and then perturb
$g_{\mu\nu}$ and $b_{\mu\nu}$ away from that. In the next section we develop the
general Hamiltonin formalism needed for such calculations.

\newpage

\subsection{Covariant parameter evolution}
\label{subsection: covariant parameter evolution}

\subsubsection{Target space Killing evolution}
\label{Target space Killing evolution}

We now set out to develop a machinery of parameter evolution
obtained from supersymmetry constraints, which will be the basis of a covariant 
perturbation theory for systems described by such constraints.

\paragraph{Parameter evolution from the constraints.}

A generator of (target space) time evolution can be obtained from constraint equations 
of the form \refer{constraints of susy system} if
$(\manifold,g)$ admits a timelike Killing vector $v_0$: The observable associated 
with the ``observer'' $v_0$ is ${\cal L}_{v_0}$ 
(and, in general, not $v_0^\mu\partial_\mu$ or $v_0^\mu\partial^c_\mu$ or the like),
since this is invariantly defined and furthermore ``gauge invariant'' 
in the sense that it commutes with the constraints:
\begin{eqnarray}
  \label{the timelike Killing vector}
  \commutator{\Dirac_\pm}{{\cal L}_{v_0}} &=& 0
  \,.
\end{eqnarray}

At this point we first assume that $\Dirac_\pm = \extd \pm \coextd$ 
are the Dirac operators associated with the ordinary, undeformed,
exterior derivative. The following construction will then be generalized
step by step to the deformed cases.

Since $v_{0}\inner\clifford_\pm$ is an invertible operator, the equivalences
\begin{eqnarray}
  \label{constraints as evolution equations}
  &&\Dirac_\pm\,\omega \;=\; 0
  \nonumber\\
  &\Leftrightarrow&
  v_{0}\inner \clifford_\mp\, \Dirac_\pm\, \omega \;=\; 0
  \nonumber\\
  &\Leftrightarrow&
  \left(v_{0}\inner\clifford_+\,\Dirac_- \pm v_{0}\inner\clifford_-\, \Dirac_+\right)\omega \;=\; 0
  \nonumber\\
  &\Leftrightarrow&
  \left(
    \antiCommutator{v_{0}\inner\clifford_+}{\Dirac_-} 
     \pm 
    \antiCommutator{v_{0}\inner\clifford_-}{\Dirac_+}
  \right)\omega \;=\;
  -
  \left(
    \commutator{v_{0}\inner\clifford_+}{\Dirac_-} 
     \pm 
    \commutator{v_{0}\inner\clifford_-}{\Dirac_+}
   \right)\omega
  \nonumber\\
  &\stackrel{\refer{one particular consequence}\refer{for the other sign}}{\Leftrightarrow}&
  \left\lbrace
    \begin{array}{rcl}
      2 (\partial_{[\mu}v_{0\nu]})\left(
       \coordCreator^\mu\coordCreator^\nu + \coordAnnihilator^\mu\coordAnnihilator^\nu
      \right)\;\omega
      &=& -
          \left(
           \commutator{v_{0}\inner\clifford_+}{\Dirac_-} 
           +
           \commutator{v_{0}\inner\clifford_-}{\Dirac_+}
         \right)\omega 
      \\
      4{\cal L}_{v_0}\;\omega &=& 
          -\left(
           \commutator{v_{0}\inner\clifford_+}{\Dirac_-} 
           -
           \commutator{v_{0}\inner\clifford_-}{\Dirac_+}
         \right)\omega 
    \end{array}
  \right.
\end{eqnarray}
hold.
The last line of  \refer{constraints as evolution equations} has the form
of a Schr\"odinger equation
\begin{eqnarray}
  \label{Dirac-Schroedinger equation}
  i{\cal L}_{v_0}\,\omega &=& {\bf H}_{v_0}\,\omega
  \,,
\end{eqnarray}
where the Hamiltonian ${\bf H}_{v_0}$ is defined by
\begin{eqnarray}
  \label{definition of the Hamiltonian}
  {\bf H}_{v_0}
  &\defas&
  \frac{i}{4}
          \left(
           \commutator{v_{0}\inner\clifford_-}{\Dirac_+}
           -
           \commutator{v_{0}\inner\clifford_+}{\Dirac_-} 
         \right)
  \nonumber\\
  &=&
  \frac{i}{2}
  \left(
    \commutator{v_0\inner \coordCreator}{\coextd} - 
    \commutator{v_0\inner \coordAnnihilator}{\extd}
  \right)
  \nonumber\\
  &=&
  \frac{i}{2}
  \left(
    v_0 \inner \clifford_- \Dirac_+ - v_0\inner \clifford_+\Dirac_-
  \right)
  +
  i{\cal L}_{v_0}
  \,.
\end{eqnarray}
For special cases this Hamiltonian is indeed well known:
For instance for flat Minkowski background it is
the sum of two copies of the ordinary Hamiltonian of the Dirac electron:
\begin{eqnarray}
  {\bf H}_{v_0}
  &=&
  \frac{1}{2}
  \left(
  \clifford_-^0 \clifford_-^i
  -
  \clifford_+^0 \clifford_+^i
  \right)
  i\partial_i
  \,,\hspace{1cm}(\mbox{for}\;g_{\mu\nu} = \eta_{\mu\nu} \;\mbox{and}\; v_0 = \partial_0)
  \,.
\end{eqnarray}
In the context of classical electromagnetism in turn it is known as the
Maxwell operator, which generates time evolution of the electromagnetic field.
(This is discussed in more detail in \fullref{p-form electromagnetism}.)
It is remarkable that a generalization of these well known Hamiltonians 
plays a role for supersymmetric quantum systems and indeed for the superstring.
Heuristically this can be understood from the fact that both the Dirac particle as well
as form field quanta appear in the massless sector of the superstring.

The Hamiltonian
\refer{definition of the Hamiltonian}
is ``time independent'' in the sense that (by \refer{Killing Lie commutes with its own cliffords})
\begin{eqnarray}
  \commutator{{\cal L}_{v_0}}{{\bf H}_{v_0}} &=& 0
  \,.
\end{eqnarray}
The left hand side of the second but last line in
\refer{constraints as evolution equations}
gives a measure for how the $v_0$-evolution of the 
$\clifford_+$ sector deviates from that of the $\clifford_-$ sector: If the
curl $\partial_{[\mu}v_{0\nu]}$ of the Killing vector vanishes (which is equivalent to
$v_0$ being covariantly constant), then 
$
           \commutator{v_{0}\inner\clifford_+}{\Dirac_-} 
           \;=\;-
           \commutator{v_{0}\inner\clifford_-}{\Dirac_+}
$
on states that satisfy the constraints, and the Hamiltonian reduces to
\begin{eqnarray}
  {\bf H}_{v_0}
  &=&
  \frac{i}{2}
  \commutator{v_{0}\inner\clifford_-}{\Dirac_+}
 \;=\;
  -\frac{i}{2}
  \commutator{v_{0}\inner\clifford_+}{\Dirac_-}
  \hspace{1cm}
  (\mbox{on-shell and for}\; 
  \partial_{[\mu}v_{0\nu]} = 0)
  \,.
\end{eqnarray}

\subparagraph{Killing-deformed Hamiltonian.}
The generalization of all this to the Killing-deformed operators $\Dirac_{k,\pm}$
\refer{k deformed Dirac operators}
is straightforward: 
The analogue of \refer{constraints as evolution equations} is
\begin{eqnarray}
  \label{constraints as evolution equations kdeformed case}
  &&\Dirac_{k,\pm}\,\omega \;=\; 0
  \nonumber\\
  &\Leftrightarrow&
  \left\lbrace
    \begin{array}{rcl}
      \left(
      2 (\partial_{[\mu}v_{0\nu]})\left(
       \coordCreator^\mu\coordCreator^\nu + \coordAnnihilator^\mu\coordAnnihilator^\nu
      \right)
      +4i v_{0\mu}k^\mu
      \right)
      \;\omega
      &=& -
          \left(
           \commutator{v_{0}\inner\clifford_+}{\Dirac_{k,-}} 
           +
           \commutator{v_{0}\inner\clifford_-}{\Dirac_{k,+}}
         \right)\omega 
      \\
      4{\cal L}_{v_0}\;\omega &=& 
          -\left(
           \commutator{v_{0}\inner\clifford_+}{\Dirac_{k,-}} 
           -
           \commutator{v_{0}\inner\clifford_-}{\Dirac_{k,+}}
         \right)\omega 
    \end{array}
  \right.
  \,.
  \nonumber\\
\end{eqnarray}
Since the left hand side of the lower line remains unchanged, the deformed Hamiltonian
${\bf H}_{k,v_0}$
is again of the form
\begin{eqnarray}
  \label{k-deformed Hamiltonian}
  {\bf H}_{k,v_0}
  &\defas&
  \frac{i}{4}
          \left(
           \commutator{v_{0}\inner\clifford_-}{\Dirac_{k,+}}
           -
           \commutator{v_{0}\inner\clifford_+}{\Dirac_{k,-}} 
         \right)
  \nonumber\\
  &\equalby{k deformed Dirac operators}&
  {\bf H}_{v_0}
  +
  \frac{1}{4}
          \left(
           \commutator{v_{0}\inner\clifford_-}{k \inner\clifford_-}
           +
           \commutator{v_{0}\inner\clifford_+}{k \inner \clifford_+} 
         \right)  
  \,.
\end{eqnarray}
If  ${\cal L}_{v_0}$ is still to commute with 
the constraints, we need, due to \refer{commutator of lie derivative with creator and annihilator}, 
to require that 
\begin{eqnarray}
  \label{requirement that energy and reparameterization commute}
  [v_0,k] &\shallbe& 0
  \,.
\end{eqnarray}
As shown in \S\fullref{loop space and B-field background} 
this condition indeed holds for the vector $k$
used for representing the superstring on loop space. 

$\,$\\ 

In order that the Hamiltonian $\bf H$ generates proper unitary
evolution we need a scalar product on states (restricted to hypersurfaces
perpendicular to the ``time'' direction induced by $v_0$) with respect to
which $\bf H$ is self-adjoint. This is the subject of the next section.

\subsubsection{Scalar product.}
\label{scalar product}
One notable point about the Hamiltonian \refer{definition of the Hamiltonian},
or its generalization  \refer{k-deformed Hamiltonian}, is that it is
\emph{anti}-hermitian with respect to the Hodge inner product $\bracket{\cdot}{\cdot}$
(\cf \refer{Hodge inner product}):
\begin{eqnarray}
  {\bf H}_{v_0}^\dag &=& -{\bf H}_{v_0}
  \,.
\end{eqnarray}
For doing quantum mechanics we therefore need to construct, as in
\refer{modified inner product}, a scalar product
$\bracket{\cdot}{\cdot}_\hermitianMetricOperator$ from the indefinite 
$\bracket{\cdot}{\cdot}$ with respect to which the Hamiltonian is a self-adjoint
operator. 

The obvious generalization of \refer{naive and simple hermitian metric operator} is
\begin{eqnarray}
  \label{general hermitian metric operator}
  \hermitianMetricOperator &\defas&
  \frac{1}{v_0\inner v_0}
  v_0\inner \clifford_- v_0\inner\clifford_+
  \,.
\end{eqnarray}
This $\hermitianMetricOperator$ obviously satisfies
\begin{eqnarray}
  \hermitianMetricOperator^2 &=& 1
  \nonumber\\
  \hermitianMetricOperator^\dag &=& \hermitianMetricOperator
\end{eqnarray}
and
\begin{eqnarray}
  \commutator{{\cal L}_{v_0}}{\hermitianMetricOperator}
  &=& 0
  \,.
\end{eqnarray}

Because of
\begin{eqnarray}
  \hermitianMetricOperator{\bf H}_{v_0}\hermitianMetricOperator
  &=&
  -{\bf H}_{v_0}
\end{eqnarray}
the $v_0$-Hamiltonian is indeed self-adjoint with respect to
$\bracket{\cdot}{\cdot}_{\hermitianMetricOperator}$:
\begin{eqnarray}
  \label{Hamiltonian self-adjoint wrt proper scalar product}
  {\bf H}_{v_0}^{\dag_\hermitianMetricOperator}
  &=&
  (\hermitianMetricOperator {\bf H}_{v_0}\hermitianMetricOperator^{-1})^\dag
  \nonumber\\
  &=&
  \hermitianMetricOperator {\bf H}^\dag_{v_0}\hermitianMetricOperator
  \nonumber\\
  &=&
  {\bf H}_{v_0}
  \,.
\end{eqnarray}

To see that this makes sense, assume that target space $\manifold$ is
\emph{static} and  foliate $\manifold$ by spacelike hypersurfaces
orthogonal to the timelike Killing vector field $v_0$. Let $t_{v_0}$ be the
coordinate that parametrizes the flow lines of $v_0$, defined by
\begin{eqnarray}
  \label{definition of hypersclice time}
  \extd\, t_{v_0} &=& \frac{1}{v_0\inner v_0}v_{0\mu}  dx^\mu
  \,,
\end{eqnarray}
so that
\begin{eqnarray}
  \commutator{{\cal L}_{v_0}}{t_{v_0}} &=& 1
\end{eqnarray}
and
\begin{eqnarray}
  \commutator{\Dirac_\mp}{t_{v_0}} &=& \frac{1}{v_0\inner v_0}v_0\inner\clifford_\pm
\end{eqnarray}
Clearly, the Hamiltonian ${\bf H}_{v_0}$ commutes with this time variable:
\begin{eqnarray}
  \label{Hamiltonian commutes with time}
  \commutator{{\bf H}_{v_0}}{t_{v_0}}
  &=&
  \commutator{
  \frac{i}{4}
  \left(
    \commutator{v_0\inner \clifford_-}{\Dirac_+}
    -
    \commutator{v_0\inner \clifford_+}{\Dirac_-}
  \right)
 }{t_{v_0}}
  \nonumber\\
 &=&
  \frac{1}{v_0\inner v_0}
  \frac{i}{4}
  \left(
    \commutator{v_0\inner \clifford_-}{v_0\inner \clifford_-}
    -
    \commutator{v_0\inner \clifford_+}{v_0\inner \clifford_+}
  \right)
  \nonumber\\
  &=&
  0
  \,.
\end{eqnarray}
The manifold $\manifold$ is foliated into ``equal time'' slices 
by the mapping
$t \mapsto \Sigma_{v_0}\of{t} \defas \set{p\in \manifold| t_{v_0}\of{p} = t}$, and
the induced metric $h$ on each leaf is
\begin{eqnarray}
  h_{\mu\nu} &=& g_{\mu\nu}  - \frac{1}{v_0\inner v_0} v_{0\mu} v_{0\nu}
  \,.
\end{eqnarray}
The determinant of the full metric tensor then splits as
\begin{eqnarray}
  \label{determinant of spatially split metric}
  \sqrt{-g} &=& \sqrt{- v_0\inner v_0}\,\sqrt{h}
  \,.
\end{eqnarray}
Since ${\bf H}_{v_0}$ generates unitary evolution from one $\Sigma_{v_0}$ to the next, 
the scalar product of physical states should be restricted to a fixed (but arbitrary)
hyperslice. Hence define
\begin{eqnarray}
   \label{spatial scalar product}
  \bracket{\cdot}{\cdot}_{v_0}
  &\defas&
  \bra{\cdot}\delta\of{t_{v_0}}\ket{\cdot}_\hermitianMetricOperator
  \nonumber\\
  &=&
  \bra{\cdot}\delta\of{t_{v_0}}\hermitianMetricOperator\ket{\cdot}
  \,.
\end{eqnarray}
Because of \refer{Hamiltonian commutes with time} the Hamiltonian is still
hermitian with respect to \refer{spatial scalar product},
\begin{eqnarray}
  {\bf H}_{v_0}^{\dag_{v_0}}
  &=&
  {\bf H}_{v_0}
  \,,
\end{eqnarray}
and hence generates a unitary evolution along $t_{v_0}$ 
(\cf p. \pageref{Propagator and physical states}), implying in particular that
\begin{eqnarray}
  {\cal L}_{v_0} \bracket{\omega}{\omega}_{v_0} &=& 0
  \hspace{1cm}
  \mbox{(for $\Dirac_\pm \omega = 0$)}
  \,.
\end{eqnarray}
It is easily checked that no problems arise when adapting this to the
$k$-deformed case: The deformed Hamiltonian
${\bf H}_{k,v_0}$ is also anti-hermitian with respect to $\bracket{\cdot}{\cdot}$ and
hermitian with respect to $\bracket{\cdot}{\cdot}_{v_0}$:
\begin{eqnarray}
  \label{eta hermiticity of k-deformed Hamiltonian}
  \left({\bf H}_{k,v_0}\right)^{\dag_\hermitianMetricOperator}
  &=&
  {\bf H}_{k,v_0}
\end{eqnarray}
(The proof is given in \fullref{proofs of self-adjointness of the Hamiltonian}).

Taking everything together, the proper scalar product reads explicitly
(\cf \refer{local inner product})
\begin{eqnarray}
  \bracket{\alpha}{\beta}_{v_0}
  &=&
  \int\limits_{\manifold,t_{v_0} = 0}
  \bra{\alpha} v_0\inner \clifford_-\,v_0\inner\clifford_+\ket{\beta}_{\rm loc}
  \frac{1}{\sqrt{-v_0\inner v_0}}\sqrt{h}\,d^{D-1}x
  \,.
\end{eqnarray}
This scalar product is related to the timelike component of 
a conserved current: For a given $\omega$ define the tensor $T^{\mu\nu}$ by 
\begin{eqnarray}
  \label{generalized stress energy tensor}
  T^{\mu\nu}
  &\defas&
  \bra{\omega}\clifford_-^\mu\clifford_+^\nu\ket{\omega}_{\rm loc}
  \,.
\end{eqnarray} 
If $\omega$ satisfies the constraint $\Dirac_\pm \omega = 0$ or
$\Dirac_{k,\pm}\omega = 0$, then this tensor is conserved,
\begin{eqnarray}
  \nabla_\mu T^{\mu\nu} &=& 0
\end{eqnarray}
and in particular
\begin{eqnarray}
  P_{v_0}^\mu &\defas& T^{\mu}{}_\nu v_0^\nu
\end{eqnarray}
is a conserved current and
\begin{eqnarray}
  \bracket{\omega}{\omega}_{v_0}
  &=&
  \int\limits_{\manifold,t_{v_0 = 0}}
  \bra{\omega}v_0\inner\clifford_- v_0\inner\clifford_+\ket{\omega}_{\rm loc}
  \;\sqrt{h}d^{D-1}x
  \nonumber\\
  &=&
  \int\limits_{\manifold, t_{v_0} = 0}
  \frac{1}{\sqrt{-v_0\inner v_0}}
  v_0\inner P_{v_0}\;
  \;\sqrt{h}d^{D-1}x
  \,.
\end{eqnarray}
In coordinates adapted to the foliation we have
\begin{eqnarray}
  v_0^\mu &=& \delta^\mu_0
  \nonumber\\
  v_{0\mu} &=& g_{\mu 0}\; = \; \delta_\mu^0 \,v_0\inner v_0
\end{eqnarray}
and hence the scalar product is time independent:
\begin{eqnarray}
  {\cal L}_{v_0}
  \bracket{\omega}{\omega}_{v_0}
  &=&
  \partial_0
  \int\limits_{\manifold, t_{v_0} = 0}
  P^0_{v_0}\;
  \;\sqrt{-v_0\inner v_0}\sqrt{h}d^{D-1}x
  \nonumber\\
  &=&
  \int\limits_{\manifold, t_{v_0} = 0}
    \partial_0
  \left(
  \sqrt{-g}
  P^0_{v_0}\right)\;
  \;d^{D-1}x  
  \nonumber\\
  &=&
  -
  \int\limits_{\manifold, t_{v_0} = 0}
    \partial_i
  \left(
  \sqrt{-g}
  P^i_{v_0}\right)\;
  \;d^{D-1}x    
  \nonumber\\
  &=&
  0
  \,.
\end{eqnarray}

When restricted to the undeformed case and to 
2-form fields these constructions reduce to relations well
known from Dirac theory and classical electromagnetism, see
\refer{Maxwell stress energy tensor}.

$\,$\\

The formalism so far gives us a unitary Hamiltonian evolution 
along a timelike Killing vector field obtained from the supersymmetry
constraints $\Dirac_\pm \omega = 0$. But according to
\refer{constraints as evolution equations} 
the associated Schr\"odinger equation contains only
part of the physical content of these constraints. Further information
is contained in the second but last line of \refer{constraints as evolution equations}. 
We will
now show that this yields a constraint on states restricted to
spacelike hypersurfaces which is compatible with our Hamiltonian
${\bf H}$.

\paragraph{Propagator and physical states.}
\label{Propagator and physical states}
It is convenient to introduce the further abbreviation
\begin{eqnarray}
  {\bf C}_{v_0}
  &\defas&
  v_0\inner\clifford_+\Dirac_-
  +
  v_0\inner\clifford_-\Dirac_+
  \,,
\end{eqnarray}
so that \refer{constraints of susy system} is equivalently rewritten as
\begin{eqnarray}
  \label{Hamiltonian form of susy constraints}
  \Dirac_\pm \omega = 0
  &\Leftrightarrow&
  \left\lbrace
  \begin{array}{rcl}
  i{\cal L}_{v_0}\omega &=& {\bf H}_{v_0}\omega
  \\
  {\bf C}_{v_0}\omega &=& 0
  \end{array}
  \right.
  \,.
\end{eqnarray}
Because of
\begin{eqnarray}
  \commutator{t_{v_0}}{{\bf C}_{v_0}} &=& 0
  \nonumber\\
  \commutator{{\cal L}_{v_0}}{{\bf C}_{v_0}} &=& 0
\end{eqnarray}
the constraint ${\bf C}_{v_0}$ must hold on each hyperslice $\Sigma_t$ seperately:
\begin{eqnarray}
  {\bf C}_{v_0}\omega &=& 0
  \nonumber\\
  \Rightarrow\;
  {\bf C}_{v_0}\left(\delta\of{t_{v_0},t}\omega\right) &=& 0
  \,.
\end{eqnarray}
In fact, every form $\omega_0$ on $\Sigma_0$, which satisfies the
spatial constraint
${\bf C}_{v_0}\omega_0 = 0$, uniquely corresponds to a state $\omega$ on all of
$\manifold$,  given by
\begin{eqnarray}
  \omega &=& 
  \exp\of{-i{\bf H}t_{v_0}}\omega_0
  \,,
\end{eqnarray}
that satisfies the full constraints \refer{Hamiltonian form of susy constraints}.
In order to see this, note that the constraint ${\bf C}_{v_0}$
\emph{commutes weakly} with the Hamiltonian ${\bf H}_{v_0}$, i.e.
up to a term that vanishes when the spatial constraints are fulfilled:
\begin{eqnarray}
  \label{proposition that commutator C and H is prop to C in main text}
  \commutator{{\bf C}_{v_0}}{{\bf H}_{v_0}}
  &=&
  \frac{1}{2i v_0\inner v_0}
  (\nabla_{[\mu} v_{\nu]})
  v_0\inner\clifford_- \clifford_-^\mu\clifford_+^\nu
  v_0\inner \clifford_+
   \;{\bf C}_{v_0}
  \,.
\end{eqnarray}
(The proof of this is given in appendix \fullref{proof that commutator C with H is prop to C}.)
Hence we have
\begin{eqnarray}
  {\bf C}_{v_0} (1-i{\bf H}_{v_0}\epsilon)\omega_0 
  &=&
  i\epsilon  \commutator{{\bf H}_{v_0}}{{\bf C}_{v_0}}\omega_0
  \nonumber\\
  &\equalby{proposition that commutator C and H is prop to C in main text}&
  0
  \,,
  \skiph{$i\epsilon  \commutator{{\bf H}_{v_0}}{{\bf C}_{v_0}}\omega_0$}
  (\mbox{for}\;{\bf C}_{v_0}\omega_0 = 0)
\end{eqnarray}
for any constant $\epsilon$.
Iterating this argument yields
\begin{eqnarray}
  {\bf C}_{v_0}\left(1-i{\bf H}_{v_0}t_{v_0}/n\right)^n\omega_0 &=& 0
  \,,\hspace{1cm}\mbox{for}\; n\in \N\;\mbox{and}\;
    {\bf C}_{v_0}\omega_0 = 0
  \,,
\end{eqnarray}
which in the limit $n\to\infty$ gives
\begin{eqnarray}
  {\bf C}_{v_0}
  \exp\of{-i{\bf H}_{v_0}t_{v_0}}\omega_0 \;=\; 0\,,
  \hspace{1cm}
   (\mbox{for}\; {\bf C}_{v_0}\omega_0 \;=\; 0)
  \,.
\end{eqnarray}
Since, by assumption, ${\cal L}_{v_0}\omega_0 = 0$, the state $\omega$
of course also satisfies the Schr\"odinger equation:
\begin{eqnarray}
  i{\cal L}_{v_0} \exp\of{-i{\bf H}_{v_0} t_{v_0}}\omega_0 
  &=&
  {\bf H}_{v_0}\exp\of{-i{\bf H}_{v_0} t_{v_0}}\omega_0 
  \,.
\end{eqnarray}

The generalization to the $k$-deformed case is again unproblematic,
since one finds in perfect analogy with 
\refer{proposition that commutator C and H is prop to C in main text} that
\begin{eqnarray}
  \label{proposition that commutator Ck and Hk is prop to Ck in main text}
  \commutator{{\bf C}_{k,v_0}}{{\bf H}_{k,v_0}}
  &=&
  \frac{1}{2i v_0\inner v_0}
  (\nabla_{[\mu} v_{\nu]})
  v_0\inner\clifford_- \clifford_-^\mu\clifford_+^\nu
  v_0\inner \clifford_+
   \;{\bf C}_{k,v_0}
  \,.
\end{eqnarray}
(The proof is given
in appendix \ref{proofs}
on 
p. \pageref{proposition that commutator Ck and Hk is prop to Ck}.)

The above constructions show that a covariant Hamiltonian  with
all the familiar properties can be
constructed in $D=1$ and $D=2$ supersymmetric systems with purely
gravitational background. We now want to generalize all this to
the case where there is additionally a non-vanishing $b$-field background.
In order to do so we make use of the fact that the supersymmetry
constraints in such backgrounds are obtained from those of the
already understood backgrounds by a deformation induced by
the deformation operator \refer{definition of b-deformation exponent}.

\subsubsection{Parameter evolution in the presence of a $B$-field}
\label{Parameter evolution in the presence of a B-field}

The parameter evolution that we are interested in requires that the
background fields be ``time independent''. Hence all background fields
must have vanishing Lie derivative along $v_0$. 
For the $b$-field this is equivalent to (\cf \refer{definition of b-deformation exponent})
\begin{eqnarray}
  \label{v0 Lie derivative commutes with b field}
  \commutator{{\cal L}_{v_0}}{{\bf W}^{(b)}} &=& 0
  \,.
\end{eqnarray}

Recall \refer{susy generators for B field background} that the $b$-field induces on $\extd$ and $\coextd$ the
deformation
\begin{eqnarray}
  \extd^{(b)} &\defas& e^{-{\bf W}^{(b)}}\extd e^{{\bf W}^{(b)}}
  \nonumber\\
  \coextd^{(b)} &\defas& e^{{\bf W}^{\dag(b)}}\coextd e^{-{\bf W}^{\dag(b)}}
  \,.
\end{eqnarray}
Since for further constructions it will be essential to have an 
analogue of \refer{definition Lie derivative operator} and
\refer{adjoint definition of Lie derivative}
and hence of \refer{constraints as evolution equations}, we define the
following deformations of the form creators and annihilators:
\begin{eqnarray}
  \label{b-field deformations of creators and annihilators}
  \coordCreator^{(b)\mu}
  &\defas&
  e^{{\bf W}^\dag}\coordCreator^\mu e^{-{\bf W}^\dag}
  \nonumber\\
  &=&
  \coordCreator^\mu + \commutator{\coordCreator^\mu}{\frac{1}{2}b_{\alpha\beta}\coordAnnihilator^\alpha
    \coordAnnihilator^\beta}
  \nonumber\\
  &=&
  \coordCreator^\mu + b^\mu{}_\beta \coordAnnihilator^\beta
  \nonumber\\
  \coordAnnihilator^{(b)\mu}
  &\defas&
  e^{-{\bf W}}\coordAnnihilator^\mu e^{{\bf W}}
  \nonumber\\
  &=&
  \coordAnnihilator^\mu + \commutator{\coordAnnihilator^\mu}{\frac{1}{2}b_{\alpha\beta}\coordCreator^\alpha
    \coordCreator^\beta}
  \nonumber\\
  &=&
  \coordAnnihilator^\mu + b^\mu{}_\beta \coordCreator^\beta  
  \,.
\end{eqnarray}
The purpose of this definition is that now the relations
\begin{eqnarray}
  \antiCommutator{v_0\inner \coordAnnihilator^{(b)}}{\extd^{(b)}}
  &=&
  e^{-{\bf W}^{(b)}}\antiCommutator{v_0\inner \coordAnnihilator}{\extd}
  e^{{\bf W}^{(b)}}
  \nonumber\\
  &=&
  e^{-{\bf W}^{(b)}} {\cal L}_{v_0}
  e^{{\bf W}^{(b)}}  
  \nonumber\\
  &\equalby{v0 Lie derivative commutes with b field}&
  {\cal L}_{v_0}
\end{eqnarray}
and
\begin{eqnarray}
  \antiCommutator{v_0\inner \coordCreator^{(b)}}{\coextd^{(b)}}
  &=&
  -{\cal L}_{v_0}
\end{eqnarray}
hold, and analogously for the $k$-deformed case:
\begin{eqnarray}
  \label{Lie Killing from deformed CAR}
  \antiCommutator{v_0\inner \coordAnnihilator^{(b)}}{\extd_k^{(b)}}
  &=&
  e^{-{\bf W}^{(b)}}
  \left(
    \antiCommutator{v_0\inner \coordAnnihilator}{\extd}
    +
    \antiCommutator{v_0\inner \coordAnnihilator}{ik\inner \coordAnnihilator}
  \right)
  e^{{\bf W}^{(b)}}
  \nonumber\\
  &=&
  {\cal L}_{v_0}  
\nonumber\\
  \antiCommutator{v_0\inner \coordCreator^{(b)}}{\coextd_k^{(b)}}
  &=&
  -{\cal L}_{v_0}  
  \,.
\end{eqnarray}
The deformed creators and annihilators satisfy
\begin{eqnarray}
  \antiCommutator{\coordCreator^{(b)\mu}}{\coordCreator^{(b)\nu}} &=& 0
  \nonumber\\
  \antiCommutator{\coordAnnihilator^{(b)\mu}}{\coordAnnihilator^{(b)\nu}} &=& 0
  \nonumber\\
  \antiCommutator{\coordAnnihilator^{(b)\mu}}{\coordCreator^{(b)\nu}} &=& 
  g^{\mu\nu} + b^\mu{}_\alpha b^{\nu\alpha}
  \nonumber\\
  &=&
  g^{(b)\mu\nu}
  \,.
\end{eqnarray}
The tensor
\begin{eqnarray}
  \label{open string metric}
  g^{(b)}_{\mu\nu}
  &\defas&
  (g - bg^{-1}b)_{\mu\nu}
\end{eqnarray}
is known in string theory as the \emph{open string metric} in the presence of
a $b$-field (\cf \cite{SeibergWitten:1999}, p.9). Note that even though it
plays a role similar to a metric tensor, we will never shift indices with
anything but the ordinary metric $g$. In particular
\begin{eqnarray}
  \label{open string metric}
  g^{(b)\mu\nu}
  &\defas&
  (g + bg^{-1}b)_{\mu^\prime\nu^\prime}g^{\mu^\prime\mu}g^{\nu^\prime\nu}
  \,.
\end{eqnarray}
The $b$-deformed analogue of the Clifford generators \refer{definition of the Clifford generators}
is of course
\begin{eqnarray}
  \label{b-deformed Clifford generators}
  \clifford_\pm^{(b)\mu}
  &\defas&
  \coordCreator^{(b)\mu} \pm \coordAnnihilator^{(b)\mu}
  \nonumber\\
  &=&
  \clifford_\pm^\mu 
  \pm
  b^\mu{}_\alpha \clifford_\pm^\alpha
  \nonumber\\
  &=&
  \left(e_a{}^\mu{} \mp b_\alpha{}^\mu\right)\clifford_\pm^\alpha
  \,,
\end{eqnarray}
satisfying
\begin{eqnarray}
  \antiCommutator{\clifford_\pm^{(b)\mu}}{\clifford_\mp^{(b)\nu}}
   &=& 0
  \nonumber\\
  \antiCommutator{\clifford_\pm^{(b)\mu}}{\clifford_\pm^{(b)\nu}}
  &=&
  \pm2 g^{(b)\mu\nu}
  \,.
\end{eqnarray}
We will often need the covariant version of the deformed Clifford generators
\refer{b-deformed Clifford generators}:
\begin{eqnarray}
   \label{{b-deformed covariant Clifford generators}}
  \clifford^{(b)}_{\mu\pm}
  &=&
  \left(
    e_\mu{}^a \pm b_\mu{}^a
  \right)
  \clifford_{a\pm}
  \,.
\end{eqnarray}
Equations \refer{b-deformed Clifford generators} and \refer{{b-deformed covariant Clifford generators}}
motivate the introduction of the $b$-deformed version of the vielbein $e_a{}^\mu$ and its inverse
$e_\mu{}^a$:
\begin{eqnarray}
  \label{b-defomrmed vielbeine}
  e^{(b)}_{\pm}{}_a{}^\mu
  &\defas&
  e_a{}^\mu \mp b_a{}^\mu
  \nonumber\\
  \label{b-defomrmed inverse vielbeine}
  e^{(b)}_{\pm}{}_\mu{}^a
  &\defas&
  e_\mu{}^a \pm b_\mu{}^a
  \,.
\end{eqnarray}
In terms of these we can write succinctly
\begin{eqnarray}
  \clifford_\pm^{(b)}
  &=&
  e^{(b)}_\pm{}_a{}^\mu \;\clifford_{\pm}^a
  \nonumber\\
  \clifford^{(b)}_{\mu\pm}
  &=&
  e_\pm^{(b)-}{}_\mu{}^a \;\clifford_{a\pm}
\end{eqnarray}

The purpose of all this is that
using \refer{Lie Killing from deformed CAR} it is now 
immediate that, in complete analogy with 
\refer{constraints as evolution equations}, we have
\begin{eqnarray}
  \label{b-deformation of Killing vector by clifford anticommutators}
  \antiCommutator{v_0\inner \clifford_+^{(b)}}{\Dirac_-^{(b)}}
  -
  \antiCommutator{v_0\inner \clifford_-^{(b)}}{\Dirac_+^{(b)}}
  &=&
  4{\cal L}_{v_0}
  \,.
\end{eqnarray}

This means that the construction \refer{definition of the Hamiltonian} of a Hamiltonian 
generator of parameter evolution carries over to the $b$-deformed case
as follows:
\begin{eqnarray}
  \label{Hamiltonian generator in the presence of a b-field background}
  {\bf H}^{(b)}_{v_0}
  &=&
  \frac{i}{4}
  \left(
  \commutator{v_0\inner \clifford_+^{(b)}}{\Dirac_-^{(b)}}
  -
  \commutator{v_0\inner \clifford_-^{(b)}}{\Dirac_+^{(b)}}
  \right)
  \nonumber\\
  &=&
  \frac{i}{2}
  \left(
    \clifford_-^{(b)}\Dirac_+^{(b)}
    -
    \clifford_+^{(b)}\Dirac_-^{(b)}
  \right)
  +
  i{\cal L}_{v_0}
  \,.
\end{eqnarray}
Noting that (by \refer{b-field deformations of creators and annihilators} 
and \refer{v0 Lie derivative commutes with b field})
\begin{eqnarray}
  \commutator{{\cal L}_{v_0}}{v_0\inner \coordCreator^{(b)}}
  \;=\; 0 
  \;=\;
  \commutator{{\cal L}_{v_0}}{v_0\inner \coordAnnihilator^{(b)}}
  \,,
\end{eqnarray}
it is easy to see (the details are given in
appendix \fullref{proofs of self-adjointness of the Hamiltonian}) 
that this Hamiltonian is self-adjoint with respect to
the scalar product induced by the appropriately deformed
Krein space operator (\cf \refer{general hermitian metric operator})
\begin{eqnarray}
  \label{hermitian metric operator for b-field background}
  \hermitianMetricOperator^{(b)}
  &\defas&
  \left(v_0\inner \clifford_-^{(b)}\right)^{-2}
  v_0\inner \clifford_-^{(b)} v_0\inner\clifford_+^{(b)}
  \,,
\end{eqnarray}
i.e.
\begin{eqnarray}
  \label{b-deformed hamiltonian is selfadjoitn wrt b-deformed metric operator}
  {\bf H}_{v_0}^{\dag_{\hermitianMetricOperator^{(b)}}}
  &=&
  {\bf H}_{v_0}
  \,.
\end{eqnarray}

But care has to be exercised, since $\hermitianMetricOperator$-hermiticity is
not sufficient for many applications. What really matters is 
hermiticity with respect to the time-reparameterization
gauge fixed scalar product \refer{spatial scalar product}
induced by
\begin{eqnarray}
  \label{time fixed bfield Krein operator}
  \hermitianMetricOperator^{(b)}_{v_0}
  &=&
  \hermitianMetricOperator^{(b)}\delta\of{t_{v_0}}
  \,.
\end{eqnarray}
An $\hermitianMetricOperator$-hermitian operator $A = A^{\dag_\hermitianMetricOperator}$
is $v_0$-hermitian if it commutes with the time coordinate $t_{v_0}$ defined by
\refer{definition of hypersclice time}. This is not the case for
the operator \refer{Hamiltonian generator in the presence of a b-field background}
(\cf \refer{Hamiltonian commutes with time}):
\begin{eqnarray}
  \commutator{{\bf H}_{v_0}^{(b)}}{t_{v_0}}
  &=&
  \frac{1}{v_0\inner v_0}
  \frac{i}{4}
  \left(
  \commutator{v_0\inner \clifford_+^{(b)}}{v_0\inner \clifford_+}
  -
  \commutator{v_0\inner \clifford_-^{(b)}}{v_0\inner \clifford_-}
  \right)
  \,.
\end{eqnarray}

This failure to be $v_0$-hermitian can be remedied by adding an appropriate
correction operator. Define
\begin{eqnarray}
  \label{modified Hamiltonian}
  \tilde {\bf H}_{v_0}^{(b)}
  &\defas&
  {\bf H}_{v_0}^{(b)}
  -
  \frac{1}{v_0\inner v_0}
  \frac{i}{4}
  \left(
    \commutator{v_0\inner \clifford_+^{(b)}}{v_0\inner \clifford_+ }
    -
    \commutator{v_0\inner \clifford_-^{(b)}}{v_0\inner \clifford_- }
  \right)
  {\cal L}_{v_0}
  \,.
\end{eqnarray}
This operator is $\hermitianMetricOperator^{(b)}_{v_0}$-hermitian 
(by the same argument as in \refer{simple principle of proof of eta-hermititcity})
and by construction commutes with $t_{v_0}$, therefore it is 
$\hermitianMetricOperator^{(b)}_{v_0}$-hermitian:
\begin{eqnarray}
  (\tilde {\bf H}_{v_0}^{(b)})^{\dag_{\hermitianMetricOperator^{(b)}_{v_0}}}
  &=&
  \tilde {\bf H}_{v_0}^{(b)}
  \,.
\end{eqnarray}
On physical states $\ket{\phi}$ this operator satisfies
\begin{eqnarray}
 \label{quasi Schroedinger equation for b-field background}
 \left(
  1
  -
  \frac{1}{v_0\inner v_0}
  \frac{1}{4}
  \left(
    \commutator{v_0\inner \clifford_+^{(b)}}{v_0\inner \clifford_+ }
    -
    \commutator{v_0\inner \clifford_-^{(b)}}{v_0\inner \clifford_- }
  \right)
 \right)
    i {\cal L}_{v_0}
  \ket{\phi}
 &=&
    \tilde {\bf H}_{v_0}^{(b)}
 \ket{\phi}
  \,.
\end{eqnarray}
We write
\begin{eqnarray}
  \label{definition of K}
  K 
  &\defas&
  \frac{1}{v_0\inner v_0}
  \frac{1}{4}
  \left(
    \commutator{v_0\inner \clifford_+^{(b)}}{v_0\inner \clifford_+ }
    -
    \commutator{v_0\inner \clifford_-^{(b)}}{v_0\inner \clifford_- }
  \right)
  \nonumber\\
  &=&
  \frac{1}{2 v_0 \inner v_0}
  \left(
    v_0\inner \clifford_+^{(b)} v_0\inner \clifford_+
    -
    v_0\inner \clifford_-^{(b)}v_0\inner \clifford_- 
  \right)
  -1
  \nonumber\\
  &=&
  v_0^\mu b_{\mu\nu} \clifford_+^\nu
  v_0\inner  \clifford_{+}
  +
  v_0^\mu b_{\mu\nu}\clifford_-^\nu
  v_0 \inner\clifford_{-}
\end{eqnarray}
for the operator on the left. This operator is $\hermitianMetricOperator^{(b)}_{v_0}$-hermitian
\begin{eqnarray}
  \label{hermiticity of K}
  K^{\dag_{\hermitianMetricOperator^{(b)}_{v_0}}}
  &=& K
  \,.
\end{eqnarray}
In terms of $K$ equation \refer{quasi Schroedinger equation for b-field background} 
becomes
\begin{eqnarray}
  \label{quasi Schroedinger equation for b-field background abbreviated form}
  (1-K)i{\cal L}_{v_0} \ket{\phi}
  &=&
  \tilde {\bf H}_{v_0}^{(b)}\ket{\phi}
  \,.
\end{eqnarray}

This is the modified form of the Schr{\"o}dinger equation that 
needs to be used whenever it is crucial that the operator on the
right hand side really is self-adjoint with respect to
$\bra{\cdot}\hermitianMetricOperator^{(b)}_{v_0}\ket{\cdot}$, 
with $\hermitianMetricOperator^{(b)}_{v_0}$ given by
\refer{time fixed bfield Krein operator},
i.e.
that it really commutes with the evolution parameter $t_{v_0}$.
This will in particular be necessary in perturbation theory
(see \S\fullref{qm perturbation theory}).

\paragraph{Parameter evolution in the presence of torsion.}

Often in the literature a $B$-field background is addressed as 
a torsion background. This is justified since, as discussed
in \S\fullref{example background b field} 
(\cf \refer{b-deformed covariant derivative operator}), the 
$B$-field induces a deformation of the covariant derivative
which makes it act like the covariant derivative with
torsion $\propto +dB$ on one spinor bundle and with torsion
$\propto -dB$ on the other. However, this deformed
operator is of course not the covariant derivative on the
exterior bundle that one would ordinarily associate
with a connection of non-vanishing torsion. Instead,
the latter is, as discussed in \S\fullref{Torsion},
given by expression \refer{covariant derivative with torsion}.

The deformation of the supersymmetry generators
associated with \refer{covariant derivative with torsion},
which one might perhaps naively associate with a ``torsion background'', 
does not arise in string theory. Nevertheless, because it is
interesting in itself, we mention that for this case, too, one
can carry out the program of \S\ref{Target space Killing evolution}:

So consider replacing the constraints \refer{constraints of susy system} by their torsion-deformed versions
\refer{torsion-perturbed Dirac operators}:
\begin{eqnarray}
  \Dirac_{T,\pm}\omega &=& 0
  \,,
\end{eqnarray}
for some non-vanishing antisymmetric torsion tensor $T_{\mu\alpha\beta}$.
Then the construction \refer{constraints as evolution equations} 
gives rise to the torsion-deformed Lie derivative operators 
\begin{eqnarray}
  {\cal L}_{T,v}
  &\defas&
  \antiCommutator
    {\extd_T}
    {v_\mu \coordAnnihilator^\mu}
   \nonumber\\
   &=&
   \antiCommutator{\extd}{v^\mu \coordAnnihilator_\mu}
   -
  \antiCommutator{T_\mu{}^\alpha{}_\beta \coordCreator^\mu\coordCreator^\beta\coordAnnihilator_\alpha}
   {{v^\mu \coordAnnihilator_\mu}}  
  \nonumber\\
  &=&
  {\cal L}_{v}
  -2
  v^\mu T_\mu{}^\alpha{}_\beta \coordCreator^\beta\coordAnnihilator_\alpha
  \,.
\end{eqnarray}
Because the term on the right is anti-hermitian for all $v$, the operator
${\cal L}_{T,v}$ still satisfies the crucial condition \refer{Lie antihermitian when v is Killing}:
\begin{eqnarray}
  ({\cal L}_{T,v})^\dag = -{\cal L}_{T,v}
  &\Leftrightarrow&
  \mbox{$v$ Killing}
  \,.
\end{eqnarray}
Furthermore the covariant derivative of a Killing vector $v_\mu$ with respect to $\omega_T$ is
still antisymmetric:
\begin{eqnarray}
  \nabla_{T,\mu}v_\nu &=& \nabla_{T,[\mu}v_{\nu]}
  \hspace{1cm}\mbox{($v$ Killing)}
  \,.
\end{eqnarray}
This is the condition that the proofs \fullref{proof that commutator C with H is prop to C}
rely on. Hence they carry over to the torsion deformed case and we can
straightforwardly generalize \refer{Dirac-Schroedinger equation},
\refer{definition of the Hamiltonian} and
\refer{proposition that commutator C and H is prop to C in main text}
to the case of non-vanishing torsion by replacing 
${\cal L}_{v_0}$ by  ${\cal L}_{T,v_0}$ and
$\Dirac_\pm$ by $\Dirac_{T,\pm}$
throughout.

\subsubsection{Perturbation of background fields}
\label{Perturbation of background fields}

We have now succeeded in constructing covariant Hamiltonian operators
for general metric and Kalb-Ramond field backgrounds. Any pertuabtion of these
background fields will induce a perturbation of this Hamiltonian operator.
Since there are some subtleties involved in calculating that perturbed
Hamiltonian from the perturbed background fields we explicitly spell out the
necessary steps in this section. The next section then shows how, given the
perturbed Hamiltonian, the first order perturbation theory of ordinary
quantum mechanics can be adapted to Schr{\"o}dinger equations of the form
\refer{quasi Schroedinger equation for b-field background abbreviated form}.

A perturbation of the background fields labelled by a perturbation
parameter $\epsilon$
\begin{eqnarray}
  g_{\mu\nu} &\to& \underbrace{g^{(0)}_{\mu\nu}}_{= g_{\mu\nu}} 
    + \sum\limits_{n=1}^\infty \underbrace{g^{(n)}_{\mu\nu}}_{\order{\epsilon^n}}
  \nonumber\\
  b_{\mu\nu} &\to& \underbrace{b^{(0)}_{\mu\nu}}_{= b_{\mu\nu}} 
    + \sum\limits_{n=1}^\infty \underbrace{b^{(n)}_{\mu\nu}}_{\order{\epsilon^n}}
\end{eqnarray}
induces a deformation of the various operators considered here.
This section briefly collects some of the relevant formulae, which will be needed
in \S\fullref{qm perturbation theory} for writing down an expression for
the first order energy shift.

In the $\sigma$-model Lagrangian the background fields act as 
coupling constants for the canonical fields $\Gamma_\pm^a, X^\mu, \partial_{X^\mu}$,
which themselves therefore receive no perturbation:
\begin{eqnarray}
  \label{fields which receive no perturbation}
  \Gamma_\pm^a &\to& \clifford_\pm^a
  \nonumber\\
  X^\mu &\to& X^\mu
  \nonumber\\
  \partial_{X^\mu} &\to& \partial_{X^\mu}
  \,.
\end{eqnarray}
Geometrically this means that while perturbing $g$ and $b$ the coordinates on
the configuration manifold are fixed, as is the chosen ONB section of the
two Clifford bundles:
\begin{eqnarray}
  \commutator{\partial_{X^\mu}}{X^\nu} &=& \delta_\mu^\nu
  \nonumber\\
  \antiCommutator{\Gamma_{a\pm}}{\Gamma^b_{\pm}} &=& \pm 2 \delta_a^b
  \,.
\end{eqnarray}
The perturbed geometry is felt by the canonical fields via the perturbation of
the vielbein
\begin{eqnarray}
  e_a{}^\mu &=& e_a^{(0)}{}^\mu + \underbrace{e_a^{(1)}{}^\mu}_{= \order{\epsilon}}
  + \order{\epsilon^2}
  \,.
\end{eqnarray}
This has to satisfy
\begin{eqnarray}
  ds^2\of{e_a,e_b} &\shallbe& \eta_{ab}
  \nonumber\\
  &=&
  \underbrace{
    ds_0^2\of{ e_a^{(0)}, e_b^{(0)} }
  }_{= \eta_{ab}}
  \nonumber\\
  &&+
  \underbrace{
  ds_1^2\of{e_a^{(0)}, e_b^{(0)}}
  +
  ds_0^2\of{e_a^{(0)}, e_b^{(1)}}
  +
  ds_0^2\of{e_b^{(0)}, e_a^{(1)}}
  }_{\order{\epsilon}\shallbe 0}
  \nonumber\\
  &&
  +
  ds_2^2\of{e_a^{(0)},e_b^{(0)}}
  +
  ds_1^2\of{e_a^{(1)},e_b^{(0)}}
  +
  ds_1^2\of{e_a^{(0)},e_b^{(1)}}
  \nonumber\\
  &&
  \underbrace{
  +
  ds_0^2\of{e_a^{(1)},e_b^{(1)}}
  +
  ds_0^2\of{e_a^{(2)},e_b^{(0)}}
  +
  ds_0^2\of{e_a^{(0)},e_b^{(2)}}
  }_{\order{\epsilon^2}\shallbe 0}
  \nonumber\\
  &&
  +
  \order{\epsilon^3}
  \,,
\end{eqnarray}
where we write $ds_n^2\of{v,w}$ for $g^{(n)}_{\mu\nu}v^\mu w ^\nu$.
Hence the first and second order perturbation of the vielbein is given by
\begin{eqnarray}
  e^{(1)}_a
  &=&
  \left(
    q^{(1)}_{ab}
    -\frac{1}{2} ds^2_{(1)}\of{e^{(0)}_a,e^{(0)}_b} 
  \right)
  (e^{b(0)})
  \nonumber\\
  e^{(2)}_a
  &=&
  \left(
    q^{(2)}_{ab}
    -\frac{1}{2} 
   \left(
     ds^2_{(2)}\of{e^{(0)}_a,e^{(0)}_b}
  +
  ds_1^2\of{e_a^{(1)},e_b^{(0)}}
  +
  ds_1^2\of{e_a^{(0)},e_b^{(1)}}
  +
  ds_0^2\of{e_a^{(1)},e_b^{(1)}} 
   \right) 
  \right)
  (e^{b(0)})
  \,,
  \nonumber\\
\end{eqnarray}
where
\begin{eqnarray}
  q_{ab} &=& -q_{ba}
\end{eqnarray}
is an arbitrary antisymmetric tensor which incorporates the gauge freedom in the
choice of vielbein.
The inverse vielbein is then
\begin{eqnarray}
  e^a{}_\mu &=&
  e_b{}^\nu \eta^{ab} g_{\mu\nu}
  \nonumber\\
  &=&
  \underbrace{
  e^{(0)}_b{}^\nu \eta^{ab} g^{(0)}_{\mu\nu}
  }_{\order{1}}
  +
  \underbrace{
  e^{(0)}_b{}^\nu \eta^{ab} g^{(1)}_{\mu\nu}
  +
  e^{(1)}_b{}^\nu \eta^{ab} g^{(0)}_{\mu\nu}
  }_{\order{\epsilon}}
  +
  \order{\epsilon^2}
  \,.    
\end{eqnarray}

For the ``structure functions'' $f_a{}^c{}_b$ of the vielbein one has
\begin{eqnarray}
  &&\commutator{e_{a}}{e_{b}}
  \;=\;
  f_{a}{}^{c}{}_{b} \, e_{c}
  \nonumber\\
  &\Rightarrow&
  \underbrace{
  \commutator{e^{(0)}_{a}}{e^{(0)}_{b}}
  }_{\order{\epsilon^0}}
  + 
  \underbrace{
  \commutator{e^{(0)}_{a}}{e^{(1)}_{b}}  
  + 
  \commutator{e^{(1)}_{a}}{e^{(0)}_{b}}
  }_{\order{\epsilon}}
  +
  \order{\epsilon^2}
  \;=\;
  \underbrace{
  f^{(0)}{}_{a}{}^{c}{}_{b} \, e^{(0)}_{c}
  }_{\order{\epsilon^0}}
  +
  \underbrace{
  f^{(1)}{}_{a}{}^{c}{}_{b} \, e^{(0)}_{c}
  +
  f^{(0)}{}_{a}{}^{c}{}_{b} \, e^{(1)}_{c}
  }_{\order{\epsilon}}
  +
  \order{\epsilon^2}
  \nonumber\\
\end{eqnarray}
and therefore their first-order perturbation is found to be
\begin{eqnarray}
  \Rightarrow
  f^{(1)}{}_{a}{}^c{}_b
  &=&
  \left(
  \commutator{e^{(0)}_{a}}{e^{(1)}_{b}}  
  + 
  \commutator{e^{(1)}_{a}}{e^{(0)}_{b}}
  -
  f^{(0)}{}_{a}{}^{c^\prime}{}_{b} \, e^{(1)}_{c^\prime}
  \right)
  \inner e^{c(0)}
  \,.
\end{eqnarray}
Now the shift in the ONB connection

\begin{eqnarray}
  \omega_{abc}
  &=&
  \frac{1}{2}
  \left(
    f_{abc}
    +
    f_{bca}
    -
    f_{cab}
  \right)
\end{eqnarray}
is immediate:
\begin{eqnarray}
  \omega^{(1)}_{abc}  
  &=&
  \frac{1}{2}
  \left(
    f^{(1)}_{abc}
    +
    f^{(1)}_{bca}
    -
    f^{(1)}_{cab}
  \right)  
  \,.
\end{eqnarray}
The ONB components of the field strength
\begin{eqnarray}
  h_{abc} &=&
  e_a{}^\mu e_\nu{}^\mu e_a{}^\rho h_{\mu\nu\rho}
\end{eqnarray}
obviously receive the correction
\begin{eqnarray}
  h^{(1)}_{abc}
  &=&
  e^{(0)}_a{}^\mu e^{(0)}_b{}^\nu e^{(0)}_c{}^\rho h^{(1)}_{\mu\nu\rho}  
  +
  e^{(1)}_a{}^\mu e^{(0)}_b{}^\nu e^{(0)}_c{}^\rho h^{(0)}_{\mu\nu\rho}  
 +
  e^{(0)}_a{}^\mu e^{(1)}_b{}^\nu e^{(0)}_c{}^\rho h^{(0)}_{\mu\nu\rho}  
 +
  e^{(0)}_a{}^\mu e^{(0)}_b{}^\nu e^{(1)}_c{}^\rho h^{(0)}_{\mu\nu\rho} 
  \,.
  \nonumber\\
\end{eqnarray}
Also the perturbation of the $b$-deformed covariant
derivative operator \refer{b-deformed covariant derivative operator}
simply reads
\begin{eqnarray}
  \gradOp_a^{(b)(1)}
  &\defas&
  \frac{1}{4}\omega^{+(1)}_{abc}
  \clifford^{b+}\clifford^{c+}
  -
  \frac{1}{4}\omega^{-(1)}_{abc}
  \clifford^{b-}\clifford^{c-}
  \,,
\end{eqnarray}
where of course
\begin{eqnarray}
  \omega^{\pm(1)} &\defas& \omega^{(1)} \pm \frac{1}{2}h^{(1)}
  \,.
\end{eqnarray}
With these ingredients the perturbation of the supercharges (the Dirac operators)
are found to be (\cf \refer{b-deformed Dirac operators})
\begin{eqnarray}
  \label{perturbed b-deformed Dirac operators}
  \Dirac_{k\mp}^{(b)(1)}
  &=&
  \clifford_\pm^a
  \left(
  \gradOp^{(b)(1)}_a
  -
  i(b^{(1)}_{a\nu} \mp g^{(1)}_{a\nu})k^\nu
  \right)
  -
  \frac{1}{12}
  h^{(1)}_{abc}
  \clifford_\pm^a\clifford_\pm^b\clifford_\pm^c
  \,.
  \nonumber\\
\end{eqnarray}
Here it is assumed that $k^\mu$ remains unperturbed, which is the
case for the superstring, where $k^\mu \to X^{\prime \mu}\of{\sigma}$.

Finally this allows to write down an expression for the
perturbation of the target-space Hamiltonian
\refer{Hamiltonian generator in the presence of a b-field background}:	
In addition  to the
modification of the supercharges \refer{perturbed b-deformed Dirac operators}
the deformed Clifford generators \refer{b-deformed Clifford generators}
will receive a correction:
\begin{eqnarray}
  v_0 \inner \clifford^{(b)}_\pm
  &=&
  v_{0,\mu}\left(e_a{}^\mu \pm b^\mu{}_a\right)\clifford_\pm^a
  \nonumber\\
  &=&
  \left[v_{0,\mu}\left(e_a{}^\mu \pm b^\mu{}_a\right)\right]^{(0)}
  \clifford_\pm^a
  +
  \left[v_{0,\mu}\left(e_a{}^\mu \pm b^\mu{}_a\right)\right]^{(1)}
  \clifford_\pm^a
  +
  \cdots
  \nonumber\\
  &\defas&
  \left[v_0 \inner \clifford_\pm^{(b)}\right]^{(0)}
  +
  \left[v_0 \inner \clifford_\pm^{(b)}\right]^{(1)}
  +
  \cdots
  \,.
\end{eqnarray}
Therefore there are two contributions to the perturbation of the
Hamiltonian \refer{Hamiltonian generator in the presence of a b-field background}:
\begin{eqnarray}
  \left[{\bf H}^{(b)}_{v_0}\right]^{(1)}
  &=&
  \frac{i}{4}
  \left(
  \commutator{[v_0\inner \clifford_+^{(b)}]^{(1}}{\Dirac_-^{(b)}}
  -
  \commutator{[v_0\inner \clifford_-^{(b)}]^{(1)}}{\Dirac_+^{(b)}}
  \right)
  \nonumber\\
  &&
  +
  \frac{i}{4}
  \left(
  \commutator{v_0\inner \clifford_+^{(b)}}{[\Dirac_-^{(b)}]^{(1)}}
  -
  \commutator{v_0\inner \clifford_-^{(b)}}{[\Dirac_+^{(b)}]^{(1)}}
  \right)
  \,.
\end{eqnarray}

\subsubsection{Perturbation theory}
\label{qm perturbation theory}

With a Hamiltonian generator of target space time evolution in hand, the
standard techniques of quantum mechanical perturbation theory can be adapted.
The differences that one has to deal with are the need for the 
Krein space operator 
$\hermitianMetricOperator^{(b)}_{v_0}$
\refer{time fixed bfield Krein operator} 
 and the presence of non-vanishing
$K$ in the modifed Schr{\"o}dinger equation 
\refer{quasi Schroedinger equation for b-field background abbreviated form},
which may (but need not) appear in the presence of non-vanishing
Kalb-Ramond backgrounds. 

So what we are interested in is finding approximate solutions to the
Eigenvalue problem
\begin{eqnarray}
  \label{generalized eigenvalue problem}
 \left[(1 - K)i{\cal L}_{v_0} - \tilde{\bf H}_{v_0}^{(b)}\right]\ket{\phi}  &=& 0
  \nonumber\\
 \Leftrightarrow\;\;\;\;\;
  \left[(1 - K)E_{n} - \tilde{\bf H}_{v_0}^{(b)}\right]\ket{\phi_n}  &=& 0
\end{eqnarray}
on the basis that a solution to 0th order in the perturbation is known
\begin{eqnarray}
  \label{oth order solution to eigenvalue problem}
  \left[(1 - K^{(0)})E^{(0)}_{n} - \tilde{\bf H}_{v_0}^{(b)(0)}\right]\ket{\phi^{(0)}_n}  &=& 0
  \,.
\end{eqnarray}
Because $\tilde{\bf H}_{v_0}^{(b)(0)}$ is hermitian with respect to
$(\hermitianMetricOperator_{v_0}^{(b)})^{(0)}$ it follows that 
the $\phi_n^{(0)}$ for different $E_n^{(0)}$ are orthogonal 
with respect to $\bra{\cdot}(\hermitianMetricOperator^{(b)}_{v_0})^{(0)}(1-K^{(0)})\ket{\cdot}$.
We shall assume that they form a complete basis. The completeness relation can then be
written in the form
\begin{eqnarray}
  \ket{\phi^{(0)}}
  &=&
  \sum\limits_n
  \frac{
    \langle{\phi_n^{(0)}}|(\hermitianMetricOperator_{v_0}^{(b)})^{(0)}(1-K^{(0)})|{\phi^{(0)}}\rangle
   }{
     \langle\phi_n^{(0)}|(\hermitianMetricOperator_{v_0}^{(b)})^{(0)}(1-K^{(0)})|{\phi_n^{(0)}}\rangle
   }
   \ket{\phi_n^{(0)}}
  \,.
\end{eqnarray}

In order to find an expression for the first order perturbation of eigenvalues and states
we multiply equation \refer{generalized eigenvalue problem} from the left by
$
  \bra{\phi_m^{(0)}}
  (\hermitianMetricOperator^{(b)}_{v_0})^{(0)}
$,
which gives
\begin{eqnarray}
  \label{intermediate step in b-deformed first order perturbation}
  \bra{\phi_m^{(0)}}
  (\hermitianMetricOperator^{(b)}_{v_0})^{(0)}
  \left[
  (1 - (K^{(0)} + K^{(1)}))E_n 
  - 
  (\tilde{\bf H}_{v_0}^{(b)})^{(0)}
  -
   (\tilde{\bf H}_{v_0}^{(b)})^{(1)}
  \right]\ket{\phi_n^{(0)} + \phi_n^{(1)}}
  \;=\;
  0 + \cdots
  \nonumber\\
\end{eqnarray}
up to terms of higher than first order. 
(Here $A^{(m)}$ is the mth order perturbation of the object $A$.)
The point of taking the scalar product with respect to the
\emph{unperturbed} operator
$\hermitianMetricOperator^{(b)(0)}_{v_0}$ (see \refer{time fixed bfield Krein operator})
is that it allows us to use the 
hermiticity 
\refer{hermiticity of K}
of $(\tilde{\bf H}_{v_0}^{(b)})^{(0)}$ with respect to this scalar product 
to apply it to the left and write
\begin{eqnarray}
  \bra{\phi_m^{(0)}}
  (\hermitianMetricOperator^{(b)}_{v_0})^{(0)}
  (\tilde{\bf H}_{v_0}^{(b)})^{(0)}
  &\equalby{oth order solution to eigenvalue problem}&
  \bra{\phi_m^{(0)}}
  (\hermitianMetricOperator^{(b)}_{v_0})^{(0)}
  (1 - K^{(0)})E_m^{(0)}
  \,.
\end{eqnarray}
The remaining occurence of ${\cal L}_{v_0}$
in $(\tilde{\bf H}_{v_0}^{(b)})^{(1)}$
can be applied to the right to give, as usual,
\begin{eqnarray}
  i{\cal L}_{v_0}\ket{\phi_n} &=& E_n\ket{\phi} \;=\; 
(E_n^{(0)} + E_n^{(1)})\ket{\phi_n^{(0)} + \phi_n^{(1)}} + \cdots
  \,.
\end{eqnarray}
Inserting this in \refer{intermediate step in b-deformed first order perturbation} gives 
\begin{eqnarray}
  &&0 + \mbox{(second order perturtbations)}
  \nonumber\\
  &=&
  \bra{\phi_m^{(0)}}
  (\hermitianMetricOperator^{(b)}_{v_0})^{(0)}
  \Bigg[
  (1 - (K^{(0)} + K^{(1)}))(E_n^{(0)} + E_n^{(1)}) 
  - 
  (1-K^{(0)})E_m^{(0)}-
  \nonumber\\
  &&\hspace{80pt}
  -
   ({\bf H}_{v_0}^{(b)})^{(1)}
  + K^{(1)}(E_n^{(0)} + E_n^{(1)})
  \Bigg]\ket{\phi_n^{(0)} + \phi_n^{(1)}}
  \nonumber\\
  &=&
  \bra{\phi_m^{(0)}}
  (\hermitianMetricOperator^{(b)}_{v_0})^{(0)}
  \left[
  (1 - K^{(0)})(E_n^{(0)} + E_n^{(1)}) 
  - 
  (1-K^{(0)})E_m^{(0)}
  -
   ({\bf H}_{v_0}^{(b)})^{(1)}
  \right]\ket{\phi_n^{(0)} + \phi_n^{(1)}}
  \,.
  \nonumber\\
\end{eqnarray}
When we now set $m=n$ this
gives the sought-after expression for the first order energy shift: 
\begin{eqnarray}
  \label{general expression for energy shift}
  E_n^{(1)}
  &=&
  \frac{
    \bra{ \phi_n^{(0)}} 
    (\hermitianMetricOperator_{v_0}^{(b)})^{(0)} 
   ({\bf H}_{v_0}^{(b)})^{(1)}
   \ket{\phi_n^{(0)}}  
   }{
    \bra{ \phi_n^{(0)}} 
    (\hermitianMetricOperator_{v_0}^{(b)})^{(0)} 
    \left(
    1-
    K^{(0)}
    \right)
   \ket{\phi_n^{(0)}}  
   }
   \,.
\end{eqnarray}

Setting $m\neq n$ instead produces an equation for the first order shift of the states
\begin{eqnarray}
  \label{pre equation for first order shift in the states}
  \bra{\phi_m^{(0)}}
  \left[
  (\hermitianMetricOperator^{(b)}_{v_0})^{(0)}
   ({\bf H}_{v_0}^{(b)})^{(1)}
  +
  E_n^{(1)}
  K^{(0)} 
  \right]
  \ket{\phi_n^{(0)} }
  &=&
  (E_n^{(0)} - E_m^{(0)})
  \bra{\phi_m^{(0)}}
  (\hermitianMetricOperator^{(b)}_{v_0})^{(0)}
  (1 - K^{(0)}) 
  \ket{\phi_n^{(1)}}
  \,,
  \nonumber\\
\end{eqnarray}
which yields (when in the degenerate case the left hand side is appropriately diagonalized as usual) 
\begin{eqnarray}
  \ket{\phi_n^{(1)}}
  &=&
  \sum\limits_{m\neq n}
  \frac{1}{E_n^{(0)} - E_m^{(0)}}
  \frac{
  \bra{\phi_m^{(0)}}
  (\hermitianMetricOperator^{(b)}_{v_0})^{(0)}
   ({\bf H}_{v_0}^{(b)})^{(1)}
  +
  E_n^{(1)}
  K^{(0)} 
  \ket{\phi_n^{(0)} }
}{
    \bra{ \phi_n^{(0)}} 
    (\hermitianMetricOperator_{v_0}^{(b)})^{(0)} 
    \left(
    1-
    K^{(0)}
    \right)
   \ket{\phi_n^{(0)}}    
}  
 \ket{\phi_n^{(0)}}
  \,.
\end{eqnarray}

Both expression 
are essentially those familiar from 
perturbation theory of elementary quantum mechanics. The appearance of the $K^{(0)}$ term
is just a correction factor due to the fact that in the presence
of a non-vanishing $b$-field the Hamiltonian must be modified
(\cf \refer{modified Hamiltonian}) by an additional term in order to commute with the
time coordinate. Heuristically this is due to the fact that
the Kalb-Ramond torsion modifies the parallel transport along $v_0$.

We can use the special nature of our covariant Hamiltonian to write in the
numerator of \refer{general expression for energy shift}
\begin{eqnarray}
  \label{consideration wrt 1st order exp value of Hamiltonian}
  (\hermitianMetricOperator^{(b)})^{(0)}
  ({\bf H}^{(b)}_{v_0})^{(1)}
  \ket{\phi^{(0)}}
  &=&
  (\hermitianMetricOperator^{(b)})^{(0)}
  \left(
    \frac{i}{2}
    \left(
      v_0\inner \clifford^{(b)}_+ \Dirac^{(b)}_-
      -
      v_0\inner \clifford^{(b)}_- \Dirac^{(b)}_+
    \right)
    - i {\cal L}_{v_0}
  \right)^{(1)}
  \ket{\phi^{(0)}}
  \nonumber\\
  &=&
  (\hermitianMetricOperator^{(b)})^{(0)}
  \left(
    \frac{i}{2}
    \left(
      v_0\inner \clifford^{(b)}_+ (\Dirac^{(b)}_-)^{(1)}
      -
      v_0\inner \clifford^{(b)}_- (\Dirac^{(b)}_+)^{(1)}
    \right)
    - i ({\cal L}_{v_0})^{(1)}
  \right)
  \ket{\phi^{(0)}}
  \nonumber\\
  &=&
  -
  \left(
  \frac{i}{2}
  \left(
    v_0\inner\clifford_-^{(b)}
    (\Dirac^{(b)}_-)^{(1)}
    +    
    v_0\inner\clifford^{(b)}_+
    (\Dirac^{(b)}_+)^{(1)}
  \right)
  +
  (\hermitianMetricOperator^{(b)})^{(0)}
  i
  ({\cal L}_{v_0})^{(1)}
  \right)
  \ket{ \phi^{(0)} }
  \,,
  \nonumber\\
\end{eqnarray}
where $(\Dirac^{(b)}_{\pm})^{(0)}\ket{\phi^{(0)}} = 0$ has been used.
This expression drastically simplifies in the light cone limit:

\subparagraph{Light cone limit.}

When there are two independent light-like Killing vectors $p$ and $k$ with 
$p\inner p = 0 = k\inner k$ and $p \inner k = 1/2$, then $v_0$ is determined by
one boost parameter $\gamma$:
\begin{eqnarray}
  v_0 &\defas&
  e^\gamma p - e^{-\gamma}k
  \,.
\end{eqnarray}
If $v_0^\mu g^{(b)}_{\mu\nu} v_0^\nu$ is independent of $\gamma$ then
in the limit $\gamma\to \infty$ the norm of any state $\ket{\phi}$ for which
the expectation value
$\bra{\phi}p\inner \clifford_+^{(b)} p\inner \clifford_-^{(b)} \ket{\phi} \neq 0$
is dominated by this expectation value and scales as $e^{2\gamma}$. 
Hence expectation values 
$\bra{\phi}A\ket{\phi}/\bra{\phi}\hermitianMetricOperator^{(b)}\ket{\phi}$
of any other operator $A$ are in the light cone limit
given by their component which scales as $e^{2\gamma}$, i.e. by
$e^{2\gamma}\lim\limits_{\gamma \to \infty}\of{e^{-2\gamma}A}$.

Comparison with \refer{consideration wrt 1st order exp value of Hamiltonian} 
then shows that in the light cone limit we have
\begin{eqnarray}
  \label{light cone evaluation of Hamiltonian shift}
  \bra{\phi^{(0)}}
   (\hermitianMetricOperator_{v_0}^{(b)})^{(0)}
   ({\bf H}_{e^{-\gamma} v_0 }^{(b)})^{(1)}
  \ket{\phi^{(0)}}
  &\stackrel{\gamma\to\infty}{\longrightarrow}&
  -
  \bra{\phi^{(0)}}
   (\hermitianMetricOperator_{v_0}^{(b)})^{(0)}
   i ({\cal L}_{e^{-\gamma} v_0 })^{(1)}
  \ket{\phi^{(0)}}  
  \,.
\end{eqnarray}
This simplification is possible due to the special nature of the Hamiltonian,
which, as discussed in \S\fullref{Target space Killing evolution}, 
differs from ${\cal L}_{v_0}$ essentially only by being expressed in terms
of commutators of the supercharges instead of anticommutators.

A similar simplification of the denominator of 
\refer{general expression for energy shift} does not
occur \emph{in general} in the light cone limit. But for instance 
for the application that will be discussed in 
\S\fullref{covariant calculation of AdS spectrum}
$K$ simply vanishes in this limit.

$\,$\\

We have thus obtained a rather simple explicit general formula 
for the first order  energy shift
(as measured along some specified Killing vector field) of the supersymmetric
system under consideration. In order to evaluate it one just needs to
plug the expressions for the perturbed 
fields and operators discussed in 
\S\fullref{qm perturbation theory} into equation 
\refer{general expression for energy shift}
(or its light cone limit 
\refer{light cone evaluation of Hamiltonian shift}). 

Although this calculation may of course become tedious, 
it is straightforward. In particular there is no need to deal
with issues of gauge fixing and second-class constraints, which may
become quite involved in non-trivial backgrounds 
(\cf \S4 of \cite{CallanLeeMcLoughlinSchwarzSwansonWu:2003}).

One practical problem of the method presented here, though, 
inevitably arises precisely due to
its covariance: The shift in the covariant momentum is not (at least not generally)
restricted to be parallel to the particular Killing vector chosen to represent the
flow of parameter time, which is the only component measured by
\refer{general expression for energy shift}. This is no problem of principle,
because the remaining spacelike momenta shifts can be computed in perturbation
theory just as well: 

The shifted momenta along Killing vectors $v_i$ $i>0$ other than the timelike vector $v_0$ are
obtained by diagonalizing the first order perturbation of the matrix
\begin{eqnarray}
  \label{shift in longitudinal momenta}
  P^i_{nm}
  &\defas&
  \frac{
    \bra{\phi_n} (\hermitianMetricOperator^{(b)}_{v_0}) i{\cal L}_{v_i}\ket{\phi_m}
  }
  {
    \bra{\phi_n} (\hermitianMetricOperator^{(b)}_{v_0}) \ket{\phi_m}
  }
  \,,
\end{eqnarray}
which involves first order shifts of the states themselves.

However, as will be discussed in \S\fullref{curvature corrections to superstring spectra} 
in the context of a special 
example, one can choose adapted vielbein fields such that some
states don't receive any curvature corrections themselves. For such states then
formula \refer{general expression for energy shift} yields already all the 
desired information.

\newpage

\section{Curvature corrections to superstring spectra}
\label{curvature corrections to superstring spectra}

The previous sections made use only of very general properties of
supersymmetric quantum systems. In the following we specialize the formalism to the
superstring and demonstrate the covariant perturbation theory 
\S\fullref{qm perturbation theory} by calculuating
curvature corrections to the superstring spectrum in the toy example 
of ${\rm AdS}_3 \times {\rm S}^3$ in its Penrose limit.

The technology to treat superstrings in the differential geometric 
framework presented here is discussed in detail in \cite{Schreiber:2004},
which is the basis for the following discussion.

The rationale behind the following example calculation is analogous to that of
\cite{ParnachevSahakyan:2002},
where the authors test a perturbation method for the bosonic string in
${\rm AdS}_3 \times {\rm S}^3$ in order to later apply it
in \cite{ParnachevSahakyan:2002b} to the non-trivial ${\rm AdS}_5 \times {\rm S}^5$
case.
 
Other perturbation techniques for the superstring in light-cone gauge are
presented in \cite{CallanLeeMcLoughlinSchwarzSwansonWu:2003,DharMandalWadia:2003}.
The point of the formalism presented here is that it does not require to fix
light-cone gauge nor even the presence of a lightlike Killing vector in target space,
even though the latter does simplify the calculations. On the other hand, it is
not yet clear how to incorporate RR-backgrounds in the present formalism
(\cf \cite{Schreiber:2004}) which would be necessary for applications in 
${\rm AdS}_5$.  

We begin in \S\ref{loop space and B-field background} by demonstrating how the 
formalims developed here 
applies to the superstring in gravitational and Kalb-Ramond backgrounds.
\S\fullref{review of AdS3} reviews some general facts related to 
the ${\rm AdS}_3 \times {\rm S}^3$ and its pp-wave Penrose limit
which are then used in \S\fullref{covariant calculation of AdS spectrum} 
for the covariant
perturbative calculation of the superstring spectrum in this background,
following the methods developed in 
\S\fullref{section: covariant parameter evolution}.
For comparison \S\fullref{exact calculation of the spectrum} 
derives the exact spectrum
and in \S\fullref{discussion of perturbative result} the result is discussed.

\subsection{Superstrings in $B$-field backgrounds with loop space formalism}
\label{loop space and B-field background}

As is shown in detail in \cite{Schreiber:2004}, the 
closed superstring fits into the general framework of 
\S\fullref{section: covariant parameter evolution}
when the configuration space 
is identified with \emph{loop space},
the space of maps from the circle ${\rm S}^1$ into target space.
This loop space is coordinatized by the embedding fields 
$X^\mu\of{\sigma} =: X^{(\mu,\sigma)}$ 
and the metric $G_{(\mu,\sigma)(\nu,\sigma^\prime)}\of{X}$ on loop space 
which is induced by the target space metric $g_{\mu\nu}$ is taken to be
\begin{eqnarray}
  G_{(\mu,\sigma)(\nu,\sigma^\prime)}\of{X}
  :=
  g\of{X\of{\sigma}}
  \delta\of{\sigma,\sigma^\prime}
  \,.
\end{eqnarray}
On the exterior bundle over loop space there act the form creation
operators ${\cal E}^{\dag (\mu,\sigma)}$ 
and form annihilation operators ${\cal E}_{(\mu,\sigma)}$ 
(which, for finite dimensional manifolds, were denoted 
$\coordCreator^\mu$ and $\coordAnnihilator_\nu$, respectively, in
\S\fullref{creation/annihilation and clifford algebra}), that,
together with the coordinates $X^{(\mu,\sigma)}$ and their
partial derivatives $\partial^c_{(\mu,\sigma)}$, have the 
canonical supercommutators
\begin{eqnarray}
  \commutator{\partial^c_{(\mu,\sigma)}}{X^{(\nu,\sigma^\prime)}}
  &=&
  \delta_\mu^\nu\,\delta\of{\sigma,\sigma^\prime}
  \nonumber\\
  \antiCommutator
    {{\cal E}_{(\mu,\sigma)}}
    {{\cal E}^{\dag (\nu,\sigma^\prime)}}
  &=&
  \delta_\mu^\nu\,\delta\of{\sigma,\sigma^\prime}
  \,,
\end{eqnarray}
with all other brackets vanishing.

Independent of the metric $g_{\mu\nu}$ on target space this metric 
on loop space has a \emph{reparameterization isometry} generated by
the vector field
\begin{eqnarray}
  K^{(\mu,\sigma)} &\defas&
  T X^{\prime \mu}\of{\sigma}
  \,,
\end{eqnarray}
where the constant $T$ is identified with the string tension. It
it is this Killing vector field which, when used in 
\refer{k-deformed exterior derivatives}, gives the 
fermionic generators of the super Virasoro
algeba in the form of (modes of) the $K$-deformed exterior
(co-)derivative on loop space:
\begin{eqnarray}
  \label{modes of deformed extd coexted on loop space}
  \extd_{K,\xi}
  &=&
  \int d\sigma\;
  \xi\of{\sigma}
  \left(
    {\cal E}^{\dag\mu}\of{\sigma}
    \partial^c_\mu\of{\sigma}
    +
    i{\cal E}_\mu\of{\sigma}
    X^{\prime \mu}\of{\sigma}
  \right)
  \nonumber\\
  \coextd_{K,\xi}
  &=&
  -
  \int d\sigma\;
  \xi\of{\sigma}
  \left(
    {\cal E}^{\mu}\of{\sigma}
    \nabla_\mu\of{\sigma}
    +
    i{\cal E}^\dag_\mu\of{\sigma}
    X^{\prime \mu}\of{\sigma}
  \right)
  \,.  
\end{eqnarray}
Here $\xi$ is any complex function on ${\rm S}^1$.

A Kalb-Ramond $B$-field background on target space
induces a deformation of the loop space
exterior derivatives as discussed in general terms in \S\fullref{example background b field}.

So consider a 2-form $B$ field on target space
\begin{eqnarray}
  B &=& \frac{1}{2}B_{\mu\nu}dx^\mu \wedge dx^\nu
\end{eqnarray}
which induces on loop space the operator
\begin{eqnarray}
  {\bf W}^{(B)}\of{X}
  &\defas&
  \frac{1}{2}
  B_{(\mu,\sigma)(\nu,\sigma^\prime)}\of{X}
  {\cal E}^{\dag (\mu,\sigma)}{\cal E}^{\dag(\nu,\sigma^\prime)}
  \nonumber\\
  &\defas&
  \int d\sigma\;
  \frac{1}{2}B_{\mu\nu}\of{X\of{\sigma}}{\cal E}^{\dag \mu}\of{\sigma}
   {\cal E}^{\dag \nu}\of{\sigma}
\end{eqnarray}
with loop-space components
\begin{eqnarray}
  B_{(\mu,\sigma)(\nu,\sigma^\prime)}\of{X}
  &=&
  B_{\mu\nu}\of{X\of{\sigma}}\delta_{\sigma,\sigma^\prime}
  \,.
\end{eqnarray}
Being the integral over a weight 1 object this operator is (classically) reparameterization
invariant
\begin{eqnarray}
  \commutator{
    {\cal L}_\xi
  }{
   {\bf W}^{(B)} 
  }
  &=&
  0
  \,.
\end{eqnarray}
The deformations \refer{susy generators for B field background} now read
(setting $T=1$ for convenience)
\begin{eqnarray}
  \label{b-field deformed loop space susy generators}
  \extd_{K,\xi}^{(B)}
  &\defas&
  \exp\of{-{\bf W}^{(B)}}
  \extd_{K,\xi}
  \exp\of{{\bf W}^{(B)}}
  \nonumber\\
  &=&
  \extd_{K,\xi} + 
  \commutator{\extd_{K,\xi}}{{\bf W}^{(b)}}
  \nonumber\\
  &=&
  \int d\sigma\;
  \xi\of{\sigma}
  \Bigg(
  {\cal E}^{\dag\mu}\of{\sigma}\gradOp_\mu\of{\sigma}
  +
  i {\cal E}_\mu\of{\sigma}X^{\prime\mu}\of{\sigma}
  \nonumber\\
  &&
  +
  \frac{1}{6}
  H_{\alpha\beta\gamma}\of{X\of{\sigma}}
  {\cal E}^{\dag \alpha}\of{\sigma}{\cal E}^{\dag \beta}\of{\sigma}
   {\cal E}^{\dag \gamma}\of{\sigma}
  -
  i{\cal E}^{\dag\mu}B_{\mu\nu}\of{X\of{\sigma}}X^{\prime\nu}\of{\sigma}
  \Bigg)
  \nonumber\\
  \coextd_{K,\xi}^{(B)}
  &\defas& (\extd_K^{(B)})^\dag
  \nonumber\\
  &=&
  \exp\of{{\bf W}^{\dag(B)}}
  \coextd_K
  \exp\of{-{\bf W}^{\dag(B)}}
  \nonumber\\
  &=&
  -
  \int d\sigma\;
  \xi\of{\sigma}
  \Bigg(
  {\cal E}^{\mu}\of{\sigma}\gradOp_\mu\of{\sigma}
  +
  i {\cal E}^\dag_\mu\of{\sigma}X^{\prime\mu}\of{\sigma}
  \nonumber\\
  &&
  +
  \frac{1}{6}
  H_{\alpha\beta\gamma}\of{X\of{\sigma}}
  {\cal E}^{\alpha}\of{\sigma}{\cal E}^{\beta}\of{\sigma}
   {\cal E}^{\gamma}\of{\sigma}
  -
  i{\cal E}^{\mu}B_{\mu\nu}\of{X\of{\sigma}}X^{\prime\nu}\of{\sigma}
  \Bigg)
  \,.
  \nonumber\\
\end{eqnarray}
(It can be checked \cite{Chamseddine:1997} that this is indeed the same result  
found by  canonical analysis of the action of the respective 1+1 dimensional
nonlinear $\sigma$-model.)

In view of equation \refer{current algebra of the formal bosonic currents}
a crucial property that needs to be checked is the algebra of the
bosonic currents.
A straightforward but tedious calculation gives the result
\begin{eqnarray} 
  \label{commutator of general pre currents in Bfield background}
  \commutator{J_{a^\pm}^{\rm bos\pm}\of{\sigma}}{J_{b^\pm}^{\rm bos\pm}\of{\sigma^\prime}}
  &=&
   -\frac{i}{T}\Big(\; 
     \delta(\sigma,\sigma^\prime) \,f_{a^\pm}{}^{c^\pm}{}_{b^\pm}\, 
     J^{\rm bos \pm}_{c^\pm}\of{\sigma} 
     \mp 
     \delta^\prime\of{\sigma,\sigma^\prime}\, 2 G_{a b}\of{\sigma}
  \nonumber\\
  &&
  \mp
  2
  \delta\of{\sigma,\sigma^\prime}X^{\prime \mu}\of{\sigma}
  \omega^\pm[e^\pm]_{\mu a^\pm a^{\prime\pm}}
  +
  {\bf R}^{(h)}_{a^\pm b^\pm}
  \;\Big)
  \,,
  \nonumber\\
\end{eqnarray}
(where ${\bf R}^{(h)}_{a^\pm b^\pm}$ is the torsion deformed
curvature operator \refer{torsion deformed curvature operator}).
This equation holds true generally for arbitrary backgrounds with the
objects $f_a{}^c{}_b\of{\sigma}$ being the ``structure functions''
of the vielbein:
\begin{eqnarray}
  f_a{}^c{}_b \, e_c
  &\defas&
  \commutator{e_a}{e_b}
  \,.
\end{eqnarray}
For the special case of SWZW background fields these of course become the 
structure constants of the group and $\omega^\pm[e^\pm]$ vanishes
\refer{vanishing of the connections with torsion in WZW models}, so that
the $J_{a^\pm}^{\rm bos \pm}$ really do satisfy the \emph{current algebra}
\begin{eqnarray}
  \commutator{J_{a^\pm}^{\rm bos\pm}\of{\sigma}}{J_{b^\pm}^{\rm bos\pm}\of{\sigma^\prime}}
  &=&
   -i\frac{1}{T}\Big(\; 
     \mp\delta(\sigma,\sigma^\prime)\, f_{a^\pm}{}^{c^\pm}{}_{b^\pm}\, 
     J^{\rm bos\pm}_{c^\pm}\of{\sigma} 
     -
     \delta^\prime\of{\sigma,\sigma^\prime}\, 2G_{a^\pm b^\pm}\of{\sigma}
  \;\Big)
  \,.
  \nonumber\\
\end{eqnarray}
This is the functional, canonical version of what is usually written as a 
CFT OPE (e.g. \cite{DiFrancescoMathieuSenechal:1997,KacTodorov:1985})
\begin{eqnarray}
  \label{current current OPE}
  J^{\rm bos}_a\of{z} J^{\rm bos}_b\of{w}
  &=&
  \frac{\frac{k^{\rm bos}}{2} \eta_{ab}}{(z-w)^2}
  +
  \frac{i f_a{}^c{}_b J^{\rm bos}_c\of{w}}{z-w}
  \,,
\end{eqnarray}
where here $k^{\rm bos}$ is the \emph{level} of the current algebra 
generated by the bosonic currents
and
$\eta_{ab}$ the Killing metric of the respective Lie group.

In order to make the relation between the functional and the CFT notation
more manifest consider the modes
\begin{eqnarray}
   j^{\rm bos}_{a,n}
  &\defas&
  -T
  \int d\sigma\; J^{{\rm bos}-}_a\of{\sigma}e^{-in\sigma}
  \nonumber\\
  \tilde j^{\rm bos}_{a,n}
  &\defas&
  -T
  \int d\sigma\; J^{{\rm bos}+}_a\of{\sigma}e^{+in\sigma}
\end{eqnarray}
which satisfy the algebra
\begin{eqnarray}
  \commutator{j^{\rm bos}_{a,m}}{j^{\rm bos}_{b,n}}
  &=&
  m\, 4\pi T g_{ab}\, \delta_{m,-n}
  +i\,
  f_a{}^c{}_b\,
  j^{\rm bos}_{c,m+n}
  \;=\;
  m\,\frac{2}{\alpha^\prime} \,  g_{ab}\, \delta_{m,-n}
    +i\,
  f_a{}^c{}_b\,
  j^{\rm bos}_{c,m+n}
  \nonumber\\
  \commutator{\tilde j_{a,m}}{\tilde j_{b,n}}
  &=&
  m\, 4\pi T  g_{ab}\, \delta_{m,-n}
  +i\,
  f_a{}^c{}_b\,
  \tilde j^{\rm bos}_{c,m+n}
  \;=\;
  m\, \frac{2}{\alpha^\prime	}  g_{ab}\, \delta_{m,-n}
  +i\,
  f_a{}^c{}_b\,
  \tilde j^{\rm bos}_{c,m+n}
  \,.
  \nonumber\\
\end{eqnarray}
Comparison with the algebra of the modes
\begin{eqnarray}
  j^{\rm bos}_{a,n} &\defas&
  \oint \frac{dz}{2\pi i}
  j^{\rm bos}_a\of{z}
  z^n
\end{eqnarray}
which reads
\begin{eqnarray}
  \label{current mode algebra}
  \commutator{j^{\rm bos}_{a,m}}
  {j^{\rm bos}_{b,n}}
  &=&
  m \,\frac{k^{\rm bos}}{2}\eta_{ab}\, \delta_{n,-m}
  + 
  i
  f_a{}^c{}_b j^{\rm bos}_{c,m+n}
\end{eqnarray}
yields the relation
\begin{eqnarray}
  \label{relation bosonic level to group manifold scale}
  g_{ab}
  &=& 
  \frac{k^{\rm bos} \alpha^\prime}{4}
  \eta_{ab}
\end{eqnarray}
between level $k^{\rm bos}$ of the algebra of \emph{bosonic} currents
and the scale of the group manifold. Since the level $k$ of the total currents is
$k = k^{\rm bos} -2 \coxeter$
this gives finally
\begin{eqnarray}
  \label{relation level to group manifold scale}
  g_{ab}
  &=&  
  \frac{(k-2 \coxeter) \alpha^\prime}{4}
  \eta_{ab}
  \,.
\end{eqnarray} 

In summary, the above yields all the tools and information needed to apply the 
methods of \S\fullref{section: covariant parameter evolution} to
superstrings backgrounds that are supported by $B$-field flux.
An example of an application in this context is the content of
the next sections.

\subsection{Review of ${\rm AdS}_3 \times {\rm S}^3$ and its Penrose limit}
\label{review of AdS3}

The supergravity solution of $Q_5$ D5-branes wrapped on a four-torus
of volume $v$ together with $Q_1$ fundamental strings parallel to the 
D5-branes reads (\cite{GiveonKutasovSeiberg:1998}, \S4)
\begin{eqnarray}
  e^{-2\Phi}
  &=&
  \frac{1}{g^2}f_5^{-1}f_1
  \nonumber\\
  H &=&
  2
  \left(
    Q_5 \epsilon_3 + \frac{g^2 Q_1}{v} f_5 f_1^{-1} \ast_6 \epsilon_3
  \right)
  \nonumber\\
  ds^2
  &=&
  f^{-1}
  \left(
    -dx_0^2 + dx_1^2
  \right)
  +
  f_5
  \left(
    dr^2 + r^2 dS_3^2
  \right)
  + 
  dT_4^2
  \,,
\end{eqnarray}
where
\begin{eqnarray}
  f_1 &=& 1 + \frac{g^2 \alpha^\prime Q_1}{v}\frac{1}{r^2}
  \nonumber\\
  f_5 &=&
  1 + \alpha^\prime Q_5 \frac{1}{r^2}
  \,.
\end{eqnarray}
In the near-horizon limit, $\frac{1}{r^2} \gg 1$, the metric becomes
that of ${\rm AdS}_3 \times {\rm S}^3 \times T^4$:
\begin{eqnarray}
  \label{near-horizon limit of NS 5/F1}
  g^{-2} \;=\; e^{-2\Phi} &=& \frac{1}{v}\frac{Q_1}{Q_5}
  \nonumber\\
  H &=& 
  2 Q_5
  \left(
    \epsilon_3 + \ast_6 \epsilon_3
  \right)
  \nonumber\\
  ds^2
  &=&
  R^2 
  \left(
  r^2
  \left(
    -d x_0^2 + dx_1^2
  \right)
  +
  \frac{1}{r^2}
  dr^2
  +
  dS_3^2
  \right)
  + dT^4
  \nonumber\\
  &=&
  R^2
  \left(
    \underbrace{
    -\cosh^2\of{\rho} dt^2
    + d\rho^2
    +
    \sinh^2\of{\rho}d\phi^2
    }_{= ds^2_{{\rm Ads}_3}}
    +
    \underbrace{
    \cos^2\of{\theta} d\psi^2
    + d\theta^2
    +
    \sin^2\of{\theta}d\chi^2
    }_{= ds^2_{ {\rm S}^3 } }    
  \right)
  +
  dT^4
  \,,
  \nonumber\\
\end{eqnarray}
with
\begin{eqnarray}
  R^2 &\defas&
  Q_5 \alpha^\prime
  \,.
\end{eqnarray}
The metric is that of the group manifold 
${\rm SL}(2,\R)\times {\rm SU}(2) \times {\rm U}\of{1}^4$ and the $B$-field
provides the parallelizing torsion, so that superstrings on this
background are described by an SWZW model (\cf \S\fullref{WZW models}).	

Higher order corrections to the supergravity solutions will 
force the radius of ${\rm AdS}_3$ to be slightly different from that
of ${\rm S}^3$, as discussed below. Therefore write the metric 
(we will ignore the trivial ${\rm T}^4$-factor in the following)
as
\begin{eqnarray}
  ds^2 &=&
    R_{\rm SL}^2
    \left(
       -\cosh^2\of{\rho}\, dt^2 
       + d\rho^2
       + \sinh\of{\rho} d\phi^2
    \right)
    +
    R_{\rm SU}^2
    \left(
       \cos^2\of{\theta}\, d\psi^2 
       + d\theta^2
       + \sin\of{\theta} d\chi^2
    \right)
  \,.
  \nonumber\\
\end{eqnarray}
A possible choice of (left/right)-invariant vielbein 
fields (following \cite{MaldacenaOoguri:2000}, eq. (9)) is  
\begin{eqnarray}
  K_3 &\defas& 
  -\frac{i}{2}\partial_t + \frac{i}{2}\partial_\phi
  \nonumber\\
  K_+ &\defas&
  \frac{1}{2}
  \left(
  e^{+i(\phi+t)}\tanh\of{\rho}\partial_t
  -
  ie^{+i(\phi+t)}\partial_\rho
  +
  e^{+i(\phi+t)}\coth\of{\rho} \partial_\phi
  \right)
  \nonumber\\
  K_- &\defas&
  \frac{1}{2}
  \left(
  -
  e^{-i(\phi+t)}\tanh\of{\rho}\partial_t
  -
  ie^{-i(\phi+t)}\partial_\rho
  -
  e^{-i(\phi+t)}\coth\of{\rho} \partial_\phi  
  \right)
  \nonumber\\
  \nonumber\\
  J_3 &\defas& 
  -\frac{i}{2}\partial_\psi - \frac{i}{2}\partial_\chi
  \nonumber\\
  J_+ &\defas&
  \frac{1}{2}
  \left(
  -
  e^{+i(\chi+\psi)}\tan\of{\rho}\partial_\psi
  -
  ie^{+i(\chi+\psi)}\partial_\theta
  +
  e^{+i(\chi+\psi)}\cot\of{\rho} \partial_\chi
  \right)
  \nonumber\\
  J_- &\defas&
  \frac{1}{2}
  \left(
  e^{-i(\chi+\psi)}\tan\of{\rho}\partial_\psi
  -
  ie^{-i(\chi+\psi)}\partial_\theta
  -
  e^{-i(\chi+\psi)}\cot\of{\rho} \partial_\chi    
  \right)
  \,.
\end{eqnarray}

\begin{eqnarray}
  K_3 &\defas& 
  -\frac{i}{2}\partial_t + \frac{i}{2}\partial_\phi
  \nonumber\\
  K_+ &\defas&
  \frac{1}{2}
  \left(
  e^{-i(\phi-t)}\tanh\of{\rho}\partial_t
  -
  ie^{-i(\phi-t)}\partial_\rho
  -
  e^{-i(\phi-t)}\coth\of{\rho} \partial_\phi
  \right)
  \nonumber\\
  K_- &\defas&
  \frac{1}{2}
  \left(
  -
  e^{+i(\phi-t)}\tanh\of{\rho}\partial_t
  -
  ie^{+i(\phi-t)}\partial_\rho
  +
  e^{+i(\phi-t)}\coth\of{\rho} \partial_\phi  
  \right)
  \nonumber\\
  \nonumber\\
  J_3 &\defas& 
  -\frac{i}{2}\partial_\psi + \frac{i}{2}\partial_\chi
  \nonumber\\
  J_+ &\defas&
  \frac{1}{2}
  \left(
  -
  e^{-i(\chi-\psi)}\tan\of{\rho}\partial_\psi
  -
  ie^{-i(\chi-\psi)}\partial_\theta
  -
  e^{-i(\chi-\psi)}\cot\of{\rho} \partial_\chi
  \right)
  \nonumber\\
  J_- &\defas&
  \frac{1}{2}
  \left(
  e^{+i(\chi-\psi)}\tan\of{\rho}\partial_\psi
  -
  ie^{+i(\chi-\psi)}\partial_\theta
  +
  e^{+i(\chi-\psi)}\cot\of{\rho} \partial_\chi    
  \right)
  \,.
\end{eqnarray}
These vectors are normalized so as to have have the standard non-vanishing Lie brackets: 
\begin{eqnarray}
  \commutator{K_3}{K_\pm} &=& \pm K_\pm
  \nonumber\\
  \commutator{K_+}{K_-} &=& -2 K_3
  \nonumber\\
  \commutator{J_3}{J_\pm} &=& \pm J_\pm
  \nonumber\\
  \commutator{J_+}{J_-} &=& +2 J_3
\end{eqnarray}
This fixes their inner products with respect to $ds^2$ to
\begin{eqnarray}
  K_3 \inner K_3 &=& \frac{R_{\rm SL}^2}{4}
  \nonumber\\
  K_+ \inner K_- &=& -2 \frac{R_{\rm SL}^2}{4}
  \nonumber\\
  J_3 \inner J_3 &=& -\frac{R_{\rm SU}^2}{4}
  \nonumber\\
  J_+ \inner J_- &=& -2 \frac{R_{\rm SU}^2}{4}
  \,. 
\end{eqnarray}
This is $\pm\frac{1}{4}R_{{\rm SL}/{\rm SU}}^2$ times the Killing metric on the 
two group manifolds.\footnote{
With 
$e_0 \defas K_3\,,\, e_1 \defas K_+, \cdots$ and
the structure constants $f_a{}^c{}_b$ defined by
$\commutator{e_a}{e_b} = f_a{}^c{}_b e_c$
we have
\begin{eqnarray}
  -\frac{1}{2}f_a{}^r{}_s f_b{}^s{}_r
  &=&
  \left[
    \begin{array}{cccccc}
      -1 & 0 & 0 & 0 & 0 & 0 \\
      0 & 0 & 2 & 0 & 0 & 0 \\
      0 & 2 & 0 & 0 & 0 & 0 \\
      0 & 0 & 0 & -1 & 0 & 0 \\
      0 & 0 & 0 & 0 & 0 & -2 \\
      0 & 0 & 0 & 0 & -2 & 0
    \end{array}
  \right]
  \,.
\end{eqnarray}
} 
According to 
\refer{relation level to group manifold scale} this means that the level of the associated
algebra of \emph{total} currents is 
\begin{eqnarray}
  k  &=& R_{\rm SL}^2/\alpha^\prime -2
  \nonumber\\
   &=& R_{\rm SU}^2/\alpha^\prime + 2
  \,,
\end{eqnarray}
which is, up to a small correction, proportional to the size of spacetime in units of the string scale.
For the calculation of the exact string spectrum on ${\rm AdS}_3 \times {\rm S}^3$
(this is discussed in \S\fullref{exact calculation of the spectrum} below) one needs 
the quadratic Casimir
\begin{eqnarray}
  C &\defas& -\eta^{ab} e_a e_b
  \nonumber\\
  &=&
  -K_3 (K_3 + 1) + K_- K_+ + J_3(J_3 + 1) + J_- J_+
  \,.
\end{eqnarray}

A particularly interesting further limit is the \emph{Penrose limit}
of the ${\rm AdS}_3 \times {\rm S}^3$ background 
(see e.g. \cite{Hikida:2003}). 
It is obtained by
concentrating on the vicinity of a lightlike geodesic going around the 
equator of the ${\rm S}^3$ factor, i.e. one with momentum proportional
to $K_3  \pm J_3$. 

In order to find the background structure in this limit introduce
the following vielbein basis adapted to this geodesic motion:
\begin{eqnarray}
  \label{adapted vielbein basis}
  F &\defas&
  \frac{1}{k}\left(J_3 - K_3\right)
  \nonumber\\
  J &\defas& J_3 + K_3
  \nonumber\\
  P_1 &\defas& \frac{1}{\sqrt{k}} K_+
  \nonumber\\
  P^\ast_1 &\defas& \frac{1}{\sqrt{k}} K_-
  \nonumber\\  
  P_2 &\defas& \frac{1}{\sqrt{k}} J_+
  \nonumber\\
  P^\ast_2 &\defas& \frac{1}{\sqrt{k}} J_-
  \,.
\end{eqnarray}
Since their non-vanishing commutators are
\begin{eqnarray}
  \commutator{J}{P_i} &=& P_i
  \nonumber\\
  \commutator{J}{P^\ast_i} &=& - P^\ast_i
  \nonumber\\
  \commutator{P_1}{P_1^\ast} &=& F - \frac{1}{k}J
  \nonumber\\
  \commutator{P_2}{P_2^\ast} &=& F + \frac{1}{k}J
  \nonumber\\
  \commutator{F}{P_i^{(\ast)}} &=& \pm \frac{1}{k}P_i^{(\ast)}
\end{eqnarray}
one sees that in the Penrose limit $J_3 - K_3 \approx k \to \infty$
with $J_3 \approx -K_3$ the Lie algebra \emph{contracts} to that
of the so-called extended Heisenberg group $H_6$ 
which describes a pp-wave background \cite{Hikida:2003}. 

\subsection{Covariant perturbative calculation of the  superstring spectrum}
\label{covariant calculation of AdS spectrum}

Our aim is to use the perturbation theory of 
\S\fullref{section: covariant parameter evolution} to calculate
(in the same spirit as \cite{ParnachevSahakyan:2002} 
but using covariant techniques and superstrings) the correction to
the sring spectrum in the small parameter $1/k$. That is,
we start with the exact spectrum of superstrings on the $H_6$
pp-wave background and then turn on curvature corrections turning
the pp-wave background into the true ${\rm AdS}_3 \times {\rm S}^3$
geometry.

The calculation involves computing the various perturbed quantities
discussed in \S\fullref{Perturbation of background fields}. Most importantly,
one finds for the metric in the adapted vielbein basis 
\refer{adapted vielbein basis}
the expansion
\begin{eqnarray}
  \eta_{ab}
  &=&
  \frac{\alpha^\prime}{2}
  \left[
    \begin{array}{cccccc}
      0 & 1 & 0 & 0 & 0 & 0 \\
      1 & -2 & 0 & 0 & 0 & 0 \\
      0 & 0 & 0 & -1 & 0 & 0 \\
      0 & 0 & -1 & 0 & 0 & 0 \\
      0 & 0 & 0 & 0 & 0 & -1 \\
      0 & 0 & 0 & 0 & -1 & 0
    \end{array}
  \right]
  +
  \frac{\alpha^\prime}{k}
  \left[
    \begin{array}{cccccc}
      0 & 0 & 0 & 0 & 0 & 0 \\
      0 & 0 & 0 & 0 & 0 & 0 \\
      0 & 0 & 0 & 1 & 0 & 0 \\
      0 & 0 & 1 & 0 & 0 & 0 \\
      0 & 0 & 0 & 0 & 0 & -1 \\
      0 & 0 & 0 & 0 & -1 & 0
    \end{array}
  \right]
  + \order{1/k^2}
  \,.
\end{eqnarray}
An examination of the $B$-deformed vielbein field \refer{b-defomrmed vielbeine}
shows that the calculation simplifies when $v_0 := e^\gamma F - e^{-\gamma}J$
for $\gamma \to \infty$ is chosen as the timelike Killing vector 
\refer{the timelike Killing vector}, since then the correction operator $K$
\refer{definition of K} \emph{vanishes}, $K=0$ and we can make use of 
formula \refer{light cone evaluation of Hamiltonian shift} to evaluate the
first order shift of string energy as measured along the Killing vector $v_0$
by computing the expectation value of the first order shift 
in the loop-space Lie derivative along $F$. By using equation 
\refer{total WZW currents are deformed Lie derivatives} one finds that this
Lie derivative is just the sum of the left- and right-moving total
SWZW \emph{currents} along $F$
\refer{definition of bsosnic current as superpartner of fermion in general formalism}:
\begin{eqnarray}
  i \left({\cal L}_F\right)
  &=&
  -i\left(J^+_{F,0} + J^-_{F,0}\right)
  \,.
\end{eqnarray}
The perturbation in this loop-space Lie derivative is most conveniently
computed using formula \refer{coordinate Killing Lie derivative}
in the appendix.
One finds
\begin{eqnarray}
  \label{shift in lightcone Lie derivative along F}
  i \left({\cal L}_F\right)^{(1)}
  &=&
  -
  \frac{1}{k} 
  \frac{1}{\alpha^\prime}
  \int d\sigma\;
  \left(
  \Gamma_{P_1,+} \Gamma_{P_1^\ast,+}
  -
  \Gamma_{P_2,+} \Gamma_{P_2^\ast,+}
  -
  \Gamma_{P_1,-} \Gamma_{P_1^\ast,-}
  +
  \Gamma_{P_2,-} \Gamma_{P_2^\ast,-}
  \right)
  \,.
  \nonumber\\
\end{eqnarray}
According to formula 
\refer{light cone evaluation of Hamiltonian shift} 
the energy shifts that we are looking for
are the expectation values of 
\refer{shift in lightcone Lie derivative along F} in the unperturbed states.
Obviously \refer{shift in lightcone Lie derivative along F} is just a
kind of fermion number-operator. To be more precise, let 
$N^{\prime {\rm fer}}_{\rm SL}$ be the number of $\Gamma_{P_1}$
excitations minus the number of $\Gamma_{P_1^\ast}$ excitations
of the string,
and similarly let $N^{\prime {\rm fer}}_{\rm SU}$ be the number of
$\Gamma_{P_2}$
excitations minus the number of $\Gamma_{P_2^\ast}$ excitations
for both the left- and the right-moving sector.
This is a measure for the fermionic contribution to the 
angular momentum of the string state with respect to $K_3$ and $J_3$
(\cf \cite{Hikida:2003} and see also the discussion 
\S\fullref{exact calculation of the spectrum} below).
By explicitly constructing the 
bosonic and fermionic 
physical DDF states for Type II strings
in the pp-wave limit of ${\rm AdS}_3 \times {\rm S}^3$
(a calculation that closely follows \cite{Hikida:2003} and will therefore
not be given here) one checks that our unperturbed states 
are indeed eigenstates with respect to $\hat N^{\prime {\rm fer}}_{\rm SL}$
and $\hat N^{\prime {\rm fer}}_{\rm SU}$.

This means that
we can finally write down the expectation values of 
\refer{shift in lightcone Lie derivative along F} in
the unperturbed states,
which are nothing but the energy shifts $E^{(1)}$ that we are looking for,
as
\begin{eqnarray}
  \label{final result}
  E^{(1)}
  &=&
  \frac{1}{k}
  \left(
    N^{\prime {\rm fer}}_{\rm SL}
    -
    N^{\prime {\rm fer}}_{\rm SU}
  \right)
  \,.
\end{eqnarray}

This is the result of our covariant perturbative calculation of the
spectrum of Type II superstrings around the pp-wave limit of
${\rm AdS}_3 \times {\rm S}^3$.

In order to check this result, the next section discusses 
the calculation of the \emph{exact}
superstring spectrum on ${\rm AdS_3}\times {\rm S}^3$. 
The result is further discussed in
\S\fullref{discussion of perturbative result}

\subsection{Exact calculation of the spectrum}
\label{exact calculation of the spectrum}

In the following a generalization of the discussion in \S 4 of 
\cite{ParnachevSahakyan:2002} is given, calculating the
lightcone energy of superstrings on ${\rm AdS}_3 \times {\rm S}^3$
to all orders in $1/k \approx  \alpha^\prime/R^2$.

We first consider the bosonic string 
(and concentrate on one chirality sector for notational brevity):
Let $-h(h+1) + j(j+1)$ be the eigenvalue of the Casimir $\eta_{ab}J^a_0 J^b_0$
of the ${\rm SL}\of{2,\R}\times {\rm SU}\of{2}$ current algebra and
let $N \in \N$ be the level of a given state. Then the $L_0$ Virasoro constraint on this
state reads 
\begin{eqnarray}
  \label{SLxSU L0 constraint}
  -\frac{h(h+1)}{k} + \frac{j(j+1)}{k} + N = a
\end{eqnarray}
for a given normal ordering constant $a$.

The eigenvalues of $h^3$ and $j^3$ of the zero modes of 
$K^3_0$ and $J^3_0$ can be written as
\begin{eqnarray}
  h^3 &=& h + N^\prime_{\rm SL}
  \nonumber\\
  j^3 &=& j + N^\prime_{\rm SU}
  \,,  
\end{eqnarray} 
where, for instance, $N^\prime_{\rm SU}$ grows by one for every 
$J^+_{-n}$ (bosonic current) excitation and
is reduced by one for every $J^-_{-n}$ excitation 
(due to $\commutator{J^3_0}{J^\pm_{n}} = \pm J^{\pm}_n$
and $\commutator{J^3_0}{\psi^\pm_{n}} = \pm \psi^{\pm}_n$).

The characteristic lightlike momenta of the $H_6$ model are,
according to \refer{adapted vielbein basis}, the light cone energy $H$ 
associated with the vector field $J$ and
transversal momentum $p_-$ associated with the vector field $F$:
\begin{eqnarray}
  \label{H6 lightlike momenta in terms of SLxSU eigenvalues}
  H &=& h^3 + j^3
  \nonumber\\
  p_- &=& \frac{1}{k}\left(h^3 - j^3\right)
  \,.
\end{eqnarray} 
Using the physical state condition \refer{SLxSU L0 constraint} we want to express these
momenta as functions of each other and of the transverse excitations:
\begin{eqnarray}
  \label{general functional relation between H and p-}
  H &=& H\of{p_-,N,N^\prime}
  \nonumber\\
  p_- &=& p_-\of{H,N,N^\prime}
  \,.
\end{eqnarray} 
Solving \refer{SLxSU L0 constraint} for $h$ and picking the positive solution yields
\begin{eqnarray}
  h &=& {\frac{-1 + {\sqrt{1 + 4\,j + 4\,{j^2} - 4\,a\,k + 4\,k\,n}}}{2}}
  \,.
\end{eqnarray}
Furthermore, 
equations \refer{H6 lightlike momenta in terms of SLxSU eigenvalues} 
solved for $j$ give, respectively:
\begin{eqnarray}
  j &=&
  \frac{
    H + H^2 - k(N-a)
   + (N^\prime_{\rm SL} + N^\prime_{\rm SU})
   \left(
     1 + 2H + N^\prime_{\rm SL} + N^\prime_{\rm SU}
   \right)
  }
  {
    2
    \left(
      1 + H + N^\prime_{\rm SL} + N^\prime_{\rm SU}
    \right)
  }
  \nonumber\\
  j &=&
  \frac{
    k(a - N - p_-)
    + N^\prime_{\rm SU} - N^\prime_{\rm SL}
    +
    \left(
      k p_- + N^\prime_{\rm SU} - N^\prime_{\rm SL}
    \right)^2
  }
  {2\left(N^\prime_{\rm SL} - N^\prime_{\rm SU} - k p_-\right)}
\end{eqnarray}
Inserting $k$ and $j$ in \refer{H6 lightlike momenta in terms of SLxSU eigenvalues}
yields the simple result
\begin{eqnarray}
  \label{exact result for lightcone spectrum H}
  H 
  &=&
  -1 + N^\prime_{\rm SL} + N^\prime_{\rm SU}
  + \frac{N-a}{p_- - (N^\prime_{\rm SL}-N^\prime_{\rm SU})/k}
  \\
  \label{exact result for lightcone spectrum p}
  \Leftrightarrow\;
  p_-
  &=&
  \frac{N - a}{ 1+ H - N^\prime_{\rm SL} - N^\prime_{\rm SU}}
  +
  \frac{N^\prime_{\rm SL} - N^\prime_{\rm SU}}{k}
  \,.
\end{eqnarray}
When the expression for $H$ is expanded to first order in $1/k$ and $N^\prime_{\rm SL} = 0$ 
we get the result known from
\cite{ParnachevSahakyan:2002}. Note that the series for $p_-$ stops already after the
first order.

The generalization to the superstring is immediate: 
The $j$ and $h$ quantum numbers are
now those associated with the \emph{bosonic} currents $J^a_{\rm bos}$ but for the
light cone generators we have to use the total currents $J^a_{\rm tot}$,
since these act a Lie
derivatives , \cf \refer{total WZW currents are deformed Lie derivatives}.
The total currents are just the sum of the bosonic and the fermionic currents
\refer{definition of bsosnic current as superpartner of fermion in general formalism}
We therefore have simply
\begin{eqnarray}
   h_3 &=& h_3^{\rm bos} + h_3^{\rm fer} + N^\prime_{\rm SL}
   \nonumber\\
   h_3 &=& h_3^{\rm bos} + h_3^{\rm fer} + N^\prime_{\rm SU}
\end{eqnarray}
and
\begin{eqnarray}
  N^\prime_{\rm SL} &=&   N_{\rm SL}^{\prime \rm bos} + N_{\rm SL}^{\prime \rm fer}
  \nonumber\\
  N^\prime_{\rm SU} &=&   N_{\rm SU}^{\prime \rm bos} + N_{\rm SU}^{\prime \rm fer}
  \,.
\end{eqnarray}
 
In summary, the first order perturbation $(p_-)^{(1)}$
of $p_-$ (the momentum associated with the lighlike Killing vector $F$) 
for fixed $H$ is
\begin{eqnarray}
  \label{exact result}
  (p_-)^{(1)}
  &=&
  \frac{1}{k}
  \left(
    N^\prime_{\rm SL} - N^\prime_{\rm SU}
  \right)
  \,.
\end{eqnarray}

\subsection{Discussion of the perturbative result}
\label{discussion of perturbative result}

Comparison of the perturbative result \refer{final result}
with the exact calculation \refer{exact result}
seems to show that the covariant perturbation theory reproduces the
fermionic contribution exactly, while it seemingly misses the 
bosonic one completely. A little reflection shows however that the comparison of these
results has to take into account the following subtlety:

In the perturbative calculations which use the lightcone gauge
(as in \cite{ParnachevSahakyan:2002}) one can fix the longitudinal
momentum $p_-$ by hand while turning on the perturbation and
calculate the shift in lightcone energy $H$ for fixed $p_-$ as in
equation \refer{exact result for lightcone spectrum H}. 

However, the covariant framework that has been presented here does not
fix any gauge and in particular does not fix any of the lightcone momenta.
This means that when the background is perturbed, the states are free 
to acquire shifts in $H$ \emph{or} in $p_-$ \emph{or} in both. But 
only the combination of both $H$ and $p_-$ has
invariant meaning, which is encoded in the relations 
\refer{exact result for lightcone spectrum H}
and \refer{exact result for lightcone spectrum p} that express
$p_-$ as a function of $H$ and the excitations of the string,
or vice versa \refer{general functional relation between H and p-}.

For a complete covariant perturbative result one therefore would need to
compute not only the energy shift along $v_0$ (i.e. $p_-$ in the above case), 
but also the shift in the other longitudinal momentum ($H$). This has already been discussed
at the end of \S\fullref{qm perturbation theory}, where it was pointed
out that the computation of the shift in the second longitudinal momentum 
is tedious, because it requires knowledge of the first order perturbation of 
the states themselves.

Here we shall be content with arguing that for fermionic states
no shift in $H$ occurs, according to \refer{shift in longitudinal momenta}.
The reason is that on the one hand side one can calculate the shift in
the loop-space Lie derivative along the vector field $J$ (which measures 
the momentum $H$ according to 
\refer{H6 lightlike momenta in terms of SLxSU eigenvalues}) to be purely
bosonic, having vanishing expectation value in the unperturbed fermionic
states. Furthermore, no shift in
the fermionic \emph{states}
can expected to give any contribution to a shift in $H$ due to
\refer{shift in longitudinal momenta},
because the fermionic states are created by the 
$\Gamma_{P_i^{(\ast)}}$ oscillators together with longitudinal terms
that ensure physicality (as in the DDF construction).
But since the $\Gamma_{P_i^{(\ast)}}$ are defined with respect to the invariant
vielbein \refer{adapted vielbein basis} they receive no correction in $1/k$, according 
to equation \refer{fields which receive no perturbation}. A fermionic state
in the pp-wave limit created by a given mode of the operator $\Gamma_{P_i^{(\ast)}}$
should hence flow to a state of the full ${\rm AdS}$ background created by
the same operator $\Gamma_{P_i^{(\ast)}}$ possibly accompanied by
different longitudinal excitations. But these do not contribute to any
inner products.
 
On the other
hand, bosonic states are created by the bosonic currents, which, according to
\refer{first general definition of the currents} are rather complicated expressions
involving products of the background metric and background $b$-field
with the elementary fields \refer{fields which receive no perturbation}.
Therefore nothing can be said in general about the first order shift for $H$ of the
bosonic states, while $H^{(1)}$ for the fermionic states should vanish.

From these considerations it follows that equation 
\refer{final result} gives
the shift of $p_-$ for fixed $H$ for fermionic states, while it tells
us nothing about the shift of $H$ for bosonic states. In conclusion then
the perturbative result \refer{final result} gives the correct 
result \refer{exact result for lightcone spectrum H} for all the cases where it applies,
which are the fermionic states.
The other cases may be treated, too, in principle, but require much more
computational effort, since they require a computation of the first order shift 
in the states themselves. This is the price to be paid due to working in
a fully covariant framework where no worldsheet gauge is fixed. Hence
we find a partial result using a relatively elegant calculation, while
the full result requires tedious work.

What then is the point of using the covariant perturbative calculation
presented here, if, as in the example discussed, the calculation of the
full result is more involved than the corresponding calculation
using lightcone gauge? There are two answers:

First, one should note that the fermionic spectrum which we obtained easily
is, according to \S\fullref{exact calculation of the spectrum}, 
an exact mirror image of the bosonic spectrum.
As long as one knows that this is the case the calculation of the
energy shift of the fermionic states, which is simple
in our framework, already yields the full information about energy shifts of
all states.

Second, the motivation for the construction
of the perturbation scheme developed in  
\S\fullref{section: covariant parameter evolution} was to find a method
that is more generally applicable
than the methods requiring lightcone gauge
are, since no lightlike Killing vector is required on target space. 
It is almost inevitable that the more general method is more
involved than the one which is adapted to special cases of high symmetry.

A more general assessment of what has been accomplished here is given
in the following section.

\section{Conclusion}

It has been shown that covariant Hamiltonian evolution operators can be
constructed in relativistic supersymmetric quantum (field) theories for
a large class of interesting backgrounds, by reformulating these theories
as generalized Dirac-K{\"a}hler systems on the exterior bundle over
their bosonic configuration space. 

The crucial insight was that 
any
system of supersymmetry constraints $\Dirac_\pm\ket{\psi} = 0$ can equivalently
be rewritten as a Schr\"odinger equation generating evolution along a
time paramater together with a constraint on hypersurfaces orthogonal to that
time parameter. In various guises this construction is well familiar from
both the Dirac particle as well as the classical Maxwell field. It
is no coincidence that these two systems are related to the supersymmetric
formalism discusssed here, since they can be regarded as two sectors of
the NSR super\emph{particle}, i.e. the point particle limit of the
NSR superstring. 
We have shown how to incorporate both sectors in one coherent formalism and how
to generalize this to backgrounds with a non-vanishing 2-form Kalb-Ramond
field and hence in particular to supersymmetric Wess-Zumino-Novikov-Witten
models. 

In doing so we made use of the fact that the supersymmetry constraints
for such backgrounds can be obtained from those for trivial backgrounds by
an algebra homomorphism which generalizes the deformations considered
by Witten in \cite{Witten:1982}. 
This is crucial, because, as we have shown, by appropriately applying 
similar deformations to all
operators which appear in the construction of the covariant Hamiltonian 
for trivial backgrounds one obtains the covariant Hamiltonian for the non-trivial
background. 

One subtlety that remains is that the Hamiltonian obtained this way,
though satisfying a formal Schr\"odinger equation,
in general no longer commutes with the time parameter coordinate. But this can be
fixed by appropriately subtracting the offending terms consistently on both
sides of the Sch\"odinger equation.

When all this is done it is rather straightforward to adapt the familiar
techniques of quantum mechanical perturbation theory: After dealing with
the indefiniteness of the Hodge inner product by employing a Krein
space operator and after taking into account the above mentioned 
correction to the Hamiltonian operator one obtains an equation for the
first order energy shift that is formally very similar to the one derived in
elementary quantum mechanics. 

Because it is of importance for the application presented in 
\S\fullref{curvature corrections to superstring spectra}
 we finally considered the case where the Hamiltonian
evolution is along a (almost) lightlike vector. It turns out that 
the special nature of the Hamiltonian considered here,
together with the presence of that Krein space operator, 
leads to a considerable simplification of the formula for the
first order energy shift in this case. 

It should be noted,
that this does not involve fixing any gauge, whatsoever, in particular
this is not related to fixing a light cone gauge. 
The methods presented here are equally valid in backgrounds which
do not posses any lightlike Killing vectors at all. 
This makes them interesting for the study of superstring theory in
arbitrary nontrivial backgrounds.

As demonstrated in 
\S\fullref{loop space and B-field background}
and \cite {Schreiber:2004}, 
the machinery developed here carries over to
the case where the underlying manifold is loop space, the configuration
space of the string. The calculation presented in 
\S\fullref{covariant calculation of AdS spectrum} 
demonstrates how
to apply the above perturbation scheme to perturbatively calculate the first
order curvature correction for superstrings close to the pp-wave limit
of ${\rm AdS}_3 \times {\rm S}^3$, as was done for the bosonic string
in light cone gauge in \cite{ParnachevSahakyan:2002}. 

It turns out that the calculation of fermionic states 
(those created by fermionic worldsheet oscillators from the ground state)
by our method fully profits from the elegance of the covariant approach,
while the first order spectrum of bosonic states requires knowledge of the 
first order shift in the states themselves, a fact that makes any
direct calculation much more tedious. This has been discussed in
detail in \S\fullref{discussion of perturbative result}. 

The natural next step would be to apply our formalism to superstring
spectra on ${\rm AdS}_5 \times {\rm S}^5$
(\cf \cite{CallanLeeMcLoughlinSchwarzSwansonWu:2003}). 
This requires the as yet
unknown incorporation of RR-background fields into the framework of 
\S\fullref{section: covariant parameter evolution}­. 
As is well known, RR-backgrounds are almost impossible
to handle in terms of $\sigma$-models and Lagrangian formalism. Therefore
it would be interesting to further analyze the deformation mechanism of
\S\fullref{deformations of the supersymmetry generators}. 
Possibly this way backgrounds can be incorporated that
defy a Lagrangian description. First steps in this direction are
discussed in \cite{Schreiber:2004} and \cite{Giannakis:2002}.

$\,$\\
\acknowledgments{
I am grateful to Robert Graham for helpful discussions 
and valuable assistance.
This work has been supported by the SFB/TR 12.
}

\newpage
\appendix
\section{Differential geometry in terms of operators on the exterior bundle}
\label{differential geometry}
\setcounter{equation}{0}
\renewcommand{\theequation}{\Alph{section}.\arabic{equation}}

\subsection{Creation/Annihilation and Clifford algebra}
\label{creation/annihilation and clifford algebra}

Consider a semi-Riemannian manifold $\left(\manifold,g\right)$ of dimension $D$
with metric $g$, which has signature $(D-s,s)$.
On the space 
$\Omega\of{\Lambda\of{\manifold}}$ (which we take to be complexified)
of a suitable class of sections of the exterior bundle $\Lambda\of{\manifold}$ 
(the bundle of differential forms of arbitrary degree) 
over this manifold,
we have the operators $\coordCreator^\mu$ of exterior multiplication,
defined by
\begin{eqnarray}
  \label{operator of exterior multiplication}
  \coordCreator^\mu \omega &\defas& dx^\mu\wedge \omega
  \,,\hspace{.5cm} \Omega\of{\Lambda\of{\manifold}} \ni
 \omega = \omega_{(0)} + \omega_{\mu_1}dx^{\mu_1} + \omega_{\mu_1\mu_2} dx^{\mu_1}\wedge dx^{\mu_2}\wedge + \cdots
  \,.
\end{eqnarray}
With respect to the usual Hodge inner product 
$\bracket{\cdot}{\cdot}$ on $\Omega\of{\Lambda\of{\manifold}}$,
\begin{eqnarray}
  \label{Hodge inner product}
  \bracket{\alpha}{\beta}
  &=&
  \int\limits_\manifold
  \bar \alpha \wedge \star \beta
  \nonumber\\
  &\defas&
  p!
  \int\limits_\manifold
  \sqrt{g}
  {\bar\alpha}_{\mu_1 \mu_2\cdots}\beta^{\mu_1 \mu_2\cdots}
  \; d^D x
\end{eqnarray}
(where $\bar \alpha$ is the complex conjugate of $\alpha$),
which defines the Hodge-$\star$ operator,
their adjoints are  $\coordAnnihilator^\mu \defas (\coordCreator^\mu)^\dag$,
and both together satisfy the canonical anticommutation relations
(CAR)
\begin{eqnarray}
  \label{ext multiplication CAR}
  \antiCommutator{\coordCreator^\mu}{\coordCreator^\nu} &=& 0
  \nonumber\\
  \antiCommutator{\coordAnnihilator_\mu}{\coordAnnihilator_\nu} &=& 0
  \nonumber\\
  \antiCommutator{\coordAnnihilator_\mu}{\coordCreator^\nu} &=& \delta_\mu^\nu
  \,.
\end{eqnarray}
With the linear combinations
\begin{eqnarray}
  \label{definition of the Clifford generators}
  \clifford_\pm^\mu &\defas& \coordCreator^\mu \pm \coordAnnihilator^\mu
\end{eqnarray}
this is isomorphic to the Clifford algebra
\begin{eqnarray}
  \antiCommutator{\clifford_\pm^\mu}{\clifford_\mp^\nu} &=& 0
  \nonumber\\
  \antiCommutator{\clifford_\pm^\mu}{\clifford_\pm^\nu} &=& \pm 2g^{\mu\nu}
  \,.
\end{eqnarray}
Every element of the Clifford algebra is mapped to a differential form by
the \emph{symbol map}
\begin{eqnarray}
  \label{symbol map}
  \left(
  \omega_{(0)} + \omega_{\mu_1}\clifford_\pm^{\mu_1} + \omega_{[\mu_1,\mu_2]}
  \clifford_\pm^{\mu_1}\clifford_\pm^{\mu_2} + \cdots\right)\ket{1}
  &=&
  \omega_{(0)} + \omega_{\mu_1}dx^{\mu_1} + \omega_{[\mu_1,\mu_2]}
  dx^{\mu_1}\wedge dx^{\mu_2} + \cdots
  \,,
\end{eqnarray}
where $\ket{1}$ denotes the constant unit 0-form. 
The \emph{local} inner product $\bracket{\alpha}{\beta}_{\rm loc}$
is defined by
\begin{eqnarray}
  \label{local inner product}
  \bracket{\alpha}{\beta}
  &=&
  \int\limits_\manifold
  \bracket{\alpha}{\beta}_{\rm loc}
  \sqrt{g}d^Dx
  \,,
\end{eqnarray}
and also serves as the projection on Clifford 0-vectors, i.e.
\begin{eqnarray}
  \label{Clifford projection}
  \bra{1}\left(
     \omega_{(0)} + \omega_{\mu_1}\clifford_\pm^{\mu_1} + \omega_{[\mu_1,\mu_2]}
    \clifford_\pm^{\mu_1}\clifford_\pm^{\mu_2} + \cdots
   \right)\ket{1}_{\rm loc}
  &\defas&
  \omega_{(0)}
  \,.
\end{eqnarray}
It has the cyclic property
\begin{eqnarray}
  \label{cyclic property of local inner product}
  \bra{1}\clifford_\pm^{a_1}\clifford_\pm^{a_2}\cdots\clifford_\pm^{a_{p}}\ket{1}_{\rm loc}
  &=&
  \bra{1}\clifford_\pm^{a_2}\cdots\clifford_\pm^{a_{p}}\clifford_\pm^{a_1}\ket{1}_{\rm loc}
  \,.
\end{eqnarray}

Using a vielbein field $e^a{}_\mu$ on $\manifold$ we write the ONB frame version of
these operators as
\begin{eqnarray}
  \onbCreator^a &\defas& e^a{}_\mu \coordCreator^\mu
  \nonumber\\
  \onbAnnihilator^a &\defas& e^a{}_\mu \coordAnnihilator^\mu
  \nonumber\\
  \clifford_\pm^a &\defas& e^a{}_\mu \clifford^{\mu}_\pm
  \,.
\end{eqnarray}

The \emph{number operator}, which measures the degree of a differential form, is defined by
\begin{eqnarray}
  \numberOperator &=& \coordCreator^\mu\coordAnnihilator_\mu
  \nonumber\\
  &=&
  \onbCreator^a\onbAnnihilator_a
  \,.
\end{eqnarray}
Note that
\begin{eqnarray}
  \commutator{\numberOperator}{\clifford_\pm^\mu} &=& \clifford_\mp^\mu
  \,.
\end{eqnarray}
A shifted version of this operator, with symmetrized spectrum, is
\begin{eqnarray}
  \label{shifted number operator}
  \frac{1}{2}\clifford^a_-\clifford_{+,a}
  &=&
  \numberOperator - D/2
  \,.
\end{eqnarray}

Often it is convenient to use a  slightly modified version of the
Hodge-$\star$ operator, namely:
\begin{eqnarray}
  \label{properties of normalized duality operator}
  \bar \star &\defas&
  i^{D(D-1)/2+s}
  \left\lbrace
    \begin{array}{ll}
      \clifford_-^{a=0} \clifford_-^{a=1} \cdots \clifford_-^{a=D-1} & \mbox{if $D$ is even}\\
      \clifford_+^{a=0} \clifford_+^{a=1} \cdots \clifford_+^{a=D-1} & \mbox{if $D$ is odd}
    \end{array}
  \right.
  \,,
\end{eqnarray}
which is conveniently normalized so as to satisfy the relations
\begin{eqnarray}
  \label{relations of barred ast}
  \left(\bar \star\right)^\dag &=& (-1)^s \bar \star
  \\
  \left(\bar \star\right)^2 &=& 1
  \\
  \label{Hodge duality of creator annihilator}
  \bar \star \, \onbCreator^a &=& \onbAnnihilator^a\, \bar \star
  \,.
\end{eqnarray}
It is related to the Hodge-$\star$ via
\begin{eqnarray}
  \label{relation Hodge to normalized Hodge}
  \bar \star
  &=&
  \star
  \;
  i^{D(D-1)/2+s}
   (-1)^{\numberOperator(\numberOperator+1)/2 + D} 
  \,.
\end{eqnarray}
We note here the simple but important relation
\begin{eqnarray}
  \bar\star \numberOperator
  &=&
  \onbAnnihilator_a\onbCreator^a \bar \star
  \nonumber\\
  &=&
  (D - \numberOperator)\bar \star
  \,.
\end{eqnarray}

For $s>0$ the inner product $\bracket{\cdot}{\cdot}$ is indefinite. Assume $s=1$,
which is the case of interest here, and $\antiCommutator{\onbAnnihilator^0}{\onbCreator^0}=-1$.
Then the operator
\begin{eqnarray}
  \label{naive and simple hermitian metric operator}
  \hermitianMetricOperator &\defas& 
  \onbCreator^0\onbAnnihilator^0 - \onbAnnihilator^0\onbCreator^0
  \nonumber\\
  &=&
  \clifford_-^{a=0}\clifford_+^{a=0}
  \,,
\end{eqnarray}
(which is self-adjoint with respect to $\bracket{\cdot}{\cdot}$:
$\hermitianMetricOperator^\dag = \hermitianMetricOperator$)
swaps the spurious sign, and the modified inner product
\begin{eqnarray}
  \label{modified inner product}
  \bracket{\cdot}{\cdot}_\hermitianMetricOperator
  &\defas&
  \bracket{\cdot}{\hermitianMetricOperator\,\cdot}
\end{eqnarray}
is positive definite and indeed a scalar product. The adjoint of
an operator $A$ with respect to $\bracket{\cdot}{\cdot}_\hermitianMetricOperator$
will be written $A^{\dag_\hermitianMetricOperator}$ and is given by
\begin{eqnarray}
    A^{\dag_\hermitianMetricOperator}
  &=&
  \left(
    \hermitianMetricOperator
    A
    \hermitianMetricOperator^{-1}
  \right)
  \nonumber\\
  &=&
  \hermitianMetricOperator^{-1}
  A^\dag
  \hermitianMetricOperator
  \,.
\end{eqnarray}
(The term $\hermitianMetricOperator^{-1}$ is here not evaluated further to allow for slightly
more general $\hermitianMetricOperator$ that will be discussed
in \S\fullref{Target space Killing evolution}, \cf \refer{general hermitian metric operator}.)

\subsection{Differential operators}
  \label{differential operators}
		Let $\gradOp_\mu$, which is the \emph{covariant derivative operator}
		with respect to the Levi-Civita-connetion $\Gamma_\mu{}^\alpha{}_\beta$ 
		of $g_{\mu\nu}$, be defined by
		\begin{eqnarray}
			\label{definition of action of covGradOp}
			\commutator{\gradOp_\mu}{f} &=& \left(\partial_\mu f\right),
			\;\;\;f\in\Lambda^0\of{\manifold}
			\nonumber\\
			\commutator{\gradOp_\mu}{\coordCreator^\alpha}
			&=&
			- \Gamma_\mu{}^\alpha{}_\beta \coordCreator^\beta
			\,.
		\end{eqnarray}
		If 
		$\omega_\mu{}^a{}_b$ is the Levi-Civita connection in the
		orthonormal vielbein frame,
		\begin{eqnarray}
			\label{definition of the Levi-Civita connection in vielbein frame}
			\omega_\mu{}^a{}_b
			&\defas&				
			e^a{}_\alpha \left(
                         \delta^\alpha{}_\beta \partial_\mu 
			+  \Gamma_\mu{}^\alpha{}_\beta
                        \right)
			(e^{-1})^\beta{}_b
			\,,
		\end{eqnarray}
		then the last line is equivalent to
		\begin{eqnarray}
			\commutator{\gradOp_\mu}{\onbCreator^a} &=& 
			- \omega_\mu{}^a{}_b \onbCreator^b
			\,. 
		\end{eqnarray}
		This way one has:
		\begin{eqnarray}
			\gradOp_\mu 
			\left(\omega_{\alpha_1\cdots\alpha_p}
			dx^{\alpha_1}\wedge\cdots\wedge dx^{\alpha_p} \right)
			&=& 
			\left(\nabla_\mu \omega_{\alpha_1\cdots\alpha_p}\right)
			dx^{\alpha_1}\wedge\cdots\wedge dx^{\alpha_p} 
			\nonumber\\
			&=&
			\left(\nabla_{[\mu} \omega_{\alpha_1\cdots\alpha_p]}\right)
			dx^{\alpha_1}\wedge\cdots\wedge dx^{\alpha_p} 
			\,.
		\end{eqnarray}
Usually one also identifies the operator version of the \emph{connection 1-form}
\begin{eqnarray}
  \fatomega^a{}_b &\defas&
  \coordCreator^\mu\omega_\mu{}^a{}_b
  \,.
\end{eqnarray}
The commutator of the covariant derivative operators with themselves gives the
\emph{Riemann curvature operator}:
\begin{eqnarray}
  \label{definition Riemann operator}
  \commutator{\gradOp_\mu}{\gradOp_{\nu}}
  &\defas&
  {\bf R}_{\mu\nu}
  \nonumber\\
  &\defas&
  R_{\mu\nu\alpha\beta}\coordCreator^\alpha\coordAnnihilator^\beta
  \,.
\end{eqnarray}

  From the covariant derivative operator one can construct two flavors of partial derivative
operators, distinguished by which of the basis forms they respect as constants, i.e.
with which set of basis forms they commute. Introducing the operators
\begin{eqnarray}
  \label{definition partial derivative operators}
  \partial_\mu 
   &\defas&
  \gradOp_\mu + \omega_\mu{}^a{}_b \onbCreator^b\onbAnnihilator_a
  \nonumber\\
  \partial^c_\mu 
   &\defas&
  \gradOp_\mu + \Gamma_\mu{}^\alpha{}_\beta \coordCreator^\beta\onbAnnihilator_\alpha\,,
\end{eqnarray}
which are, according to \refer{definition of the Levi-Civita connection in vielbein frame}, related as
\begin{eqnarray}
  \label{coordinate derivative in terms of ONB derivative}
  \partial_\mu^c
  &=&
  \partial_\mu -
  e^a{}_\alpha \left(\partial_\mu e^\alpha{}_b\right) \coordCreator^b\coordAnnihilator_a
  \,,
\end{eqnarray}
one finds
		\begin{eqnarray}
                        \label{proprties of partial op}
			\commutator{\partial_\mu}{f} &=& \left(\partial_\mu f\right),
			\;\;\;f\in\Lambda^0\of{\manifold}
			\nonumber\\
			\commutator{\partial_\mu}{\onbCreator^a}
			&=&
			0
			\nonumber\\
			\commutator{\partial_\mu}{\onbAnnihilator_a}
			&=&
			0      
			\,.
		\end{eqnarray}
and
		\begin{eqnarray}
                        \label{properties partial c op}
			\commutator{\partial^c_\mu}{f} &=& \left(\partial_\mu f\right),
			\;\;\;f\in\Lambda^0\of{\manifold}
			\nonumber\\
			\commutator{\partial^c_\mu}{\coordCreator^\alpha}
			&=&
			0
			\nonumber\\
			\commutator{\partial^c_\mu}{\coordAnnihilator_\alpha}
			&=&
			0      
			\,.
		\end{eqnarray}
(Note the position of the indices in the last two lines.)
 By acting with the partial derivative operators on an arbitrary form in a given basis
one also verifies that for both the expected relations
\begin{eqnarray}
  \commutator{\partial_\mu}{\partial_\nu} &=& 0
  \nonumber\\
  \commutator{\partial^c_\mu}{\partial^c_\nu} &=& 0
\end{eqnarray}
hold.
Using \refer{definition partial derivative operators}, \refer{proprties of partial op}, and
\refer{properties partial c op} it is now easy to establish the transformation properties of
all creators and annihilators starting from \refer{definition of action of covGradOp}:
\begin{eqnarray}
  \commutator{\gradOp_\mu}{\onbCreator_a} &=& + \omega_\mu{}^b{}_a \onbCreator_b
  \nonumber\\
  \commutator{\gradOp_\mu}{\onbAnnihilator^a} &=& -\omega_\mu{}^a{}_b \onbAnnihilator^b
  \nonumber\\
  \commutator{\gradOp_\mu}{\onbAnnihilator_a} &=& +\omega_\mu{}^b{}_a \onbAnnihilator_b
  \nonumber\\
  \commutator
    {\gradOp_\mu}
    {\coordCreator_\alpha}
  &=& + \Gamma_{\mu}{}^\beta{}_{\alpha}\coordCreator_\beta
  \nonumber\\
  \commutator{\gradOp_\mu}{\coordAnnihilator^\alpha} &=& -\Gamma_\mu{}^\beta{}_\alpha \coordAnnihilator^\alpha
  \nonumber\\
  \commutator{\gradOp_\mu}{\coordAnnihilator_\alpha} &=&
   + \Gamma_{\mu}{}^\beta{}_{\alpha}\coordAnnihilator_\beta
  \,.
\end{eqnarray}
That is, all basis operators transform as they should according to the index they carry.

Note that in particular we can now write
\begin{eqnarray}
  \label{gradop in terms of cliffords}
  \gradOp_\mu
  &=&
  \partial_\mu - \omega_{\mu}{}^a{}_b \onbCreator^b \onbAnnihilator_a
  \nonumber\\
  &=&
  \partial_\mu + \omega_{\mu ab}\onbCreator^a \onbCreator^b
  \nonumber\\
  &=&
  \partial_\mu + \frac{1}{4}\omega_{\mu ab}
   \left(
    \clifford_+^a \clifford_+^b + \clifford_-^a \clifford_-^b
   \right)
  \,.
\end{eqnarray}

Another useful fact is that $\partial_\mu$ and $\gradOp_\mu$ commute with the duality operation:
\begin{eqnarray}
  \label{derivatives that commute with duality}
  \commutator{\partial_\mu}{\bar \star} &=& 0
  \nonumber\\
  \commutator{\gradOp_\mu}{\bar \star} &=& 0
  \,,
\end{eqnarray}
which follows straightforwardly by using the respective definitions.

With respect to the Hodge inner product the adjoint of $\partial_\mu$ is
\begin{eqnarray}
  \label{adjoint relation of partial}
  \left(\partial_\mu\right)^\dag &=& -\frac{1}{\sqrt{|g|}}\partial_\mu \sqrt{|g|}
  \,.
\end{eqnarray}
On the other hand the operator $\partial^c_\mu$ satisfies no such simple formula. 
Using the antisymmetry of
$\omega_{\mu a b} = \omega_{\mu [a b]}$ one finds from \refer{adjoint relation of partial}
and \refer{definition partial derivative operators} the analogous relation
\begin{eqnarray}
  \label{adjoint relation of gradOp}
  \left(\gradOp_\mu\right)^\dag = - \frac{1}{\sqrt{|g|}}\gradOp_\mu \sqrt {|g|}
  \,.
\end{eqnarray}

Next, it is of interest to have differential operators without free indices, which
map forms to forms. Such are
		obtained by contracting $\gradOp_\mu$ with some Grassmann
		or Clifford operator:

\paragraph{Exterior derivative.}
		The \emph{exterior derivative} is defined by
		\begin{eqnarray}
			\label{definition of the exterior derivative}
			\extd &\defas& \coordCreator^\mu\gradOp_\mu
			\,.
		\end{eqnarray}
		Due to the special symmetry of the Levi-Civita connection in
		the coordinate basis, the exterior derivative here has the
		simple action
		\begin{eqnarray}
			\extd \omega_{\mu_1\cdots\mu_p}dx^{\mu_1}\wedge\cdots\wedge dx^{\mu_p}
			&=&
			\partial_{[\nu}\omega_{\mu_1\cdots\mu_p]}
				dx^\nu\wedge dx^{\mu_1}\wedge\cdots\wedge dx^{\mu_p}
			\,.
		\end{eqnarray}
  This can be made manifest by noting that
  \begin{eqnarray}
   \label{extd as a derivative commuting with coordinate forms}
    \extd &=& \coordCreator^\mu \partial^c_\mu
    \,,
  \end{eqnarray}
  which follows by the definition of $\partial^c_\mu$ in \refer{definition partial derivative operators}
  and the symmetry $\Gamma_\mu{}^\alpha{}_\beta = \Gamma_{(\mu}{}^\alpha{}_{\beta)}$.
(Another way to say the same is 
\begin{eqnarray}
  \label{extd commutes with coordinate forms}
  \antiCommutator{\extd}{\coordCreator^\mu} &=& 0
  \nonumber\\
  \Leftrightarrow
  \antiCommutator{\extd}{\onbCreator^a} &=& -\coordCreator^\mu\omega_{\mu}{}^a{}_b\onbCreator^b
  \,.
\end{eqnarray}
The second line is known as the \emph{first structure equation} for vanishing torsion.)
		Therefore $\extd$ is
		\emph{nilpotent}:
		\begin{eqnarray}
			\extd^2 &=& \coordCreator^\mu\coordCreator^\nu \partial^c_\mu \partial^c_\nu
                        \nonumber\\
                        &=& 0
			\,.
		\end{eqnarray}
(Using instead the covariant derivative shows that $\coordCreator^\mu \coordAnnihilator^\nu 
\commutator{\gradOp_\mu}{\gradOp_\nu} = 0 $ and hence 
(\cf \refer{definition Riemann operator})
${\bf R}_{[\mu\nu]} = 0$.)

   Furthermore it obviously satisfies the \emph{graded Leibniz rule}:
\begin{eqnarray}
  \label{Leibniz rule of continuous extd}
  \extd\; \coordCreator^{\mu_1}\cdots \coordCreator^{\mu_p}\omega_{\mu_1\cdots \mu_p}
  &=& 
  \coordCreator^\mu\coordCreator^{\mu_1}\cdots \coordCreator^{\mu_p}(\partial_\mu\omega_{\mu_1\cdots \mu_p})
  +
  (-1)^p \coordCreator^{\mu_1}\cdots \coordCreator^{\mu_p}\omega_{\mu_1\cdots \mu_p}\; \extd
  \,.
\end{eqnarray}
This makes it easy to compute its adjoint: Let $\beta$ be any $p$-form and $\alpha$ any $D-p$-form
then
\begin{eqnarray}
  \bracket{\extd \alpha}{\beta}
  &=&
  \int \left(\extd \alpha\right) \wedge \star \beta
  \nonumber\\
  &\equalby{Leibniz rule of continuous extd}&
  -(-1)^{D-p} \int \alpha  \wedge \extd \star \beta
  \nonumber\\
  &\equalby{relation Hodge to normalized Hodge}&
  -i^{-D(D-1)/2 - s}(-1)^{p(p-1)/2}
  \int \alpha  \wedge \bar \star \bar \star \extd \bar \star \beta
  \nonumber\\
  &\equalby{relation Hodge to normalized Hodge}&
  -\int \alpha  \wedge \star \bar \star \extd \bar \star \beta
  \nonumber\\
  &=&
  -\bracket{\alpha}{\bar \star \extd \bar\star \,\beta}
  \,.  
\end{eqnarray}
Hence
\begin{eqnarray}
  \label{duality relation d extd}
  \coextd &=& -\bar \star\, \extd \,\bar \star
  \,. 
\end{eqnarray}
Using \refer{Hodge duality of creator annihilator} this gives explicitly
		\begin{eqnarray}
			\label{definition exterior coderivative in intro}
			\coextd &=&
			-\coordAnnihilator^\mu\gradOp_\mu
			\,.
		\end{eqnarray}
		We will mostly refer to this ``inner'' derivative as the \emph{exterior co-derivative}.		
		It acts on $p>0$-forms as the covariant divergence:
		\begin{eqnarray}
			\label{explicit form of coextd action on forms}
			\coextd
			\omega_{\mu_1\cdots\mu_p}dx^{\mu_1}\wedge\cdots\wedge dx^{\mu_p}
			&=&
			-p
			\left(\nabla_\mu \omega^\mu{}_{\alpha_2\cdots\alpha_p}\right)
			dx^{\alpha_2}\wedge\cdots\wedge dx^{\alpha_p} 
			\,.
		\end{eqnarray}
		The exterior co-derivative, being the adjoint of a nilpotent operator,
		is itself nilpotent:
		\begin{eqnarray}
			\coextd^2 &=& 0
			\,.
		\end{eqnarray}

It is obvious, that
\begin{eqnarray}
  \commutator{\numberOperator}{\extd} &=& \extd
  \nonumber\\
  \commutator{\numberOperator}{\coextd} &=& -\coextd
  \,.
\end{eqnarray}

\subsection{Dirac, Laplace-Beltrami, and spinors}
\label{Dirac, Laplace-Beltrami, and spinors}

		The operator
		\begin{eqnarray}
       \label{definition Dirac operator}
			\Dirac_\pm &\defas& \extd \pm \coextd
			\nonumber\\
			&=&
			\clifford^\mu_\mp \gradOp_\mu
			\,,
		\end{eqnarray}
		is called the \emph{Dirac operator} on $\Omega\of{\manifold}$. 
		Its square
		\begin{eqnarray}
			\label{definition Laplace-Beltrami operator}
			\pm\fatDelta &\defas&
			\Dirac^2_\pm 
			\nonumber\\
			&=& \left(\extd \pm \coextd\right)^2
			\nonumber\\
			&=& \pm\antiCommutator{\extd}{\coextd}
		\end{eqnarray}
		is known as the \emph{Laplace-Beltrami operator},
  which explicitly reads
\begin{eqnarray}
  \label{Weitzenboeck formula}
  \fatDelta
  &=&
  \Dirac_+^2
  \nonumber\\
  &=&
  -
  \left(
    g^{\mu\mu^\prime}\nabla_\mu\nabla_{\mu^\prime}
    +
    \Gamma_\mu{}^{\mu^\prime\mu}\nabla_{\mu^\prime}
    -
     R_{\mu\mu^\prime\kappa\lambda}e^{\dag\mu}e^{\dag\kappa}e^{\nu}e^{\lambda}
    -
    R_{\mu\lambda}e^{\dag\mu}e^{\lambda}
  \right)  
  \,.
\end{eqnarray}
This expression is known as the \emph{Weitzenb\"ock formula}
(\cf for instance \cite{BerlineGetzlerVergne:1992}, p.130,
or \cite{FroehlichGrandjeanRecknagel:1997}, eqs. (4.33),(4.45)).
{{
The Dirac and Laplace-Beltrami operators obviously satisfy
\begin{eqnarray}
  \label{adjointness of dirac and LapBelt}
  (\Dirac_\pm)^\dag &=& \pm \Dirac_\pm
  \nonumber\\
  \fatDelta^\dag &=& \fatDelta
  \,.
\end{eqnarray}

\paragraph{Spinors.}
\label{spinors}
The following briefly indicates some aspects concerning spinors
as viewed from the exterior geometry perspective, and how our algebraic notation relates to the more
commonly used matric representations.

The
Clifford bivectors $\frac{1}{2}\clifford_\pm^{ab}\defas \frac{1}{2}\clifford_\pm^{[a} \clifford_\pm^{b]}$
form a representation of the Lie algebra ${\rm so}\of{d-s,s}$ and generate the 
\emph{spin group} of the Clifford algebra, whose elements are of the form
\begin{eqnarray}
  \label{a rotor}
  R_\pm = \exp\of{\rho_{[ab]}\clifford_\pm^{ab}}
  \,. 
\end{eqnarray}
A Clifford element of the form $\psi_\pm = \rho R_\pm$, with $\rho$ a scalar, 
is sometimes called a \emph{Dirac-Hestenes} state
(e.g. \cite{RodriguesDeSouzaVazLounesto:1996}). Applying $R_\pm$ to a primitive
projector $P$ of the Clifford algebra yields the spinor representation $\psi_\pm P$ of
the group ${\rm SO}(d-s,s)$.

Now the exterior bundle can be viewed as the product of two spinor bundles.
The spin groups of the two Clifford algebras $\clifford_\pm$ act, respectively,
from the left and from the right on the Clifford elements associated with an 
element of the exterior bundle:

This is easily seen by considering, 
as in \refer{symbol map}, the Clifford-representation of 
an arbitrary (inhomogeneous) form 
$\fatomega = \omega_{(0)} + \omega_{\mu}dx^\mu + \cdots
= \underbrace{\left(\omega_{(0)} + \omega_{\mu} \clifford_\pm^\mu + \cdots\right)
  }_{\defas \Omega_\pm}\ket{0}$
and acting on it with the generators $\clifford_\pm^{ab}$ of the two commuting
spinor groups:
\begin{eqnarray}
  \clifford_\mp^{ab}\Omega_\pm\ket{0}
  &=&
  \Omega_\pm\clifford_\mp^{ab}\ket{0}
  \nonumber\\
  &=&
  \Omega_\pm\clifford_\pm^{ab}\ket{0}
  \,.
\end{eqnarray}
In this sense one of $\clifford_\pm^{ab}$ acts from the left, the other from the right on
the symbol map pre-images of an element of the exterior bundle. 

To make this more explicit consider elements 
of $\Gamma\of{\Lambda\of{\manifold}}$ of the form
\begin{eqnarray}
  \label{forms as bispinors}
  {\fatomega_\pm} &\defas& \psi_\pm \hat O {\tilde \psi_\pm} \ket{0}
  \,,
\end{eqnarray}
where $\hat O_\pm$ is a constant $\pm$-Clifford element:
\begin{eqnarray}
  &&\hat O_\pm \;=\; \in {\rm Cl}_\pm
  \nonumber\\
  &&\commutator{\partial_\mu}{\hat O_\pm} \;=\; 0\,,
\end{eqnarray}
and where $\tilde \cdot$ is the linear operation of 
\emph{Clifford reversion} which reverses the order of
Clifford generators and takes the complex conjugate of the coefficient:
\begin{eqnarray}
  \tilde{}\left(\rho \clifford_\pm^{a_1 a_2 \cdots a_p}\right)
  &\defas&
  \rho^\ast\clifford_\pm^{a_p \cdots a_2 a_1}
  \,.
\end{eqnarray}
Acting on such such a state with a spin group element $R_\pm$ gives
\begin{eqnarray}
  \label{left and right action on spinors}
  R_\pm \fatomega_\pm &=&
  \left(R_\pm\psi_\pm\right)\hat O \tilde {\psi_\pm}\ket{0}
  \nonumber\\
  \tilde R_{\mp}
  \fatomega_\pm &=&
  \psi_\pm\hat O \widetilde {(R_\pm\psi_\pm)}\ket{0}
  \,.
\end{eqnarray}

To see how this goes together with the usual way of writing spinors as represented
on some vector space note that
\begin{eqnarray}
  R_\pm \clifford_\pm^a \tilde{R_\pm}
  &=&
  \Lambda^a{}_b \clifford_\pm^b
  \,,
\end{eqnarray}
as usual. By the cyclic property \refer{cyclic property of local inner product}
\begin{eqnarray}
  \Lambda^{ab}
  &=&
  \bra{0}
    \psi_\pm \clifford_\pm^a\tilde {\psi_\pm}
    \;\clifford_\pm^b
  \ket{0}_{\rm loc}
  \nonumber\\
  &=&
  \bra{0}
    \tilde {\psi_\pm}
    \clifford_\pm^b
    \psi_\pm \;\clifford_\pm^a
  \ket{0}_{\rm loc}  
\end{eqnarray}
this implies
\begin{eqnarray}
  \label{conjugate spinor adjoint relation}
  \tilde{\psi_\pm}\clifford_\pm^a \psi_\pm
  &=&
  \clifford_\pm^b \Lambda_b{}^a
  \,.
\end{eqnarray}
Hence the construction \refer{forms as bispinors} produces the differential forms 
\begin{eqnarray}
  \psi \clifford_\pm^{a_1\cdots a_p}\tilde {\psi_\pm}\ket{0}
  &=&
  \psi\clifford_\pm^{[a_1}\tilde {\psi_\pm}\cdots
  \psi\clifford_\pm^{a_p]}\tilde {\psi_\pm}\cdots  \ket{0}
  \nonumber\\
  &=&
  \Lambda^{a_1}{}_{[b_1}\cdots\Lambda^{a_p}{}_{b_p]}
  \clifford_\pm^{b_1\cdots b_p}\ket{0}
  \nonumber\\
  &=&
  \Lambda^{a_1}{}_{[b_1}\cdots\Lambda^{a_p}{}_{b_p]}
  dx^{b_1}\wedge\cdots \wedge dx^{b_p}
  \,,
\end{eqnarray}
where we set $\rho = 1$ for brevity. Now let $\phi_\alpha = (R_\pm \phi_0)_\alpha$ be
the usual representation of the rotor $R_\pm$ as a spinor on a $2^{[d/2]}$-dimensional vector space, then
the coeffcients of the above differential form are obtained by means of the usual
expression:\footnote{
  Hence the component analogue of \refer{exterior covd seperates into spinor covd} is
  \begin{eqnarray}
    \nabla_\mu
    \left(
      \bar \phi_0
        \clifford_{+ a_1\cdots a_p}
      \phi
    \right)
    &=&
    \partial_\mu
    \left(
      \bar \phi_0
        \clifford_{+ a_1\cdots a_p}
      \phi
    \right)
    -
    \omega_\mu{}^{b_1}{}_{a_1}
    \left(
      \bar \phi_0
        \clifford_{+ b_1\cdots a_p}
      \phi
    \right)    
    -
    \cdots
    -
    \omega_\mu{}^{b_p}{}_{a_p}
    \left(
      \bar \phi_0
        \clifford_{+ a_1\cdots b_p}
      \phi
    \right)    
    \nonumber\\
    &=&
    \partial_\mu
    \left(
      \bar \phi_0
        \clifford_{+a_1\cdots a_p}
      \phi
    \right)
    -
    \left(
      \bar \phi_0
        \commutator
          {\frac{1}{4}\omega_{\mu a b}\clifford_+^a \clifford_+^b}
          {\clifford_{+ a_1\cdots a_p}}
      \phi
    \right)
    \nonumber\\
    &=&
    \overline
    {\left(
      \nabla_\mu^S \phi_0
    \right)}
     \clifford_{+a_1\cdots a_p}
      \phi  
    +
      \bar \phi_0
        \clifford_{+a_1\cdots a_p}
      \nabla_\mu^S\phi      
\end{eqnarray}
}
\begin{eqnarray}
  \bar \phi \gamma_{\pm\,b_1\cdots b_p} \phi
  &=&
  \bar \phi_0 \tilde {\psi_\pm} \gamma_{\pm\,b_1\cdots b_p} \psi_\pm \phi_0
  \nonumber\\
  &\equalby{conjugate spinor adjoint relation}&
  \underbrace{
  \bar \phi_0 \gamma_{\pm,a_1\cdots a_p}
  \phi_0
  }_{= {\rm const}}
  \;
  \Lambda^{a_1}{}_{[b_1}\cdots\Lambda^{a_p}{}_{b_p]}
  \,,
\end{eqnarray}
where now all Clifford elements refer to their matrix representation and
$\bar \phi$ is the Dirac adjoint of $\phi$.

The covariant derivative operator \refer{definition of action of covGradOp}
$\gradOp_\mu = \partial_\mu + \omega_{\mu a b}\onbCreator^a\onbAnnihilator^b
= \partial_\mu + \frac{1}{4}\left(\clifford_+^a\clifford_+^b-\clifford_-^a\clifford_-^b\right)$  
splits into a sum of covariant derivative operators 
\begin{eqnarray}
  \label{spinor covariant derivative}
  \gradOp_\mu^{S\pm}
  &\defas&
  \partial_\mu \pm \frac{1}{4}\omega_{\mu a b}\clifford_\pm^a\clifford_\pm^b
\end{eqnarray} 
that act on the two spinor bundles seperately:
\begin{eqnarray}
  \label{exterior covd seperates into spinor covd}
  \gradOp_\mu\left(
  \psi_+ \hat O_+ \tilde {\psi_+}
  \right)
  \ket{0}
  &=&
  \left(
    \gradOp_\mu^S \psi
  \right)
  \hat O
  \tilde {\psi_+}
  \ket{0}
  +
  \psi_+
  \hat O
  \widetilde
  {\left(
    \gradOp_\mu^S \psi
  \right)}
  \ket{0}
  \,.
\end{eqnarray}

But such a splitting does not take place for the Dirac operator 
\refer{definition Dirac operator} on the exterior bundle.
Due to \refer{gradop in terms of cliffords}
the Dirac operators \refer{definition Dirac operator} mix the two Clifford algebra
representations $\clifford_\pm$.

The equation $\Dirac_\pm \psi \;=\; (\extd \pm \coextd)\psi \;=\; 0$ 
is known as the (massless, free)
\emph{K{\"a}hler equation} (see \cite{BennTucker:1987}, \S 8.3). Due to the
above considerations it is equivalent (up to degeneracy) to the ordinary
(massless, free) Dirac equation on spinors (instead of on differential forms) 
only when the left (+) and right (-) Clifford algebras don't mix, which 
occurs only for $\omega_{abc} = 0$ if no other background fields are turned on, i.e. for
a flat spacetime background. But actually in string theory a
generalization of the operators $\Dirac_\pm$ does play the role of the
Dirac operator for spinors. This is possible, because the presence of
further background fields will modify $\Dirac_\pm$ in a way that cancels
the spurious terms and thus restores their ``chirality'' (in the CFT sense)
(\cf \S\fullref{WZW models}).

\subsection{Lie derivative}
\label{Lie derivative}

  From $\extd$ one recovers a directional derivative ${\cal L}_v$
along a vector field $v = v^\mu\partial_\mu$ by performing a ``contraction'':
\begin{eqnarray}
  \label{definition Lie derivative operator}
  {\cal L}_v &\defas&
  \antiCommutator 
   {\extd}{\coordAnnihilator_\mu v^\mu}
  \,.
\end{eqnarray}
This is the \emph{Lie derivative} on differential forms along $v$. More explicitly it
reads
\begin{eqnarray}
  \label{Lie in terms of partial derivatives}
  \antiCommutator 
   {\extd}{\coordAnnihilator_\mu v^\mu}
  &=&
  \antiCommutator 
   {\coordCreator^\mu \partial^c_\mu}{\coordAnnihilator_\mu v^\mu}  
  \nonumber\\
  &=&
  v^\mu \partial^c_\mu + (\partial_\mu v^\nu)\coordCreator^\mu\coordAnnihilator_\nu
  \,,
\end{eqnarray}
or alternatively
\begin{eqnarray}
  \label{Lie deriv operator in ONB form}
  \antiCommutator 
   {\extd}{\onbAnnihilator_\mu v^\mu}
  &=&
  \antiCommutator 
   {\coordCreator^\mu \gradOp_\mu}{\onbAnnihilator_\mu v^\mu}  
  \nonumber\\
  &=&
  v^\mu \gradOp_\mu + (\nabla_\mu v^\nu)\coordCreator^\mu\coordAnnihilator_\nu
  \,.
\end{eqnarray}
The form \refer{Lie in terms of partial derivatives} is convenient for checking that
\begin{eqnarray}
  \label{exterior Lie derivative commutator}
  \commutator{{\cal L}_{v}}{{\cal L}_w} &=& {\cal L}_{[v,w]}
\end{eqnarray}
and
\begin{eqnarray}
  \label{commutator of lie derivative with creator and annihilator}
  \commutator{
     {\cal L}_v
  }
  {w^\mu \coordAnnihilator_\mu }
  &=&
  [v,w]^\mu \coordAnnihilator_\mu
  \nonumber\\
  \commutator
    {{\cal L}_v}
    {w_\mu\coordCreator^\mu }
    &=&
    ({\cal L}_v w)_\mu \coordCreator^\mu
  \,,
\end{eqnarray}
while
\refer{Lie deriv operator in ONB form} is convenient for computing the adjoint:
\begin{eqnarray}
  \label{adjoint of Lie derivative}
  \left({\cal L}_v\right)^\dag
  &=&
  -\frac{1}{\sqrt{g}}\gradOp_\mu \sqrt{g} v^\mu
  +
  (\nabla_\nu v^\mu)\coordCreator_\mu \coordAnnihilator^\nu
  \nonumber\\
  &=&
  -{\cal L}_v
  -
  (\nabla_\mu v^\mu)
  +
  2(\nabla_{(\mu} v_{\nu)})
  \coordCreator^\mu \coordAnnihilator^\nu
  \,.
\end{eqnarray}
Obviously the Lie derivative ${\cal L}_v$ is skew-self-adjoint if and only if
\begin{eqnarray}
  \nabla_{(\mu}v_{\nu)} &=& 0
  \nonumber\\
  \Rightarrow
  \nabla_\mu v^\mu &=& 0
  \,,
\end{eqnarray}
i.e. if and only if $v$ is a Killing vector field:
\begin{eqnarray}
  \label{Lie antihermitian when v is Killing}
  ({\cal L}_v)^\dag = -{\cal L}_v
  &\Leftrightarrow&
  \mbox{$v$ is Killing}
  \,.
\end{eqnarray}
From its definition 
\refer{definition Lie derivative operator} and the duality relations
\refer{relations of barred ast} and 
\refer{duality relation d extd} it follows furthermore that the adjoint can be
expressed as
\begin{eqnarray}
  \label{adjoint and Hodge of Lie derivative}
  {\cal L}_v^\dag 
  &=&
  -\bar \star {\cal L}_v  \bar\star
  \,.
\end{eqnarray}
From this we find the equivalence
\begin{eqnarray}
  \mbox{$v$ is Killing} &\Leftrightarrow& \commutator{{\cal L }_v}{\bar \star} = 0
  \,.
\end{eqnarray}
Hence for a Killing vector $v$ it follows from taking the adjoint of 
\refer{definition Lie derivative operator} that
\begin{eqnarray}
  \label{adjoint definition of Lie derivative}
  \antiCommutator{\coextd}{\coordCreator_\mu v^\mu} &=& -{\cal L}_v
  \,
  \hspace{1cm}(v\;\mbox{Killing})
  \,.
\end{eqnarray}
One particular consequence is, that
\begin{eqnarray}
  \label{one particular consequence}
  \antiCommutator{v_\mu\clifford_+^\mu}{\,\Dirac_-}
  -
  \antiCommutator{v_\mu\clifford_-^\mu}{\,\Dirac_+}
  &=&
  4{\cal L}_v
  \hspace{1cm}(v\;\mbox{Killing})
  \,,
\end{eqnarray}
which will be rather useful later on. For the other sign one gets
\begin{eqnarray}
  \label{for the other sign}
  \antiCommutator{v_\mu\clifford_+^\mu}{\,\Dirac_-}
  +
  \antiCommutator{v_\mu\clifford_-^\mu}{\,\Dirac_+}
  &=&
  2
  (\nabla_{[\mu} v_{\nu]})
  \left(\coordCreator^\mu\coordCreator^\nu + 
    \coordAnnihilator^\mu\coordAnnihilator^\nu
  \right)
  \hspace{1cm}(v\;\mbox{Killing})
  \,.
\end{eqnarray}
Also note that for $v$ Killing one has
\begin{eqnarray}
  \label{Killing Lie commutes with its own cliffords}
  \commutator{{\cal L}_v}{v\inner \clifford_\pm}
  &=& 0\hspace{1cm}\mbox{($v$ Killing)}
  \,.
\end{eqnarray}

Another useful fact is that
the partial derivative operators $\partial^c_\mu$, defined in \refer{definition partial derivative operators},
are obviously (using \refer{extd as a derivative commuting with coordinate forms}) Lie derivatives:
\begin{eqnarray}
  \label{partial coordinate derivative as anticommutator with coordannihilator}
  \partial^c_\mu &=& \antiCommutator{\extd}{\coordAnnihilator_\mu}
\end{eqnarray}
i.e.
\begin{eqnarray}
  \label{coordinate derivative as Lie derivative}
  \partial_\mu^c &=& {\cal L}_{\partial_\mu}
  \nonumber\\
  &\equalby{coordinate derivative in terms of ONB derivative}&
  \partial_\mu
  -\left(e^a{}_\alpha \partial_\mu e^\alpha{}_b\right) \onbCreator^b \onbAnnihilator_a  
  \,.
\end{eqnarray}
(This is to be contrasted with the Lie derivative along an ONB basis vector
$v = e_a$:
\begin{eqnarray}
  {\cal L}_{e_a}
  &=&
  \partial_a + 2 \omega_{abc}\onbCreator^b \onbAnnihilator^c
  \,.) 
\end{eqnarray}
If $\partial_\mu$ is a Killing Lie derivative then 
(according to \refer{coordinate derivative as Lie derivative} and
\refer{Lie deriv operator in ONB form} )
the term $\left(e_a{}_\alpha \partial_\mu e^\alpha{}_b\right)$ is antisymmetric in
$a$ and $b$ and we can write
\begin{eqnarray}
  \label{coordinate Killing Lie derivative}
  {\cal L}_{\partial_\mu}
  &=&
  \partial_\mu
  +
  \left(e_a{}_\alpha \partial_\mu e^\alpha{}_b\right)\onbCreator^a\onbAnnihilator^b
  \hspace{3cm}
  \mbox{($\partial_\mu$ Killing)}
  \nonumber\\
  &=&
  \partial_\mu
  +
  \frac{1}{4}
  \left(e_a{}_\alpha \partial_\mu e^\alpha{}_b\right)
  \left(
    \clifford_+^a \clifford_+^b - \clifford_-^a \clifford_-^b
  \right)
  \,.
\end{eqnarray}

Accordingly the exterior derivative \refer{extd as a derivative commuting with coordinate forms} 
can also be written as
\begin{eqnarray}
  \extd &=& \coordCreator^\mu {\cal L}_{\partial_\mu}
\end{eqnarray}
and hence with \refer{duality relation d extd} 
and \refer{adjoint and Hodge of Lie derivative}
the exterior coderivative can also be written as
\begin{eqnarray}
  \coextd
  &=&
  \coordAnnihilator^\mu {\cal L}^\dag_\mu
  \,.
\end{eqnarray}

There is in general no Lie derivative on spinors, but along a Killing vector
field there is\footnote{\cf p. 195 of \cite{Tucker:1986}}:
Rewriting \refer{Lie deriv operator in ONB form} in terms of 
Clifford generators yields
\begin{eqnarray}
  {\cal L}_{v}
  &=&
  v^\mu\gradOp_\mu
  +
  \frac{1}{4}(\nabla_\mu v_\nu)(\clifford_+^\mu + \clifford_-^\mu)
    (\clifford_+^\nu - \clifford_-^\nu)
 \nonumber\\
  &=&
  v^\mu \gradOp_\mu
  +
  \frac{1}{4}(\nabla_\mu v_\nu)
  \left(
    \clifford_+^\mu\clifford_+^\nu - \clifford_-^\mu\clifford_-^\nu
  \right)
  +
  \frac{1}{4}(\nabla_\mu v_\nu)
  \left(
    \clifford_-^\mu\clifford_+^\nu + \clifford_-^\nu\clifford_+^\mu
  \right)
  \,.
\end{eqnarray}
The condition that the last term vanishes is obviously
$\nabla_{(\mu} v_{\nu)} = 0$, i.e. that $v$ is Killing. Hence in this case
the Lie derivative on differential forms is
\begin{eqnarray}
  \mbox{$v$ Killing }\;\Leftrightarrow\; {\cal L}_v
  &=&
  v^\mu \gradOp_\mu
  +
  \frac{1}{4}(\nabla_\mu v_\nu)
  \left(
    \clifford_+^\mu\clifford_+^\nu - \clifford_-^\mu\clifford_-^\nu
  \right)
\end{eqnarray}
and splits into two Lie derivatives 
\begin{eqnarray}
  \label{Killing spinor Lie derivative}
  {\cal L}_v^{{\rm S}_\pm}
  &=&
  v^\mu \gradOp_\mu^{{\rm S}_\pm}
  \pm
  \frac{1}{4}(\nabla_\mu v_\nu)
  \clifford_\pm^\mu\clifford_\pm^\nu
  \hspace{1 cm}\mbox{($v$ Killing)}
\end{eqnarray}
on spinors (\cf \refer{spinor covariant derivative}).
If we decree that the partial derivative operator in $\gradOp^{{\rm S}_\pm}$
acts only on the (respectively) left or right spinor bundle (\cf  the factorization
\refer{forms as bispinors}), then this allows us to succinctly write
\begin{eqnarray}
  \label{spinor splitting of Killing lie}
  {\cal L}_v
  &=&
  {\cal L}_v^{{\rm S}_+} + {\cal L}_v^{{\rm S}_-}
\end{eqnarray}
with
\begin{eqnarray}
  \label{spinor Lie commutator}
  \commutator{{\cal L}_v^{{\rm S}_+}}{{\cal L}_w^{{\rm S}_-}}
  &=& 0
  \,,
\end{eqnarray}
which, together with \refer{exterior Lie derivative commutator}, implies\footnote{
 It should be noted, though, that \refer{exterior Lie derivative commutator}
 holds for arbitrary $v$,$w$, while \refer{spinor Lie commutator}
 makes sense only for $v$ and $w$ both Killing, since otherwise
 $\gradOp^{{\rm S}_\pm}_v$ isn't even defined.
}
\begin{eqnarray}
  \commutator{{\cal L}_v^{{\rm S}_\pm}}{{\cal L}_w^{{\rm S}_\pm}}
  &=&
  {\cal L}_{[v,w]}^{{\rm S}_\pm}
  \,.
\end{eqnarray}

An intersting special case that plays a paramount role in the context of Lie groups  
is that where there existes an orthonormal frame in which the Killing vector $v$ has
constant components 
$v^s = \delta^s_a$.
In this case \refer{Killing spinor Lie derivative} gives
\begin{eqnarray}
  {\cal L}^{{\rm S}}_{v}
  &=&
  \gradOp_a^{{\rm S}\pm}
  \pm
  \frac{1}{4}
  \omega_{abc}\clifford_\pm^b\clifford_\pm^c
  \nonumber\\
  &=&
  \partial_a \pm \frac{1}{2}\omega_{abc}\clifford_\pm^b\clifford_\pm^c
  \hspace{1cm}\mbox{(for $v^s = \delta^s_a$)}\,.
\end{eqnarray}
But this is equal to the covariant derivative along $v$ with respect to the connection
with torsion
$\omega^\prime_{abc} = \omega_{abc} + T_{abc}$, where
the torsion tensor $T_{abc} = \omega_{abc}$ in this frame.

\subsection{Torsion.}
\label{Torsion}
Let
\begin{eqnarray}
  T_{\mu\alpha\beta}
  &=&
  T_{[\mu\alpha\beta]}
\end{eqnarray}
be a totally antisymmetric torsion tensor, and consider the 
connection with torsion $\omega_T$ given by
\begin{eqnarray}
  \omega_{T,\mu}{}^a{}_b
  \defas
  \omega_\mu{}^a{}_b + T_\mu{}^a{}_b
\end{eqnarray}
with the associated connection 1-form operator
\begin{eqnarray}
  \fatomega_T{}^a{}_b
  &\defas&
  \coordCreator^\mu
  \omega_{T,\mu}{}^a{}_b
  \,,
\end{eqnarray}
where $\omega$ is, as above, the (torsionless) Levi-Civita connection in the orthonormal frame.
The associated covariant derivative operator is
\begin{eqnarray}
  \label{covariant derivative with torsion}
  \gradOp_{T,\mu}
  &\defas&
  \partial_\mu 
  -(\omega_{\mu}{}^a{}_b + T_{\mu}{}^a{}_b)
  \coordCreator^b\coordAnnihilator_a
  \nonumber\\
  &=&
  \partial^c_\mu 
  -(\Gamma_{\mu}{}^\alpha{}_\beta + T_{\mu}{}^\alpha{}_\beta)
  \coordCreator^\beta\coordAnnihilator_\alpha
  \,,
\end{eqnarray}
whose adjoint is still of the form \refer{adjoint relation of gradOp}:
\begin{eqnarray}
  \gradOp_{T,\mu}^\dag
  &=&
  -\frac{1}{\sqrt{|g|}}\gradOp_{T,\mu}\sqrt{|g|}
  \,.
\end{eqnarray}
The operator of exterior multiplication with the torsion 2-form is
\begin{eqnarray}
  {\bf T}^\alpha
  &\defas&
  T_\mu{}^\alpha{}_\beta\coordCreator^\mu\coordCreator^\beta
  \,.
\end{eqnarray}
Perturbing the exterior derivative with this operator gives
\begin{eqnarray}
  \extd_T &=&
  \extd - {\bf T}^\alpha \coordAnnihilator_\alpha
  \nonumber\\
  &=&
  \coordCreator^\mu \gradOp_{T,\mu}
  \,.
\end{eqnarray}
Note that\footnote{\cf e.g. \cite{BennTucker:1987},\S6.4 }
\begin{eqnarray}
  \antiCommutator{\extd}{\onbCreator^a} &=& 
  {\bf T}^a + \antiCommutator{\extd_T}{\onbCreator^a}
  \nonumber\\
  &=&
  {\bf T}^a
  -\fatomega_T{}^a{}_b\onbCreator^b
  \,.
\end{eqnarray}
(This is the ``first structure equation'' in the presence of torsion, \cf \refer{extd commutes with coordinate forms}.)
Taking the adjoint gives
\begin{eqnarray}
  \label{adjoint of torsion deformed extd}
  \coextd_T \;\defas\; (\extd_T)^\dag
  &=&
  \coextd - {\bf T}^{\dag\alpha} \coordCreator_\alpha
  \nonumber\\
  &=&
  -\coordAnnihilator^\mu \gradOp_{T,\mu}
  \,,
\end{eqnarray}
where
\begin{eqnarray}
  {\bf T}^{\dag\alpha}
  &=&
  -T_\mu{}^\alpha{}_{\beta}\coordAnnihilator^\mu\coordAnnihilator^\beta
  \,.
\end{eqnarray}
The torsion-perturbed Dirac operators are
\begin{eqnarray}
  \label{torsion-perturbed Dirac operators}
  \Dirac_{T,\pm}
  \;\defas\;
  \extd_T \pm \coextd_T
  &=&
  \clifford_\mp^\mu
  \gradOp_{T,\mu}
  \nonumber\\
  &=&
  \Dirac_\pm
  -
  \clifford_\mp^\mu
  T_{\mu}{}^\alpha{}_\beta
  \coordCreator^\beta\coordAnnihilator_\alpha
  \,.
\end{eqnarray}
and, due to \refer{adjoint of torsion deformed extd}, they are still 
(see \refer{adjointness of dirac and LapBelt}) (anti-)self-adjoint:
\begin{eqnarray}
  (\Dirac_{T,\pm})^\dag &=& \pm\Dirac_\pm
  \,.
\end{eqnarray}

We mention some further common vocabulary associated with torsion 
(\cf \S2.2 of \cite{SadriSheikh-Jabbari:2003}):
The \emph{Riemann curvature operator with torsion} is defined by
\begin{eqnarray}
  \label{Riemann tensor with torsion}
  {\bf R}^{{\rm T}}_{\mu\nu}
  &=&
  \commutator{\gradOp_{{\rm T},\mu}}{\gradOp_{{\rm T,\nu}}}
  \,.
\end{eqnarray}
If a torsion tensor exitst for which these operators and hence the Riemann curvature tensor with torsion
\begin{eqnarray}
  R^{{\rm T}}_{\mu\nu\alpha\beta}
  &=&
  R_{\mu\nu\alpha\beta}
  + 
  2
  \left(
    \nabla_{[\mu}T_{\nu] \alpha\beta}
    +
    T_{[\mu| \alpha \gamma|} T_{\nu]}{}^\gamma{}_\beta
  \right)
\end{eqnarray}
vanishes, the manifold is said to be \emph{parallelizable}. If furthermore a
vielbein frame covariantly constant with respect to $\Gamma_{\rm T}$ exists the 
manifold is said to be \emph{absolute parallizable} (which implies ordinary
parallelizability). 

The associated Ricci tensor with torsion is 
\begin{eqnarray}
  R^{\rm T}_{\mu\nu}
  &=&
  R_{\mu\nu}
  -
  \nabla_\alpha T^\alpha{}_{\mu\nu}
  +
  T_{\mu\alpha\beta}T_\nu{}^{\alpha\beta}
  \,.
\end{eqnarray}
The existence of a torsion making this tensor vanish is called \emph{Ricci parallelizability}.

\subsection{Conformal transformations.}
\label{conformal (Weyl) transformations}
Assume that the manifold $\manifold$ is equipped with two metric tensors $g_{\mu\nu}$,
$\tilde g_{\mu\nu}$ related by
\begin{eqnarray}
  \tilde g_{\mu\nu}\of{p} &=& e^{2\Phi\of{p}}g_{\mu\nu}\of{p}
\end{eqnarray}
for some real function $\Phi : \manifold \to \R$. 
In the following all objects associated with $\tilde g_{\mu\nu}$ are written
under a tilde, $\tilde \cdot$, while all other objects are associated with $g_{\mu\nu}$.

The coordinate basis forms are obviously related by
\begin{eqnarray}
  {\tilde \coordCreator^\mu} &=& e^{-\Phi}\coordCreator^\mu
  \nonumber\\
  {\tilde {\coordAnnihilator_\mu}} &=& e^{\Phi}\coordAnnihilator_\mu
\end{eqnarray}
and we may choose
\begin{eqnarray}
  {\tilde {\onbCreator^a}} &=& \onbCreator^a
  \nonumber\\
  {\tilde {\onbAnnihilator_a}} &=& \onbAnnihilator_a
  \,.
\end{eqnarray}
Also obvious is the transformation of $\partial_\mu^c$:
\begin{eqnarray}
  {\tilde {\partial_\mu^c}}
  &=&
  \partial_\mu^c
  +
  (\partial_\mu\Phi)\numberOperator
  \,,
\end{eqnarray}
because this is what satisfies the definition \refer{properties partial c op}.
With \refer{extd as a derivative commuting with coordinate forms} it follows that
\begin{eqnarray}
  \label{conformal extd}
  \tilde \extd &=& 
  {\tilde {\coordCreator^\mu}}
  {\tilde {\partial_\mu^c}}
  \nonumber\\
  &=&
  e^{-\Phi}\left(\extd + \commutator{\extd}{\Phi}\numberOperator\right)
  \,.
\end{eqnarray}

The conformally transformed Lie derivative operators are also readily found,
for instance from \refer{definition Lie derivative operator}:
\begin{eqnarray}
  {\tilde {\cal L}_v}
  &=&
  \antiCommutator
    {\tilde \extd}
    {\tilde {\coordAnnihilator_\mu}v^\mu}
  \nonumber\\
  &=&
  \antiCommutator
    {e^{-\Phi}\left(\extd + \coordCreator^\nu(\partial_\nu\Phi)\numberOperator\right)}
    {e^{\Phi}{\coordAnnihilator_\mu}v^\mu}  
  \nonumber\\
  &=&
  {\cal L}_v + v^\mu(\partial_\mu\Phi)\numberOperator
  \,.
\end{eqnarray}

The relation between the above operators and their conformal transformations
is in fact a similarity transformation:
\begin{eqnarray}
  \label{similarity transformations of conformal trafos}
  {\tilde{\coordCreator^\mu}}
  &=&
  e^{-\Phi\numberOperator} \,\coordCreator^\mu\, e^{\Phi\numberOperator}
  \nonumber\\
  {\tilde{\coordAnnihilator_\mu}}
  &=&
  e^{-\Phi\numberOperator} \,\coordAnnihilator_\mu\, e^{\Phi\numberOperator}
  \nonumber\\
  {\tilde {\partial_\mu^c}}
  &=&
  e^{-\Phi\numberOperator} \,\partial_\mu^c\, e^{\Phi\numberOperator}
  \nonumber\\
  {\tilde \extd} &=&
  e^{-\Phi \numberOperator}\,\extd\,e^{\Phi \numberOperator}
  \nonumber\\
  {\tilde {{\cal L}_v}} &=&
  e^{-\Phi \numberOperator}\,{\cal L}_v\,e^{\Phi \numberOperator}
  \,.
\end{eqnarray}
But this is not true for every operator:

To find $\tilde \coextd$ one can for instance use 
\refer{duality relation d extd} 
and write
\begin{eqnarray}
  \tilde \coextd 
  &=&
  -\bar \ast \, \tilde \extd\,\bar \ast
  \nonumber\\
  &=&
  -\bar \ast \, e^{-\Phi\numberOperator} \extd\,e^{\Phi\numberOperator} \bar \ast
  \nonumber\\
  &=&
  - \, e^{-\Phi(D-\numberOperator)}\,\bar\ast\, \extd\,\bar\ast\, e^{\Phi(D-\numberOperator)} 
  \nonumber\\
  &=&
  e^{-\Phi(D-\numberOperator)}\,
  \coextd
  \, e^{\Phi(D-\numberOperator)} 
  \,,
\end{eqnarray}
or
\begin{eqnarray}
  \label{conformal coextd}
  {\tilde {\coextd}}
  &=&
  e^{-\Phi}
  \left(
    \coextd
    -
    \commutator{\coextd}{\Phi}(D-\numberOperator)
  \right)
  \,.
\end{eqnarray}
This is also a similarity transformation, but a different one. However, it coincides
with that in \refer{similarity transformations of conformal trafos} 
when evaluated on forms with eigenvalue $n$ of $\numberOperator$
equal to $n = D/2$.
An immediate consequence of this result is that, for $D$ even and when acting
on forms $\ket{\psi}$ of degree $n = D/2$, 
the equations
\begin{eqnarray}
  \extd \ket{\psi} &=& 0
  \nonumber\\
  \coextd\ket{\psi} &=& 0
\end{eqnarray}
are conformally invariant in the sense that, with
\begin{eqnarray}
  {\tilde {\ket{\psi}}} &\defas& e^{-\Phi D/2}\ket{\psi}
  \,,
\end{eqnarray}
they are equivalent to
\begin{eqnarray}
  {\tilde\extd} {\tilde{\ket{\psi}}} &=& 0
  \nonumber\\
  {\tilde\coextd} {\tilde{\ket{\psi}}} &=& 0
  \,.
\end{eqnarray}

A special case of this is the fact that ordinary classical 
source free electromagnetism
in 4 dimensions is conformally invariant.

\newpage

\section{Proofs}
\label{proofs}
\setcounter{equation}{0}

\paragraph{Proofs of self-adjointness of the Hamiltonian}
\label{proofs of self-adjointness of the Hamiltonian}

In the following the proofs of the self-adjointness of various versions
of the Hamiltonian generator are given.

\begin{itemize}

\item \underline{Ordinary case:}
Equation \refer{Hamiltonian self-adjoint wrt proper scalar product} states
that
\begin{eqnarray}
  \label{restatement that Hamiltonian self-adjoint wrt proper scalar product}
  {\bf H}_{v_0}^{\dag_{\hermitianMetricOperator}}
  &=&
  {\bf H}_{v_0}
  \,.
\end{eqnarray}
\textit{Proof:}
The proof is probably easiest when using for ${\bf H}_{v_0}$ the representation
\begin{eqnarray}
  {\bf H}_{v_0}
  &=&
  \frac{i}{2}
  \left(
    v_0\inner\clifford_- \Dirac_+
    -
    v_0\inner\clifford_+ \Dirac_-
  \right)
  +i{\cal L}_{v_0}
\end{eqnarray}
(see \refer{definition of the Hamiltonian}).
Essential are furthermore the facts
\begin{eqnarray}
  \label{intermediate step for proof of proper hermiticity}
   \antiCommutator{\Dirac_+ }{v_0\inner \clifford_+}  
    - 
  \antiCommutator{\Dirac_-}{ v_0\inner \clifford_- }
  &=&
  \clifford_-^\mu\clifford_+^\nu(\nabla_\mu v_{0\nu})
  -
  \clifford_+^\mu\clifford_-^\nu(\nabla_\mu v_{0\nu})
  \nonumber\\
  &=& 0
\end{eqnarray}
(due to the antisymmetry $\nabla_\mu v_{0\nu} = \nabla_{[\mu}v_{0\nu]}$ of the
covariant derivative of the Killing vector $v_0$)
as well as
\begin{eqnarray}
  &&(i{\cal L}_{v_0})^\dag \;=\;i{\cal L}_{v_0}
  \,,
  \nonumber\\
  &&\commutator{{\cal L}_{v_0}}{\hermitianMetricOperator} \;=\; 0
 \,.
\end{eqnarray}
Using this, one finds
\begin{eqnarray}
  \label{proof of hermititcity of ordinary Hamiltonian}
  &&
  {\bf H}_{v_0}^{\dag_{\hermitianMetricOperator}}
  \nonumber\\
  &=&
  \left(\hermitianMetricOperator {\bf H}_{v_0}\hermitianMetricOperator^{-1}\right)^\dag
  \nonumber\\
  &=&
  \hermitianMetricOperator{\bf H}^\dag_{v_0}\hermitianMetricOperator
  \nonumber\\
  &\equalby{definition of the Hamiltonian}&
  -\frac{i}{2}
  \hermitianMetricOperator
  \left(
    v_0\inner \clifford_- \Dirac_+ - v_0\inner\clifford_+ \Dirac_-
  \right)^\dag
  \hermitianMetricOperator
  +
  i{\cal L}_{v_0}
  \nonumber\\  
  &=&
  \frac{i}{2}\frac{1}{v_0\inner v_0}v_0\inner \clifford_- v_0\inner \clifford_+
    \left(
       \Dirac_+ v_0\inner \clifford_-  - \Dirac_- v_0\inner \clifford_+
    \right)
  v_0\inner \clifford_- v_0\inner \clifford_+
  \frac{1}{v_0\inner v_0}
  + 
  i{\cal L}_{v_0}
  \nonumber\\
  &=&
  -\frac{i}{2}\frac{1}{v_0\inner v_0}
    v_0\inner \clifford_- v_0\inner \clifford_+
  \left(
       \Dirac_+ v_0\inner \clifford_+  
    - 
  \Dirac_- v_0\inner \clifford_- 
  \right)
  + 
  i{\cal L}_{v_0}
  \nonumber\\
  &\equalby{intermediate step for proof of proper hermiticity}&
  \frac{i}{2}
  \frac{1}{v_0\inner v_0}
    v_0\inner \clifford_- v_0\inner \clifford_+
  \left(
       v_0\inner \clifford_+  \Dirac_+ 
    - 
   v_0\inner \clifford_- \Dirac_-
  \right)
  + 
  i{\cal L}_{v_0}
  \nonumber\\
  &=&
  \frac{i}{2}
  \left(
    v_0\inner \clifford_- \Dirac_+
    -
    v_0\inner \clifford_+ \Dirac_-
  \right)
  +i{\cal L}_{v_0}
  \nonumber\\
  &=&
  {\bf H}_{v_0}
  \,.
\end{eqnarray}

\underline{Stronger version}:

In the special case where $v_0$ is covariantly constant a stronger version of this result holds.\footnote{
The interest in the following discussion lies in the fact that it generalizes to the case of non-vanishing
$b$-field background.
}

The Hamiltonian \refer{definition of the Hamiltonian} naturally decomposes into a 
left and a right part
\begin{eqnarray}
  {\bf H }_{v_0}^{{\rm L}/{\rm R}}
  &\defas&
  \pm
  \frac{i}{4}
  \commutator
  {v_0\inner \clifford_\mp}
  {\Dirac_{\pm}}
  \,. 
\end{eqnarray}
If $v_0$ is covariantly constant then these two operators are seperately $\hermitianMetricOperator$-hermitian:
\begin{eqnarray}
  \left(
    {\bf H }_{v_0}^{{\rm L}/{\rm R}}
  \right)^{\dag_\hermitianMetricOperator}
  &=&
    {\bf H }_{v_0}^{{\rm L}/{\rm R}}
  \hspace{1cm}
  \mbox{if $\gradOp_\mu v_0 = 0$}
  \,.
\end{eqnarray}
{\it Proof}:
The proof is just a special case of \refer{proof of hermititcity of ordinary Hamiltonian}, making
use of the fact that 
\begin{eqnarray}
  \antiCommutator{\Dirac_\pm}{v_0\inner \clifford_\pm} &=& 0
  \,,
\end{eqnarray}
which is a direct consequence of the assumption that $\gradOp_\mu v_0 = 0$:
\begin{eqnarray}
  \label{simple principle of proof of eta-hermititcity}
  \left(
  \frac{i}{4}
  \commutator
  {v_0\inner \clifford_\mp}
  {\Dirac_{\pm}}
  \right)^{\dag_\hermitianMetricOperator}
  &=&
  -
  \frac{i}{4}
  \hermitianMetricOperator
  \commutator
  {v_0\inner \clifford_\mp}
  {\Dirac_{\pm}}
  \hermitianMetricOperator
  \nonumber\\
  &=&
  \pm
  \frac{i}{4}
  \frac{1}{v_0\inner v_0}
  v_0\inner\clifford_\pm
  \commutator
  {v_0\inner \clifford_\mp}
  {\Dirac_{\pm}}  
  v_0\inner\clifford_\pm
  \nonumber\\
  &=&
  \frac{i}{4}
  \commutator
  {v_0\inner \clifford_\mp}
  {\Dirac_{\pm}}  
  \,.
\end{eqnarray}

\item \underline{$k$-deformed case:}
Equation \refer{eta hermiticity of k-deformed Hamiltonian} 
states that the same is true for the $k$-deformed
Hamiltonian:
\begin{eqnarray}
  {\bf H}_{k,v_0}^{\dag_\hermitianMetricOperator}
  &=&
  {\bf H}_{k,v_0}
  \,.
\end{eqnarray}
\textit{Proof}:
It suffices to note that the analogue of 
\refer{intermediate step for proof of proper hermiticity} also holds for the
$k$-deformed case. The proof then goes through as above.

\item \underline{Background $b$-field:}
In the case of a non-vanishing $b$-field the Hamiltonian reads
\begin{eqnarray}
  {\bf H}_{v_0}
  &=&
  \frac{i}{2}
  \left(
    v_0 \inner \clifford_-^{(b)} \Dirac^{(b)}_+
   -
    v_0 \inner \clifford_+^{(b)} \Dirac^{(b)}_-
  \right)
  +
  i{\cal L}_{v_0}
  \,.
\end{eqnarray}

Equation \refer{b-deformed hamiltonian is selfadjoitn wrt b-deformed metric operator} 
states that this operator is self-adjoint with respect to
$\hermitianMetricOperator^{(b)} = 
\frac{1}{
  \left(
    v_0\inner \clifford_-^{(b)}
  \right)^2
} v_0\inner \clifford_-^{(b)} v_0\inner \clifford_+^{(b)}$.

\textit{Proof:}
The analogue of 
\refer{intermediate step for proof of proper hermiticity}
is still true:
\begin{eqnarray}
  \label{intermediate step for proof of proper hermiticity in presence of b-field}
   &&
   \antiCommutator{\Dirac^{(b)}_+ }{v_0\inner \clifford^{(b)}_+}  
    - 
  \antiCommutator{\Dirac_-^{(b)}}{ v_0\inner \clifford_-^{(b)} }
  \nonumber\\
  &=&
  \antiCommutator{e^{-{\bf W}^{(b)}}\extd e^{{\bf W}^{(b)}}
   + e^{{\bf W}^{(b)\dag}}\coextd e^{-{\bf W}^{(b)\dag}}}
   {e^{{\bf W}^{(b)\dag}}v_0\inner \coordCreator e^{-{\bf W}^{(b)\dag}}
    +e^{-{\bf W}^{(b)}} v_0\inner\coordAnnihilator \,e^{{\bf W}^{(b)}}
    }
  \nonumber\\
  &&-
  \antiCommutator{e^{-{\bf W}^{(b)}}\extd e^{{\bf W}^{(b)}}
   - e^{{\bf W}^{(b)\dag}}\coextd e^{-{\bf W}^{(b)\dag}}}
   {e^{{\bf W}^{(b)\dag}}v_0\inner \coordCreator e^{-{\bf W}^{(b)\dag}}
    -e^{-{\bf W}^{(b)}} v_0\inner\coordAnnihilator \,e^{{\bf W}^{(b)}}
    }
  \nonumber\\
  &=&
  2
  \antiCommutator{
    e^{-{\bf W}^{(b)}}\extd e^{{\bf W}^{(b)}}
   }
   {
     e^{-{\bf W}^{(b)}} v_0\inner\coordAnnihilator \,e^{{\bf W}^{(b)}}
   }
  +2
  \antiCommutator{
    e^{{\bf W}^{(b)\dag}}\coextd e^{-{\bf W}^{(b)\dag}}
  }
  {
    e^{{\bf W}^{(b)\dag}}v_0\inner \coordCreator e^{-{\bf W}^{(b)\dag}}
  }
  \nonumber\\
  &=&
  2{\cal L}_{v_0} - 2{\cal L}_{v_0} \;=\;
  0
  \,.
\end{eqnarray}
Also, the Lie derivative along $v_0$ still commutes with 
$\hermitianMetricOperator^{(b)}$:
\begin{eqnarray}
  \commutator{{\cal L}_{v_0}}{v_0\inner \clifford_\pm^{(b)}}
  &\stackrel{\commutator{{\cal L}_{v_0}}{{\bf W}^{(b)}}=0}{=}&
  0
  \,.
\end{eqnarray}
Therefore all ingredients are present to prove 
\refer{b-deformed hamiltonian is selfadjoitn wrt b-deformed metric operator} analogously 
to \refer{restatement that Hamiltonian self-adjoint wrt proper scalar product}.

\end{itemize}

\paragraph{Proof that ${\bf C}$ weakly commutes with ${\bf H}$.}
\label{proof that commutator C with H is prop to C}

The fact that the Hamiltonian generator respects the spatial constraint
is proven for various cases.

\begin{itemize}

\item \underline{Ordinary case:}

Equation \refer{proposition that commutator C and H is prop to C in main text}
on p. \pageref{proposition that commutator C and H is prop to C in main text}
states that
\begin{eqnarray}
  \label{proposition that commutator C and H is prop to C}
  \commutator{{\bf C}_{v_0}}{{\bf H}_{v_0}}
  &=&
  \frac{1}{2i v_0\inner v_0}
  (\nabla_{[\mu} v_{\nu]})
  v_0\inner\clifford_- \clifford_-^\mu\clifford_+^\nu
  v_0\inner \clifford_+
   \;{\bf C}_{v_0}
  \,.
\end{eqnarray}
\textit{Proof:}
First we can rewrite the commutator as
\begin{eqnarray}
  \label{first step in the second proof}
  \commutator{{\bf C}_{v_0}}{{\bf H}_{v_0}}
  &=&
  \commutator{{\bf C}_{v_0}}{-i{\cal L}_{v_0} + {\bf H}_{v_0}}
  \nonumber\\
  &=&
  \frac{1}{4i}
  \commutator
    {v_0\inner \clifford_+ \Dirac_- + v_0\inner \clifford_- \Dirac_+ }
    {v_0\inner \clifford_+ \Dirac_- - v_0\inner \clifford_- \Dirac_+ }
  \nonumber\\
  &=&
  \frac{1}{2i}
  \commutator{v_0\inner \clifford_- \Dirac_+}{v_0\inner \clifford_+ \Dirac_-}
  \nonumber\\
  &=&
  \frac{1}{2i}
  \left(
  v_0\inner\clifford_- \clifford_-^\mu\clifford_+^\nu(\nabla_\mu v_\nu)\Dirac_-
  -
  v_0\inner\clifford_+ \clifford_+^\mu\clifford_-^\nu(\nabla_\mu v_\nu)\Dirac_+
  \right)
  \,.
\end{eqnarray}
Since ${\bf C}_{v_0}$ and ${\bf H}_{v_0}$ both commute with $t_{v_0}$
this expression also commutes with $t_{v_0}$. This is still obvious in the third
line,
\begin{eqnarray}
  \commutator
    {\commutator{v_0\inner \clifford_-\Dirac_+}{v_0\inner \clifford_+\Dirac_-}}
    {t_{v_0}}
  &=&
    \commutator
      {v_0\inner\clifford_-\frac{1}{v_0\inner v_0}v_0\inner\clifford_-}
      {v_0\inner \clifford_+\Dirac_-}
    +
    \commutator
     {v_0\inner \clifford_-\Dirac_+}    
     {v_0\inner\clifford_+\frac{1}{v_0\inner v_0}v_0\inner\clifford_+}
  \nonumber\\
   &=&
    \commutator
      {-1}
      {v_0\inner \clifford_+\Dirac_-}
    +
    \commutator
     {v_0\inner \clifford_-\Dirac_+}    
     {1}
  \nonumber\\
  &=& 0
  \,.
\end{eqnarray}
but it is a nontrivial condition in the fourth line:
\begin{eqnarray}
  \label{something that should vanish}
 &&
 \commutator{
 \left(
  v_0\inner\clifford_- \clifford_-^\mu\clifford_+^\nu(\nabla_\mu v_\nu)\Dirac_-
  -
  v_0\inner\clifford_+ \clifford_+^\mu\clifford_-^\nu(\nabla_\mu v_\nu)\Dirac_+
  \right)
  }
  {t_{v_0}}
 \nonumber\\
 &=& 
  \frac{1}{v_0\inner v_0}
  \left(
  v_0\inner\clifford_- \clifford_-^\mu\clifford_+^\nu(\nabla_{[\mu} v_{\nu]})v_0\inner\clifford_+
  -
  v_0\inner\clifford_+ \clifford_+^\mu\clifford_-^\nu(\nabla_{[\mu} v_{\nu]})v_0\inner\clifford_-
  \right)
  \nonumber\\
  &=&
  -\frac{1}{2}\frac{1}{v_0\inner v_0}(\nabla_{[\mu} v_{\nu]})
  \commutator{v_0\inner \clifford_-}{ \clifford_-^\mu}    
  \commutator{ v_0\inner \clifford_+}{\clifford_+^\nu}
  \,,
\end{eqnarray}
where the last line follows from explicitly evaluating the respective terms:
\begin{eqnarray}
  &&(\nabla_{[\mu}v_{\nu]})
  v_0\inner \clifford_- \clifford_-^\mu\clifford_+^\nu v_0\inner \clifford_+
  \nonumber\\
 &=&
  \frac{1}{4}
  (\nabla_{[\mu}v_{\nu]})
  \left(
     \antiCommutator{v_0\inner \clifford_-}{ \clifford_-^\mu}
      +
     \commutator{v_0\inner \clifford_-}{ \clifford_-^\mu}
  \right)
  \left(
    \antiCommutator{\clifford_+^\nu}{ v_0\inner \clifford_+}
    +
    \commutator{\clifford_+^\nu}{ v_0\inner \clifford_+}
  \right)
  \nonumber\\
  &=&
  \frac{1}{4}
  (\nabla_{[\mu}v_{\nu]})
  \left(
     -2 v_0^\mu
      +
     \commutator{v_0\inner \clifford_-}{ \clifford_-^\mu}
  \right)
  \left(
    2v_0^\nu
    +
    \commutator{\clifford_+^\nu}{ v_0\inner \clifford_+}
  \right)  
  \nonumber\\
  &=&
  \frac{1}{4}
  (\nabla_{[\mu}v_{\nu]})
  \left(
     2 v_0^\mu
    \commutator{ v_0\inner \clifford_+}{\clifford_+^\nu}
     -
    2v_0^\mu
     \commutator{v_0\inner \clifford_-}{ \clifford_-^\nu}
    -
    \commutator{v_0\inner \clifford_-}{ \clifford_-^\mu}    
    \commutator{ v_0\inner \clifford_+}{\clifford_+^\nu}
  \right)  
  \nonumber\\
  \nonumber\\
  &&
  (\nabla_{[\mu}v_{\nu]})
  v_0\inner \clifford_+ \clifford_+^\mu\clifford_-^\nu v_0\inner \clifford_-
  \nonumber\\
 &=&
  \frac{1}{4}
  (\nabla_{[\mu}v_{\nu]})
  \left(
     \antiCommutator{v_0\inner \clifford_+}{ \clifford_+^\mu}
      +
     \commutator{v_0\inner \clifford_+}{ \clifford_+^\mu}
  \right)
  \left(
    \antiCommutator{\clifford_-^\nu}{ v_0\inner \clifford_-}
    +
    \commutator{\clifford_-^\nu}{ v_0\inner \clifford_-}
  \right)
  \nonumber\\
  &=&
  \frac{1}{4}
  (\nabla_{[\mu}v_{\nu]})
  \left(
     2 v_0^\mu
      +
     \commutator{v_0\inner \clifford_+}{ \clifford_+^\mu}
  \right)
  \left(
    -2v_0^\nu
    +
    \commutator{\clifford_-^\nu}{ v_0\inner \clifford_-}
  \right)  
  \nonumber\\
  &=&
  \frac{1}{4}
  (\nabla_{[\mu}v_{\nu]})
  \left(
     2 v_0^\mu
    \commutator{ v_0\inner \clifford_+}{\clifford_+^\nu}
     -
    2v_0^\mu
     \commutator{v_0\inner \clifford_-}{ \clifford_-^\nu}
    +
    \commutator{v_0\inner \clifford_-}{ \clifford_-^\mu}    
    \commutator{ v_0\inner \clifford_+}{\clifford_+^\nu}
  \right)
  \,.  
\end{eqnarray}
We will now use the fact that \refer{something that should vanish} vanishes
to prove \refer{proposition that commutator C and H is prop to C}:
From the definitions
\begin{eqnarray}
  {\bf C}_{v_0}  
  &\defas&
  v_0\inner \clifford_+ \Dirac_- + v_0\inner \clifford_- \Dirac_+
  \nonumber\\
  4\left({\cal L}_{v_0} + i{\bf H}_{v_0}\right)
  &\defas&
  v_0\inner \clifford_+ \Dirac_- - v_0\inner \clifford_- \Dirac_+
\end{eqnarray}
it follows that
\begin{eqnarray}
  \label{writing D constraints in terms of Hamiltonian constraints}
  \Dirac_- 
   &=& 
   \frac{1}{2 v_0\inner v_0}
   v_0\inner \clifford_+
   \left({\bf C}_{v_0} + 4\left({\cal L}_{v_0} + i{\bf H}_{v_0}\right)\right)
  \nonumber\\
  \Dirac_+
   &=& 
   -
   \frac{1}{2 v_0\inner v_0}
   v_0\inner \clifford_-
   \left({\bf C}_{v_0} - 4\left({\cal L}_{v_0} + i{\bf H}_{v_0}\right)\right)  
  \,.
\end{eqnarray}
Using this to replace $\Dirac_\pm$ in \refer{first step in the second proof} gives
the desired result:
\begin{eqnarray}
  &&\commutator{{\bf C}_{v_0}}{{\bf H}_{v_0}}
  \nonumber\\
  &=&
  \frac{1}{2i}
  \left(
  v_0\inner\clifford_- \clifford_-^\mu\clifford_+^\nu(\nabla_{[\mu} v_{\nu]})\Dirac_-
  -
  v_0\inner\clifford_+ \clifford_+^\mu\clifford_-^\nu(\nabla_{[\mu} v_{\nu]})\Dirac_+
  \right)
  \nonumber\\
  &\equalby{writing D constraints in terms of Hamiltonian constraints}&
  \frac{1}{4i v_0\inner v_0}
  (\nabla_{[\mu} v_{\nu]})
  \left(
  v_0\inner\clifford_- \clifford_-^\mu\clifford_+^\nu
  v_0\inner \clifford_+
   \left({\bf C}_{v_0} + 4\left({\cal L}_{v_0} + i{\bf H}_{v_0}\right)\right)
  +
  v_0\inner\clifford_+ \clifford_+^\mu\clifford_-^\nu
   v_0\inner \clifford_-
   \left({\bf C}_{v_0} - 4\left({\cal L}_{v_0} + i{\bf H}_{v_0}\right)\right)  
  \right)  
  \nonumber\\ 
  &\equalby{something that should vanish}&
  \frac{1}{2i v_0\inner v_0}
  (\nabla_{[\mu} v_{\nu]})
  v_0\inner\clifford_- \clifford_-^\mu\clifford_+^\nu
  v_0\inner \clifford_+
   \;{\bf C}_{v_0}
\end{eqnarray}

\item \underline{$k$-deformed case:}

Equation \refer{proposition that commutator Ck and Hk is prop to Ck}
states the analogous relation for the $k$-deformed operators:
\begin{eqnarray}
  \label{proposition that commutator Ck and Hk is prop to Ck}
  \commutator{{\bf C}_{k,v_0}}{{\bf H}_{k,v_0}}
  &=&
  \frac{1}{2i v_0\inner v_0}
  (\nabla_{[\mu} v_{\nu]})
  v_0\inner\clifford_- \clifford_-^\mu\clifford_+^\nu
  v_0\inner \clifford_+
   \;{\bf C}_{k,v_0}
  \,.
\end{eqnarray}
\textit{Proof:}
Because
\begin{eqnarray}
  \antiCommutator{\Dirac_{k,\pm}}{v\inner \clifford_\pm}
  &=&
  \clifford_\mp^\mu\clifford_\pm^\nu(\nabla_\mu v_nu)
  \,,
\end{eqnarray}
just as in the undeformed case, the proof completely parallels that given above.

\end{itemize}

\section{Example: Parameter evolution in classical electromagnetism}
\label{p-form electromagnetism}
\setcounter{equation}{0}

As an example of the general constructions in 
\S\fullref{Target space Killing evolution}
we demonstrate how the Hamiltonian $\bf H$ and the spatial
constraint $\bf C$ \refer{Hamiltonian form of susy constraints} 
look like in the special case where
$\Dirac_\pm\fatomega = 0$ are the Maxwell equations of sourceless
classical electromagnetism.

The Faraday 2-form is
\begin{eqnarray}
  {\bf F} &=& 
  \extd A
  \nonumber\\
  &=&
  \frac{1}{2}F_{\mu\nu}dx^\mu\wedge dx^\nu
  \nonumber\\
  &=&
  \left(\nabla A_0 - \dot A\right)_i \,dx^i\wedge dx^0
  +
  \left({\rm rot} A\right)_j \,\frac{1}{2}\epsilon^j{}_{kl}dx^k\wedge dx^l
  \nonumber\\
  &=&
  {\bf E}\wedge dt + {\bf B}
  \nonumber\\
  &=&
  E_i dx^i\wedge dx^0
  +
  B_i \frac{1}{2}\epsilon^i{}_{jk} dx^j\wedge dx^k
  \,.
\end{eqnarray}
For Minkowski space $g = \eta$ its dual reads
\begin{eqnarray}
  \star {\bf F}
  &=&
  E_i \,\frac{1}{2}\epsilon^j_{jk}dx^j\wedge dx^k
  +
  -B_i \, dx^i\wedge dx^0
  \,.
\end{eqnarray}
The constraints $\extd F = 0 = \coextd F$ hence give
\begin{eqnarray}
  0 &=& \extd {\bf F} \nonumber\\&=&
  \partial_j E_i \,dx^j\wedge dx^i\wedge dx^0
  +
  \partial_0 B_i \frac{1}{2}\epsilon^i{}_{jk} dx^0\wedge dx^j\wedge dx^k
  +
  \partial_i B^i \,dx^1\wedge dx^2\wedge dx^3
  \nonumber\\
  &=&
  \left({\rm rot}E + \dot B\right)_j\,\frac{1}{2}\epsilon^j{}_{kl}
  dx^k\wedge dx^l \wedge dx^0 
  +
  ({\rm div}B) \,dx^1\wedge dx^2\wedge dx^3
\nonumber\\
  0 &=& \extd \star {\bf F} \nonumber\\&=&
  \left(\dot E - {\rm rot}B \right)_j\,\frac{1}{2}\epsilon^j{}_{kl}
  dx^k\wedge dx^l \wedge dx^0 
  +
  ({\rm div}E) \,dx^1\wedge dx^2\wedge dx^3
  \,,
\end{eqnarray}
the components of which are the Maxwell equations.

The vector
\begin{eqnarray}
  v_0 &=& \partial_0
\end{eqnarray}
is a timelike Killing vector on Minkowski space	time. 
The associated Clifford element is
\begin{eqnarray}
 v_0\inner \clifford_\pm
  &=&
  -\clifford_\pm^0
 \,,
\end{eqnarray} 
and the Hamiltonian generator \refer{definition of the Hamiltonian} along $v_0$ is
\begin{eqnarray}
  {\bf H}_{v_0}
  &=&
  \frac{i}{2}
  \left(
    (-\clifford_-^0)\clifford_-^i \partial_i
    -
    (-\clifford_+^0)\clifford_+^i \partial_i
  \right)
  \,.
\end{eqnarray}
Its action on 2-forms is given by:
\begin{eqnarray}
  {\bf H}_{v_0} {\bf F}
  &=&
  \frac{i}{2}
  \left(
    (-\clifford_-^0)\clifford_-^i \partial_i
    -
    (-\clifford_+^0)\clifford_+^i \partial_i
  \right)
  \,
  {F_{\mu\nu}}\clifford_-^\mu\clifford_-^\nu\ket{0}
  \nonumber\\
  &=&
  -\frac{i}{2}
  \partial_iF_{\mu\nu}\commutator{\clifford_-^0\clifford_-^i}{\clifford_-^\mu\clifford_-^\nu}\ket{0}  
  \nonumber\\
  &=&
  -\frac{i}{2}
  \partial_i E_j\commutator{\clifford_-^0\clifford_-^i}{\clifford_-^j\clifford_-^0}\ket{0}  
  -
  \frac{i}{2}
  \partial_i B_j\frac{1}{2}
  \epsilon^j{}_{kl}\commutator{\clifford_-^0\clifford_-^i}{\clifford_-^k\clifford_-^l}\ket{0}  
  \nonumber\\
  &=&
  -i\left({\rm rot}E\right)_k \frac{1}{2}\epsilon^k{}_{lm}\clifford_-^l\clifford_-^m  \ket{0}
  +
  i
  \partial_i B_j\epsilon^{ji}{}_{l}\clifford_-^0\clifford_-^l\ket{0}    
  \nonumber\\
  &=&
  -i\left({\rm rot}E\right)_k \frac{1}{2}\epsilon^k{}_{lm}\clifford_-^l\clifford_-^m  \ket{0}
  +
  i
  \left({\rm rot}B\right)_j\clifford_-^j\clifford_-^0\ket{0}    
  \,.
  \nonumber\\
\end{eqnarray}
Therefore the evolution equation \refer{Dirac-Schroedinger equation} is here equivalent to
the two Maxwell equations which contain time derivatives:
\begin{eqnarray}
  i {\cal L}_{v_0} {\bf F} &=& {\bf H}_{v_0}{\bf F}
  \nonumber\\
  \Leftrightarrow
  \dot {\bf E}\wedge dt + \dot {\bf B}
  &=&
  ({\rm rot}{\bf B}) \wedge dt - {\rm rot} {\bf E}
  \,,
\end{eqnarray}
while the spatial constraint \refer{Hamiltonian form of susy constraints} is equivalent to the
remaing two Maxwell equations:
\begin{eqnarray}
  0 &=& {\bf C}_{v_0}{\bf F}
  \nonumber\\
  &=&
  -
  \left(
    \clifford_-^0\clifford_-^i \partial_i
    +
    \clifford_+^0\clifford_+^i \partial_i
  \right)
  \,
  {F_{\mu\nu}}\clifford_-^\mu\clifford_-^\nu\ket{0}
  \nonumber\\
  &=&
  -\partial_iF_{\mu\nu}\antiCommutator{\clifford_-^0\clifford_-^i}{\clifford_-^\mu\clifford_-^\nu}\ket{0}  
  \nonumber\\
  &=&
  -\partial_i E_j\antiCommutator{\clifford_-^0\clifford_-^i}{\clifford_-^j\clifford_-^0}\ket{0}  
  -
  \partial_i B_j\frac{1}{2}
  \epsilon^j{}_{kl}\antiCommutator{\clifford_-^0\clifford_-^i}{\clifford_-^k\clifford_-^l}\ket{0}  
  \nonumber\\
  &=&
  2
  \partial_i E^i\ket{0}
  -
  \partial_i B_j
  \epsilon^j{}_{kl}\clifford_-^0\clifford_-^{[i}}{\clifford_-^k\clifford_-^{l]}\ket{0}  
  \nonumber\\
  &=&
  2
  \partial_i E^i\ket{0}
  -
  \partial_i B^i
  \epsilon_{jkl}\clifford_-^0\clifford_-^{[j}}{\clifford_-^k\clifford_-^{l]}\ket{0}  
  \nonumber\\
  &=&
  2\, ({\rm div}E)
  - 6\, ({\rm div}B) \;dx^0\wedge dx^1\wedge dx^2\wedge dx^3 
  \,.
\end{eqnarray}
When the 2-forms in 4-dimensional Minkowski space are represented as 
6-dimensional column vectors, the set of equations \refer{Hamiltonian form of susy constraints}
is therefore precisely the well known evolution equation of electrodynamics, as for
instance discussed in \cite{Leis:1985}.

The construction of the scalar product in \S\fullref{scalar product} 
also reproduces well
known facts when applied to classical electromagnetism. In particular the
tensor $T^{\mu\nu}$ in equation \refer{generalized stress energy tensor} becomes 
the Maxwell stress-energy tensor:
\begin{eqnarray}
  \label{Maxwell stress energy tensor}
  T^{\mu\nu} &=& \frac{1}{2}\bra{{\bf F}}\clifford_-^\nu\clifford_+^\mu\ket{{\bf F}}_{\rm loc}
  \nonumber\\
  &=&
  -\frac{1}{2}\bra{1}{{\bf F}}\clifford_-^\mu{\bf F}\clifford_-^\nu\ket{1}_{\rm loc}
  \nonumber\\
  &=&
  F^{\mu\lambda}F^{\nu}{}_\lambda - \frac{1}{4}g^{\mu\nu}F_{\gamma\lambda}F^{\gamma\lambda}
  \,.
\end{eqnarray}

\section{Lie groups and algebras.}
\label{Lie groups and algebras}
\setcounter{equation}{0}

Some well known relations are assembled below for references in the main text.
We mostly follow the notation in \S11.4 of \cite{Polchinski:1998}.

The Lie algebra generators $T_a$, satisfying
\begin{eqnarray}
  \commutator{T_a}{T_b} &=& f_a{}^c{}_b T_c
\end{eqnarray}
and
\begin{eqnarray}
  &&\commutator{T_a}{\commutator{T_b}{T_c}}
  +
  \commutator{T_b}{\commutator{T_c}{T_a}}
  +
  \commutator{T_c}{\commutator{T_a}{T_b}}
  \;=\;
  0
\nonumber\\
\Leftrightarrow
  \label{cyclic property of structure constants}
  &&f_{[a}{}^e{}_{|d|} f_b{}^d{}_{c]}
  \;=\; 0
\end{eqnarray}
are represented on themselves by the adjoint action
\begin{eqnarray}
  {\rm ad}T_a\of{T_b} &=& \commutator{T_a}{T_b}
  \;=\;f_a{}^c{}_b\, T_c
\end{eqnarray}
with coefficient matrices
\begin{eqnarray}
  {\rm ad}\of{T_a}^c{}_b
  &=& f_a{}^c{}_b
  \,.
\end{eqnarray}
The Killing form serves as the metric tensor
\begin{eqnarray}
  \label{Killing form}
  \eta_{ab}
  &=&
  -\frac{1}{2\coxeter}{\rm tr}\of{{\rm ad}T_a {\rm ad}T_b}
  \nonumber\\
  &=&
  -\frac{1}{2\coxeter}{\rm tr}\of{f_a{}^c{}_s f_b{}^s{}_d}
  \nonumber\\
  &=&
  -\frac{1}{2\coxeter}f_{a}{}^t{}_s f_b{}^s{}_t  
  \,,
\end{eqnarray}
where $\coxeter$ is the dual Coxeter number.

By left- and right-translation the $T_a$ generate two commuting vielbein
fields (\cf \cite {FroehlichGawedzki:1993}, \S 4.2) 
$e_a^{\pm} = e^\pm_a{}^\mu \partial_\mu$:
\begin{eqnarray}
  e^+_a &=& g^{-1} (\partial_a g)
  \nonumber\\
  e^-_a &=& (\partial_a g)g^{-1}
  \,.
\end{eqnarray}
By default we refer to the (-) vielbein when the index is omitted:
\begin{eqnarray}
  T_a &\defas& e_a \;\defas\; e^-_a
  \,.
\end{eqnarray}

The Levi-Civita connection in this basis is 
(\cf \cite{FroehlichGrandjeanRecknagel:1997}, (4.70))
\begin{eqnarray}
  \label{Lie group LC connection}
  \omega_{abc} &=& \frac{1}{2}f_{abc}
  \,.
\end{eqnarray}
Therefore the covariant derivative operator 
\refer{gradop in terms of cliffords} and the 
spinor version \refer{spinor covariant derivative} read
\begin{eqnarray}
  \gradOp_a &=& T_a + \omega_{abc}\onbCreator^b\onbAnnihilator^c
  \nonumber\\
  &=&
  T_a
  +
  \frac{1}{2}f_{abc}\onbCreator^b\onbAnnihilator^c
  \nonumber\\
  \gradOp^{S\pm}_a &=& T_a \pm \frac{1}{4}\omega_{abc}\clifford_\pm^b\clifford_\pm^c
  \nonumber\\
  &=&
  T_a
  \pm
  \frac{1}{8}f_{abc}\clifford_\pm^b\clifford_\pm^c
  \,,
\end{eqnarray}
and the spinor Lie derivatives \refer{Killing spinor Lie derivative}
along the group's Killing vectors (the vielbein components) are
\begin{eqnarray}
  \label{spinor Killing lie derivative on group manifolds}
  {\cal L}^{{\rm S}_\pm}_{a}
  &=&
  \partial_a \pm \frac{1}{2}\omega_{abc}\clifford_\pm^a\clifford_\pm^b
  \,.
\end{eqnarray}
Since the connection terms satisfy
\begin{eqnarray}
  \label{commutator of biliniear fermionic currents on goups manifolds}
  \commutator{\frac{1}{4}f_{ast}\clifford_\pm^s\clifford_\pm^t}
  {\frac{1}{4}f_{bqr}\clifford_\pm^q\clifford_\pm^r}
  &=&
  \pm
  f_{a}{}^c{}_b
  \left(
    \frac{1}{4}  f_{csr} \clifford_\pm^s\clifford_\pm^r
  \right)
\end{eqnarray}
these manifestly represent the group's Lie algebra:
\begin{eqnarray}
  \commutator{{\cal L}^{{\rm S}_{\pm}}_a}{{\cal L}^{{\rm S}_{\pm}}_b}
  &=&
  f_a{}^c{}_b{{\cal L}^{{\rm S}_{\pm}}_c}
  \,.
\end{eqnarray}
The spinor Lie derivatives along ´the invariant Killing vectors of the group manifold
have the following commutators with various other objects.
\begin{eqnarray}
  \commutator{{\cal L}^{{\rm S}_+}_a}{\clifford_{+b}}
  &=&
  f_a{}^c{}_b \clifford_{+c}
  \nonumber\\
  \commutator{{\cal L}^{{\rm S}_+}_a}
    {\frac{1}{2}\omega_{bst}\clifford_+^s\clifford_+^t}
  &=&
  f_a{}^c{}_b \frac{1}{2}\omega_{cst}\clifford_+^s\clifford_+^t
  \nonumber\\
  \commutator{{\cal L}^{{\rm S}_+}_a}
    {\partial_b}
  &=&
  f_a{}^c{}_b \partial_c
  \,.
\end{eqnarray}
By the same argument familiar from the construction of the quadratric Casimir,
with any two objects $A_a$, $B_a$ that tranform this way an invariant under 
the action of ${\cal L}^{{\rm S}_+}_a$ can be constructed:
\begin{eqnarray}
  \label{invariants of the Lie action of the spinor Lie derivative on groups}
  \commutator{{\cal L}^{{\rm S}_+}_a}{g^{bc}A_b B_c}
  &=&
  g^{bc}f_a{}^d{}_b  A_d B_c
  +
  g^{bc} f_a{}^d{}_c A_b B_d
  \nonumber\\
  &=&
  0
  \,.
\end{eqnarray}

The Riemann curvature operator on the group manifold is
(\cf \cite{SpindelSevrinTroostVanProeyen:1988}, (A.26))
\begin{eqnarray}
  \label{curvature tensor of Lie group}
  {\bf R}_{ab} &=&
  \frac{1}{4}
  f_{abs} f^s{}_{cd}
  \onbCreator^c
  \onbAnnihilator^d
  \nonumber\\
  &=&
  \omega_{abs}\omega^s{}_{cd}
  \onbCreator^c
  \onbAnnihilator^d
  \,.
\end{eqnarray}
And the Ricci tensor and curvature scalar are
\begin{eqnarray}
  \label{Ricci tensor on group manifolds}
  R_{ab} &=&
  \frac{1}{4}
  f_{ars}f_b{}^{rs}
  \nonumber\\
  &=&
  \omega_{ars}\omega_b{}^{rs}
  \nonumber\\
  &\equalby{Killing form}&
  \frac{\coxeter}{2}g_{ab}
\end{eqnarray}
and
\begin{eqnarray}
  \label{scalar curvature on group manifolds}
  R &=& \frac{\coxeter\, d}{2}
  \,.
\end{eqnarray}

The two invariant vielbein fields $e^\pm$ on the group manifold are characterized
by the property that they are parallel with respect to the metric compatible connection
with torsion $T_{\mu\nu\gamma} = \pm f_{\mu\nu\gamma}$:
\begin{eqnarray}
  \label{parallelity of the invariant vielbeins}
  \nabla_\mu^\pm e^{\pm}_a
  &=&
  0
  \,,
\end{eqnarray}
or equivalently
\begin{eqnarray}
  \label{Lie connection and torsion}
  \omega[e^\pm]_{a^\pm b^\pm c^\pm} &=& \mp T_{a^\pm b^\pm c^\pm}
  \,.
\end{eqnarray}

\newpage
\bibliography{std}

\end{document}